\documentclass[12pt,a4paper]{article}
\pdfoutput=1
\usepackage{jheppub}
  
\usepackage[utf8x]{inputenc}
\usepackage{ucs}
\usepackage{fontenc}

\usepackage{multicol}
\usepackage{framed}
\usepackage[section]{placeins}   
\usepackage{dcolumn,makecell,booktabs,multirow}   
\usepackage[shortlabels]{enumitem}  
\usepackage{float}

\usepackage{amsmath,amssymb,mathtools,empheq,slashed,braket,array}
\usepackage{color}
\usepackage{graphicx}
\usepackage{setspace}
\usepackage{subfigure}
\usepackage{hyperref}

\usepackage{textcomp,starfont,bbold,bm}

\usepackage[printonlyused]{acronym}  
\usepackage[table]{xcolor}
\usepackage[compat=1.1.0]{tikz-feynman}
\usepackage{environ}        
\hypersetup{
    colorlinks,
    citecolor=blue,
    filecolor=blue,
    linkcolor=blue,
    urlcolor=blue
}

\usepackage{accents}

\newcommand{\leri}[1]{\left(#1 \right)}

\makeatletter
\AtBeginDocument{
	\renewcommand*{\AC@hyperlink}[2]{
		\begingroup
		\hypersetup{hidelinks}
		\hyperlink{#1}{#2}
		\endgroup
	}
}
\makeatother

\DeclarePairedDelimiter\abs{\lvert}{\rvert}
\DeclarePairedDelimiter\norm{\lVert}{\rVert}

\makeatletter
\let\oldabs\abs
\def\abs{\@ifstar{\oldabs}{\oldabs*}}

\let\oldnorm\norm
\def\norm{\@ifstar{\oldnorm}{\oldnorm*}}
\makeatother

\begin{document}


\newacro{SM}{Standard Model}
\acrodef{CTTP}{colored twisted top partner}
\acrodef{QCD}{Quantum Chromodynamics}
\acrodef{EM}{Electromagnetism}
\acrodef{LHC}{Large Hadron Collider}
\acrodef{DY}{Drell-Yan}
\acrodef{PDF}{parton distribution function}
\acrodef{HSCP}{heavy stable charged particle}
\acrodef{MCHSP}{multiply-charged heavy stable particle}
\acrodef{COM}{Center of Mass}
\acrodef{TOF}{time of flight}
\acrodef{ID}{inner detector}
\acrodef{MS}{muon system}
\acrodef{SSB}{Spontaneous Symmetry Breaking}
\acrodef{CMS}{Compact Muon Solenoid}
\acrodef{ATLAS}{A Toroidal LHC ApparatuS}
\acrodef{VBF}{vector boson fusion}
\acrodef{EW}{electroweak}
\acrodef{LO}{leading order}
\acrodef{RPC}{resistive plate chamber}
\acrodef{PACT}{pattern comparator trigger}
\acrodef{CSC}{cathode strip chamber}
\acrodef{HCAL}{hadronic calorimeter}
\acrodef{CL}{confidence level}
\acrodef{NLO}{next to leading order}

\title{Bounds and Prospects for Stable Multiply Charged Particles at the LHC}

\affiliation[a]{Department of Physics and Astronomy,
University of Sussex,\\ Brighton, BN1 9QH, UK }
\affiliation[b]{Department of Particle Physics and Astrophysics,
Weizmann Institute of Science,\\ Rehovot 7610001, Israel \vspace{2mm} }

\author[a]{Sebastian J\"{a}ger,}
\author[a]{Sandra Kvedarait\.e,}
\author[b]{Gilad Perez,}
\author[b]{Inbar Savoray}

\emailAdd{S.Jaeger@sussex.ac.uk}
\emailAdd{S.Kvedaraite@sussex.ac.uk}
\emailAdd{gilad.perez@weizmann.ac.il}
\emailAdd{inbar.savoray@weizmann.ac.il}

\abstract{Colored and colorless particles that are stable on collider scales and carry exotic electric charges, so-called \acp{MCHSP}, exist in extensions of the Standard Model, and can include the top partner(s) in solutions of the hierarchy problem. To obtain bounds on color-triplets and color-singlets of charges up to $|Q|=8$, we recast searches for signatures of two production channels: the ``open" channel -- where the particles are pair-produced above threshold, and are detectable in dedicated LHC searches for stable multiply charged leptons,  and the ``closed" channel -- where a particle-antiparticle pair is produced as a bound state, detectable in searches for a diphoton resonance. We recast the open lepton searches by incorporating the relevant strong-interaction effects for color-triplets. In both open and closed production, we provide a careful assessment of photon-induced processes using the accurate LUXqed \acs{PDF}, resulting in substantially weaker bounds than previously claimed in the literature for the colorless case. Our bounds for colored \acp{MCHSP} are shown for the first time, as the LHC experiments have not searched for them directly. Generally, we obtain nearly charge-independent lower mass limits of around $970$ GeV (color-triplet scalar), $1200$ GeV (color-triplet fermion), and $880-900$ GeV (color-singlet fermion) from open production, and strongly charge-dependent limits from closed production. In all cases there is a cross-over between dominance by open and closed searches at some charge. We provide prospective bounds for $\sqrt{s}=13 \mbox{~TeV}$ LHC searches at integrated luminosities of $39.5$ fb${}^{-1}$,  $100$ fb${}^{-1}$, and $300$ fb${}^{-1}$. Moreover, we show that a joint observation in the open and the closed channels allows to determine the mass, spin, color, and electric charge of the particle.}

\maketitle
\acresetall	

\section{Introduction}

Extensions of the \ac{SM} often contain particles that are stable, or sufficiently long-lived to be effectively stable on the time and distance scales relevant to collider
experiments. Examples include the lightest supersymmetric particle if $R$-parity is approximately or exactly conserved (see 
\cite{Martin:1997ns} for a review) and particles in certain composite Higgs models~\cite{Agashe:2004ciDM}.
It is possible that such a particle has exotic and possibly large electric charge; we will refer to this as a \ac{MCHSP}.

\begin{figure}[b!]
	\centering
	\subfigure[]{\includegraphics[scale=1]{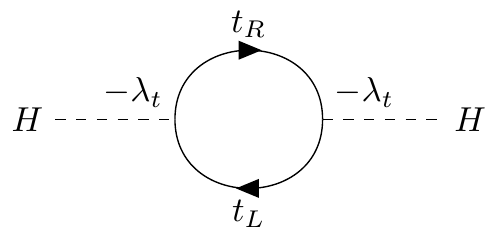}\label{diagram:toploop}}
	\subfigure[]{\includegraphics[scale=1]{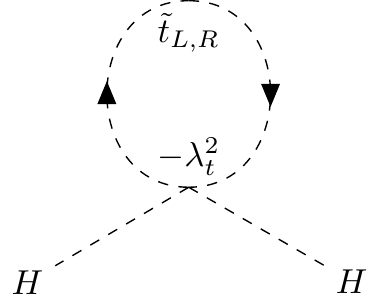}\label{diagram:scalarloop}}
	\subfigure[]{\includegraphics[scale=1]{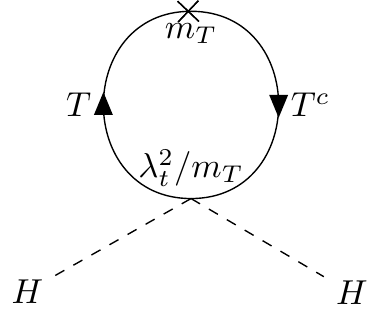}\label{diagram:fermionloop}}
	\caption{(a) Divergent top loop correction to the Higgs mass. (b) Loop contribution of a scalar top-partner. (c) Loop contribution of a fermion top-partner. The diagrams are taken from~\cite{CTTP}.}
	\label{top_loops}
\end{figure}

Within the context of the naturalness problem (see e.g~\cite{naturalness}), such \ac{MCHSP} can cure the quadratic divergence in the Higgs mass parameter; this has recently been realized in the framework of \acp{CTTP}~\cite{CTTP}. The \ac{CTTP} can take the form of a spin-0 or spin-$1/2$
color-triplet of arbitrary electric charge. The divergence cancellation occurs between the top loop in Fig.~\ref{diagram:toploop}, and a scalar \ac{CTTP} loop (Fig.~\ref{diagram:scalarloop}) or a fermion loop (Fig.~\ref{diagram:fermionloop}). The \ac{CTTP} is stable either due to an (approximate) accidental $U(1)$ symmetry, conserving partner-number, or due to an (approximate) $Z_2$ symmetry, under which the \ac{CTTP} is odd and all SM particles are even.
In fact, \acp{CTTP} of charges different from $Q=2/3+n$ or  $Q=-(1/3+n)$, where n is a non-negative integer, are not allowed to decay to \ac{SM} particles altogether~\cite{Colored}. Consequently, exotically charged top partners are likely to be stable or long lived. 

Motivated by the above, we will consider color-triplet particles with arbitrary electric charges, and refer to them as \acp{CTTP}, or ``partners", irrespective of whether they are connected to naturalness or not.
An important implication of their long lifetime is the presence of a near-threshold, positronium-like bound state. In the top partner case, this is known as the partnerium \cite{CTTP}, and we will use this term to denote the bound state in the generic case. The colored particle-antiparticle pair is bound by both a Coulomb-like \ac{QCD} potential and by \ac{EM}, with the latter becoming important for large charges. Since partnerium carries no conserved charge, it is free to annihilate into \ac{SM} particles, leaving potentially detectable signatures, the most relevant of which, for our purposes, is a diphoton resonance.

In addition to the bound-state production (referred to as ``closed"), the stable (or long-lived) partner can be pair-produced above threshold (referred to as ``open"), leaving tracks in all detector layers and eventually escaping without an observed decay. Color-triplet top partners with charges different than 2/3 have not been directly searched for at the \acs{LHC}, and are largely unconstrained. In this work, we obtain current bounds on exotically-charged scalar and fermion \acp{CTTP}, considering both open pair production and partnerium signatures. We also obtain prospective bounds, for future \acs{LHC} searches, at several integrated luminosities and \ac{COM} energy of 13~TeV. We choose to focus on multiply-charged ($|Q|>1$) color-triplet top partners, which are expected to exhibit an interesting interplay between the two channels, especially given their sizable partnerium-annihilation to a pair of photons. In addition, we consider color-singlet fermion \acp{MCHSP}, referred to as lepton-like particles. In this case, the bound state is purely \ac{EM}, referred to as ``leptonium". We restrict ourselves to $SU(2)_\text{weak}$ singlets, both for colored and colorless \acp{MCHSP}. 

The remainder of the paper is organized as follows. In Section \ref{sec:opensearches} we discuss the open-production signatures of \acp{MCHSP}, and consider the existing run-I ($\sqrt{s}=8 \mbox{~TeV}$) LHC searches for color-neutral stable particles with large electric charges. In order to recast these searches for colored particles, and to update their results for colorless particles, we compute the production cross sections and the detection efficiencies for both spinless and spin-$1/2$ color-triplets, and for colorless fermions, all with charges $Q$ in the range $1\leq|Q|\leq8$ and masses $m$ in the range $100 \,\mbox{GeV} \leq m \leq 3\, \mbox{TeV}$. We validate our methodology against the published efficiencies in the colorless case. We also obtain the required components for the prospective $\sqrt{s}=13$~TeV searches. Section~\ref{sec:partnerium} reviews the pertinent aspects of the bound state signatures, in particular the resonant-production cross section of a diphoton final state. Section~\ref{sec:currentbounds} contains our main findings, in the form of current lower limits on the masses of colored and color-neutral particles. For the color-neutral case, we obtain weaker constraints than a recent paper, albeit stronger than the bounds originally obtained by CMS; we trace these discrepancies to the photon-induced component of the signal and stress the importance of an appropriate choice of the photon \ac{PDF}. In Section~\ref{sec:projections}, we present projected bounds for \acs{LHC} searches at $\sqrt{s}=13 \mbox{~TeV}$, for integrated luminosities of $39.5 \mbox{~fb}^{-1}$, $100 \mbox{~fb}^{-1}$, and $300 \mbox{~fb}^{-1}$, taking into account the scaling of pileup. We briefly discuss how by combining an open-production effective cross section measurement and a diphoton resonance observation one can determine the mass, spin, electric charge and color charge of the particle. Our conclusions can be found in Section \ref{sec:conclusions}.

\section{Stable Multiply-Charged Particles at the LHC}\label{sec:opensearches}

Our first goal is to obtain constraints on \acp{CTTP} from their signatures as stable particles, produced above threshold. So far, there have been no \acs{LHC} searches designated for color-triplet \acp{MCHSP}. However, there have been experimental searches for other kinds of \aclp{HSCP}, which could be potentially recast to apply to \acp{CTTP}. 

The stable fermion and scalar color-triplet partners are expected to hadronize to form "R-hadrons", similarly to quarks and squarks~\cite{interactions_hadronizing}. Searches for stable R-hadrons have been carried out both in ATLAS~\cite{RHadATLAS13,RHadATLAS8,RHadATLAS7}, and in CMS~\cite{StableCMS13, StableCMS7and8,StableCMS7} for \ac{COM} energies of 7, 8 and 13~TeV. However, these searches are designated for stops and gluinos, and thus optimized for unit-charged R-hadrons. Applying such searches for multiply-charged R-hadrons could bear a significant loss of the discovery potential.

Searches for multiply-charged color-singlet fermions account for the difficulties concerning the detection of \acp{MCHSP}. These searches were conducted by ATLAS for particles with charges of 2-6~\cite{atlas_multi_8}, and conducted by CMS for  particles with charges of 1-8~\cite{StableCMS7and8}. Both searches were analyzed for $\sqrt{s}=8$~TeV, but have yet to be updated for $\sqrt{s}=13$~TeV. Results for a $Q=2$ lepton-like particle have been published by CMS for $\sqrt{s}=13$~TeV, following an analysis that uses the same discriminators as for R-hadrons~\cite{StableCMS13}. However, the resulting bound was less stringent than the one derived from the designated search for multiply-charged lepton-like particles, carried out for $\sqrt{s}=8$~TeV.

As the aforementioned searches were carried out for colorless fermions only, heavy stable \acp{CTTP} are still essentially unconstrained. While multiply-charged scalar and fermion \acp{CTTP} are expected to share a lot of phenomenological traits with multiply-charged leptons, \acs{QCD}-induced processes for color-triplets still need to be accounted for. First, one should consider the appropriate production mechanism, both for cross section and for efficiency calculations. Second, the hadronization of the colored particle-pair might yield two differently charged R-hadrons, and thus change the event acceptance. Moreover, nuclear energy loss and charge-changing effects~\cite{interactions_hadronizing} might further reduce the efficiency of the search. Therefore, the existing analyses are not sufficient for obtaining bounds on stable \acp{CTTP}. 

Furthermore, the previous analyses for colorless fermions might be lacking. As shown in the re-analyses of the ATLAS search~\cite{atlas_multi_8} in~\cite{Leptonium}, the bounds on multiply-charged particles are sensitive to the treatment of photo-induced processes, which were not included in the original \acs{LHC} analyses. However, the \ac{PDF} used in~\cite{Leptonium} has been shown to have large uncertainties for the photon \ac{PDF} and thus also for the photon luminosity~\cite{LUX,NNPDF,Aad:2016zzw}. This translates into large uncertainties on the previously obtained bounds. A more accurate determination of the photon \ac{PDF} using $ep$ scattering data was proposed in ref.~\cite{LUX,Manohar:2017eqh}, resulting in significantly smaller errors, which are at the $1\%$ level over a large range of momentum fractions. For these reasons, we would like to reanalyze the signatures of \acp{MCHSP} using the resulting LUXqed \ac{PDF}~\cite{LUX}.

This motivates us to recast a search for lepton-like \acp{MCHSP}, in order to apply its observations to fermion and scalar \acp{CTTP}, and to update the bounds on lepton-like particles. The rest of this section is dedicated to describing our recast procedure.

We chose to recast the most recent CMS search for lepton-like particles with charges of 1-8~\cite{StableCMS7and8}\footnote{The corresponding ATLAS search~\cite{atlas_multi_8} resulted in similar bounds, and should have the same qualitative efficiency behavior, however it was only applied to $Q \leq 6$.}. Since the search is a counting experiment, essentially blind to mass and charge, it is imposing a universal upper~limit on the product of the cross section and the efficiency, $\sigma \cdot \epsilon$. This ``effective cross section" upper limit is then compared to its theoretical prediction for each signal benchmark, described below, to obtain the upper bounds on the signal mass. In the next sections, we discuss our calculations of the cross sections and efficiencies separately, which are later combined to obtain the theoretical effective cross sections. As the search is only available for $\sqrt{s}=7\&8$~TeV, we calculate the bounds based on the observed result at $\sqrt{s}=8$~TeV, and estimate the expected bounds for $\sqrt{s}=13$~TeV.

For convenience, our signal benchmarks are based on the charges already considered in the original search. Namely, color-singlets with integer charges $\abs{Q_\text{LLP}}=1-8$ and color-triplets that hadronize to acquire such charges, initially charged as: $5/3 \leq Q_\text{CTTP}\leq 23/3$ and $-22/3\leq Q_\text{CTTP}\leq-4/3$, in increments of one. We did not include charges of~$-1/3\text{~and~}2/3$ in our analysis, as those were better studied in stable R-hadrons searches. Charges of 26/3 and -25/3 were disregarded due to their sizable hadronization fraction to $\abs{Q_\text{R-hadron}}=9$ particles, that were not included in the original search. It has been shown in \cite{Leptonium,ProbingKats:2012ym} that particles with such large charges can still be treated perturbatively as long as the coupling is sufficiently small and the energy domain is well below the Landau pole. This is ensured when $\alpha Q^2\lesssim \mathcal{O}(1)$. As the theory loses perturbativity for  $\alpha Q^2\gtrsim \mathcal{O}(1)$, our predictions could not be straightforwardly extrapolated for $Q\gtrsim10$. Since both the observations and the selections of the search are common to all masses and charges, one can easily interpolate our results for any intermediate charge. 

The masses of the signal benchmarks were determined in a similar fashion. Since the original search considered masses of $100-1000$~GeV, lepton-like particles of the same masses were generated in a Monte-Carlo simulation, described in the following, in order to estimate the accuracy of the efficiency calculation. Bounds were calculated for particles of masses $500-3000$~GeV.

\subsection{Recalculating Production Cross Sections}\label{production_cross_section}

The pair-production cross section of \acp{CTTP} is calculated by summing the contributions from the $gg,g\gamma$ and $\gamma\gamma$ \ac{VBF} production channels, as well as from the $q\bar{q}$ \ac{DY} production channel, mediated by $g,\gamma \text{ or }Z$. The calculation of the pair-production cross section of lepton-like particles accounts for production both by photon-fusion and by a \ac{DY} process mediated by $\gamma\text{ or }Z$. In contrast to both the original search~\cite{StableCMS7and8} and to a re-interpretation of the ATLAS search~\cite{atlas_multi_8} in~\cite{Leptonium}, all cross sections below are calculated with the LUXqed \ac{PDF} set (LUXqed17\_plus\_PDF4LHC15\_nnlo\_100)~\cite{LUX,Manohar:2017eqh}. We use MadGraph5~\cite{mg5} to calculate the parton-level cross section at \acs{LO}. The resulting cross sections are presented in Appendix~\ref{cross_section_plots}.

\begin{figure}[t!]
	\centering
	\includegraphics[scale=0.43]{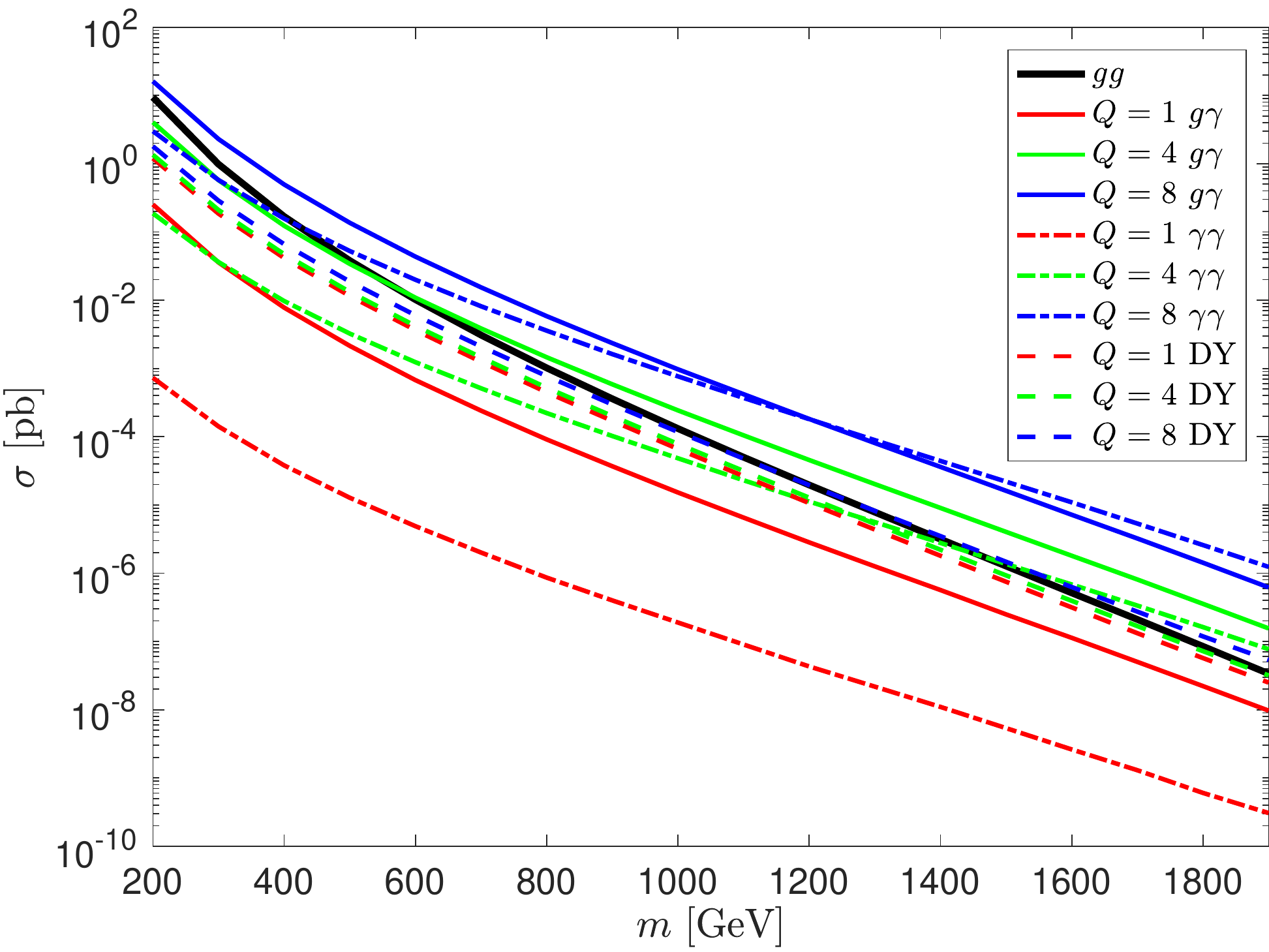}
	\caption{Different subprocesses for pair-production of a scalar \ac{CTTP} with charges of $Q=1,4,8$.}
	\label{subprocesses_scalar}
\end{figure}

The relative importance of the different production channels is highly affected by the \ac{PDF} of the incoming partons. Photon-induced charge-dependent \ac{VBF} processes are suppressed by the smallness of the photon \ac{PDF}, while charge-independent gluon-fusion processes benefit from the large \ac{PDF} of the gluon. Since the ratio between the gluon \ac{PDF} and the photon \ac{PDF} is slightly smaller at higher energies, a large charge-dependent contribution could eventually overcome the \acp{PDF} imbalance. Thus, as shown in Fig~\ref{subprocesses_scalar}, heavier particles with large charges will mostly be produced by photon-inclusive, highly charge-dependent processes, and lighter particles with small charges will mostly be produced by charge-independent processes.

We use Pythia8~\cite{pythia82,pythia64} to perform showering and hadronization. As can be seen in Table~\ref{R-hadron_fractions}, hadronized partners mainly have charges of $\pm(Q+1/3)$ and $\pm(Q-2/3)$, with only a negligible fraction of $\pm(Q+4/3)$ R-hadrons. Since hadronization of the heavy partner and anti-partner takes place mostly independently, they may hadronize into two differently charged R-hadrons.

\begin{table}[b!]
	\centering
	\begin{tabular}{|l|r|}
		\hline
		R-hadron & Fraction (\%)  \\
		\hline
		\hline
		$R^{Q+1/3}$    & 28.25 \\
		$R^{Q-2/3}$    & 21.50 \\
		$R^{Q+4/3}$   & 0.25 \\
		$\bar{R}^{-(Q+1/3)}$ & 26.75 \\
		$\bar{R}^{-(Q-2/3)}$ & 23.00 \\
		$\bar{R}^{-(Q+4/3)}$ & 0.25 \\
		\hline
	\end{tabular}
	\caption{Fractions of produced R-hadrons with specific charges, obtained using MadGraph and Pythia simulation of partner pair-production and hadronization. }
	\label{R-hadron_fractions}
\end{table}

\subsection{Efficiency Calculation} \label{sec:efficiency}
Since we do not have access to the full CMS detector simulation, we defined a set of selection criteria to account for detection efficiencies. Using our efficiency calculation, with the production mechanism described in~\cite{StableCMS7and8}, we aim to reproduce the mass bounds obtained by CMS for lepton-like particles within $15\%$ accuracy. A similar accuracy should be maintained as we calculate the bounds on the masses of \acp{CTTP}, and of lepton-like particles produced as in Section~\ref{production_cross_section}. We account for the online, offline and final selections criteria, as will be explained in the following paragraphs. Even though our treatment is somewhat rough, we will see it is more than satisfactory for obtaining mass bounds, as they are only weakly affected by efficiencies. 

\subsubsection{Procedure} 
The online selection for the search~\cite{StableCMS7and8} consists of an $E_T^\text{miss}$ trigger and/or a muon trigger. To pass the $E_T^\text{miss}$ trigger, an event should be assigned $E_T^\text{miss} \geq 150$~GeV as measured in the calorimeter. This criterion is useful to some extent for particles that were not reconstructed as muons, but we expect it to have a negligible contribution to the overall efficiency, since the offline and final selections essentially require a muon candidate. 

We therefore focus on simulating the muon trigger as our online selection. To pass the muon trigger requirements, an event must have at least one particle reconstructed as a muon. The muon candidate must have  $\eta \leq 2.1$, and $p_{T_\text{meas}}\geq40$~GeV as measured in the \ac{ID}. The transverse momentum is measured from the curvature radius of the particle's track,  $r$, under a magnetic field, $B$, which follows 
\begin{align}
r=\dfrac{p_T}{0.3\cdot Q\cdot B}\,.
\end{align}
However, the reconstruction algorithm assumes $Q=1$, and so the measured $p_T$ is $p_{T_\text{meas}}=p_{T_\text{truth}}/Q$. This effectively requires the truth-level transverse momentum to satisfy $p_{T_\text{truth}} \geq  Q\cdot \leri{40 ~\text{GeV}}$, thus reducing the efficiency for large charges and small masses. 

In addition, triggering particles must be fast enough to have both their \ac{ID} and \ac{MS} tracks in the same bunch crossing~\cite{StableCMS72012}. Since the \acs{LHC} collisions were planned to occur every 25~ns, slow particles that reach the \ac{MS} more than 25~ns after a $\beta=1$ particle, will be associated with the wrong bunch crossing and thus will not have a matching \ac{ID} track~\cite{thesis}. An additional \ac{RPC} muon trigger was applied for $\eta\leq1.6$, allowing candidates to reach the \ac{MS} up to 50 ns later than a $\beta=1$ particle~\cite{RPCtrigger}.

\ac{RPC}-triggered particles must have a minimum of four \ac{RPC} hits (three if not geometrically possible) within the trigger time window~\cite{RPCtrigger,RPCLS1}. A similar requirement also holds for particles triggered by the \acp{CSC} positioned at $\eta\geq 1.6$, as the \ac{CSC} trigger relies on three different track segments to reconstruct $p_T$~\cite{HLTtriggers}. These constraints effectively define a minimal distance, denoted as $x_\text{trigger}$, that candidates must travel within the trigger time window, as function of $\eta$.

In order to calculate the time required for a candidate to travel the distance necessary for triggering, denoted as $t_\text{TOF}$, one must account for the ionization energy loss in the \ac{HCAL} and in the \ac{MS}. Following the Bethe-Bloch formula~\cite{PDG}, the ionization energy loss rate decreases with the velocity of the particle and quadratically increases with its charge. Therefore, the timing requirement is expected to be crucial for \acp{MCHSP}, that are both produced with smaller velocities and significantly slowed down, or even stopped, by ionization energy loss. 

\begin{figure}[b!]
	\centering
	\includegraphics[scale=0.43]{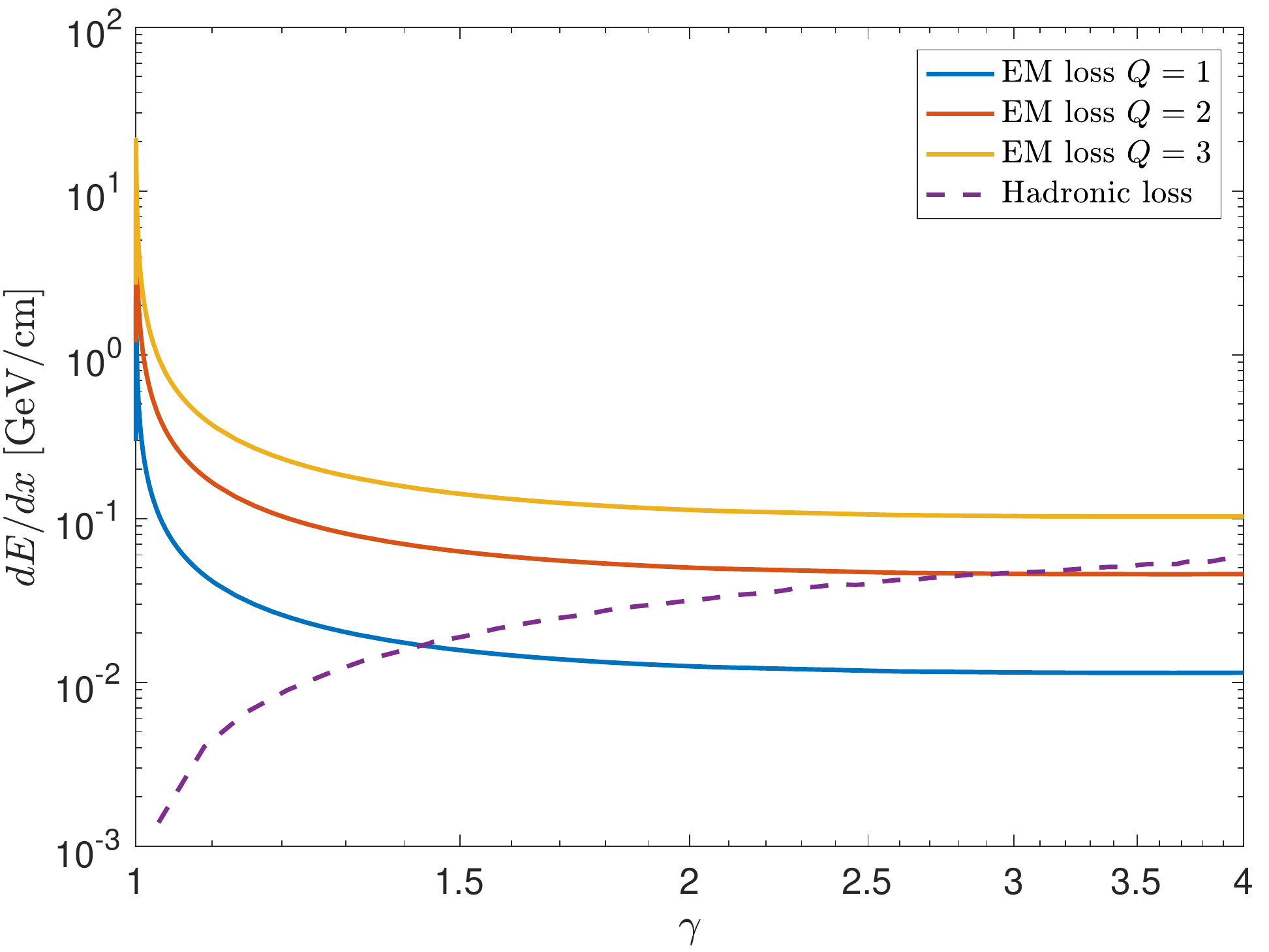}
	\caption{Energy loss per distance traveled in iron as a function of  $\gamma$. \textit{Solid}~-~ionization energy loss for $Q=1,2,3$~\cite{iron_energy_loss}. \textit{Dashed}~-~average nuclear energy loss for a hadronized stable stop~\cite{hadrons_thesis}.}
	\label{ionization_energy_loss}
\end{figure}

Heavy R-hadrons may also undergo nuclear interactions with matter, causing additional energy loss and  potentially altering the quark content of the R-hadron, resulting in a charge change~\cite{interactions_hadronizing}. However, as can be seen in Fig.~\ref{ionization_energy_loss},  for slow particles with large charges, nuclear energy loss is quite negligible compared to ionization energy loss, and hence could be ignored. Since we did not have access to a reliable simulation of charge-changing processes, we could not account for them in our analysis. As we would expect these processes to cause some efficiency loss, it would be desirable to include them in a full experimental study. The calculation of $t_\text{TOF}$ is further explained in Appendix.~\ref{TOF}. 

Candidates in events passing the online selection are subject to an offline selection specified in Tables 1-2 of~\cite{StableCMS7and8}, applied at particle level. Our offline efficiency calculation is rather limited, and only explicitly includes $p_T$ and isolation criteria, as described in lines 4-5 of Table~\ref{efficiency_simulating_table}. An additional selection requires the particle to be reconstructed as a global muon~\cite{global_muon}, filtering out particles that were not identified as muons at the muon trigger level. Therefore, we replaced the global muon selection by only accepting candidates that individually satisfy the online muon trigger requirements, as defined above. This assumption is further justified in Appendix~\ref{global_muon}. Since we cannot account for the remaining criteria without a full detector simulation, we use the values quoted in Tables C1-C16 of~\cite{thesis} as multiplicative factors for the offline efficiency calculation. A factor for each signal mass and charge is calculated by 
\begin{align}
\epsilon^\text{sim}_\text{offline}=\frac{\epsilon_\text{offline}}{\epsilon_\text{global muon}\cdot \epsilon_{p_T}\cdot \epsilon_\text{isolation}}\,,
\end{align}
where $\epsilon_\text{offline}$ is the fraction of particles passing the offline selection, out of all particles from events that passed the online selection. The efficiencies $\epsilon_\text{global muon}\text{, }\epsilon_{p_T}\text{, }\epsilon_\text{isolation}$ correspond to the fractions of particles passing the global muon, $p_T$ and $\sum _{R\leq 0.3} p_T$ requirements, respectively, out of the particles passing all selections imposed prior to them (online selection included). The aforementioned values were given  in~\cite{thesis} for lepton-like particles of charges 1-8 and masses of 100-1000~GeV. Since they vary weakly with mass, we use $m=1000$~GeV efficiencies for all $m\geq 1000$~GeV particles.

Lastly, the signal region is determined by the final selection criteria, presented in Table 3 (line 4) of~\cite{StableCMS7and8}. We include the $1/\beta \geq 1.2$ selection in our criteria, designed to identify slow particles, and calculate it using the \ac{TOF} defined in Eq.~\ref{tof_eq}. Since we cannot recreate the $I_\text{as}$ selection, we expect our efficiency to be overestimated for unit-charge particles. However, particles with larger charges are not affected~\cite{thesis}.

Our efficiency calculation may require adjustment for $\sqrt{s}=13$~TeV. In the absence of \acp{MCHSP} searches at $\sqrt{s}=13$~TeV, we have to make certain assumptions about how the selection criteria will change. The choice of $p_T$ thresholds is taken from the $\sqrt{s}=13\text{~TeV}$ search for unit-charged \aclp{HSCP}~\cite{StableCMS13}, since the corresponding $\sqrt{s}=8$~TeV searches for multiply-charged and unit-charged particles had the same $p_T$ requirements. We had no reliable estimate of how the offline and the final selections might be modified for 13 TeV. We therefore kept them the same as in 8 TeV searches, noting that the offline efficiencies given in~\cite{thesis} for the 7~TeV and the 8~TeV runs show only a weak dependence on the masses and \ac{COM} energies. 

The efficiency calculation steps and criteria are summarized in Table~\ref{efficiency_simulating_table}. Events that pass those criteria are assumed $100\%$ efficiency, as our calculation does not account for trigger inefficiencies and other hardware effects. The final efficiencies for the signal benchmarks mentioned above are given in Appendix~\ref{efficiency_values}.

\renewcommand{\arraystretch}{1.8}
\begin{table}[t!]
	\centering
	\begin{tabular}{|c|c|c|}
		\hline
		& \textbf{8~TeV} & \textbf{13~TeV}\\ \hline
		\multirow{3}{*}{\textbf{Online}} &$p_T\geq Q\cdot 40$~GeV & 	$p_T\geq Q\cdot 50$~GeV\\ \cline{2-3}
		& \multicolumn{2}{c|}{$\abs{\eta} \leq 2.1$}\\ \cline{2-3}
		& \multicolumn{2}{c|}{$t_\text{TOF}-\dfrac{x_\text{trigger}}{c} \leq$ 50~ns (25~ns)}\\ \hline
		\multirow{2}{*}{\textbf{Offline}}&$p_T\geq Q\cdot 45$~GeV & 	$p_T\geq Q\cdot 55$~GeV\\ \cline{2-3}
		& \multicolumn{2}{c|}{$\sum _{R\leq 0.3} p_T \leq 50$~GeV}\\ \hline
		\textbf{Final} &  \multicolumn{2}{c|}{$\frac{c\cdot t_\text{TOF}}{x_\text{trigger}} \geq 1.2$}\\ \hline
		\multirow{2}{*}{\textbf{Multiplicative Factor}} & \multicolumn{2}{c|}{$\epsilon^\text{sim}_\text{offline}(Q,m)$\,, $m\leq 1000$~GeV}\\
		& \multicolumn{2}{c|}{$\epsilon^\text{sim}_\text{offline}(Q,1000)$\,, $m>1000$~GeV}\\ \hline
	\end{tabular}
	\caption{Simplified efficiency calculation steps and criteria used in this analysis. Each step is applied only to candidates passing the selections in the steps above it. The online timing requirement is 50 ns for $|\eta| \leq 1.6$ and 25 ns for $|\eta|>1.6$. The multiplicative factor accounts for the offline selection criteria, which are not explicitly simulated, and instead the efficiencies associated with them are taken from~\cite{thesis}. More details in text.}
	\label{efficiency_simulating_table}
\end{table}

\subsubsection{Validation}
\begin{figure}[h!]
	\centering
	\subfigure[]{
		\includegraphics[scale=0.37]{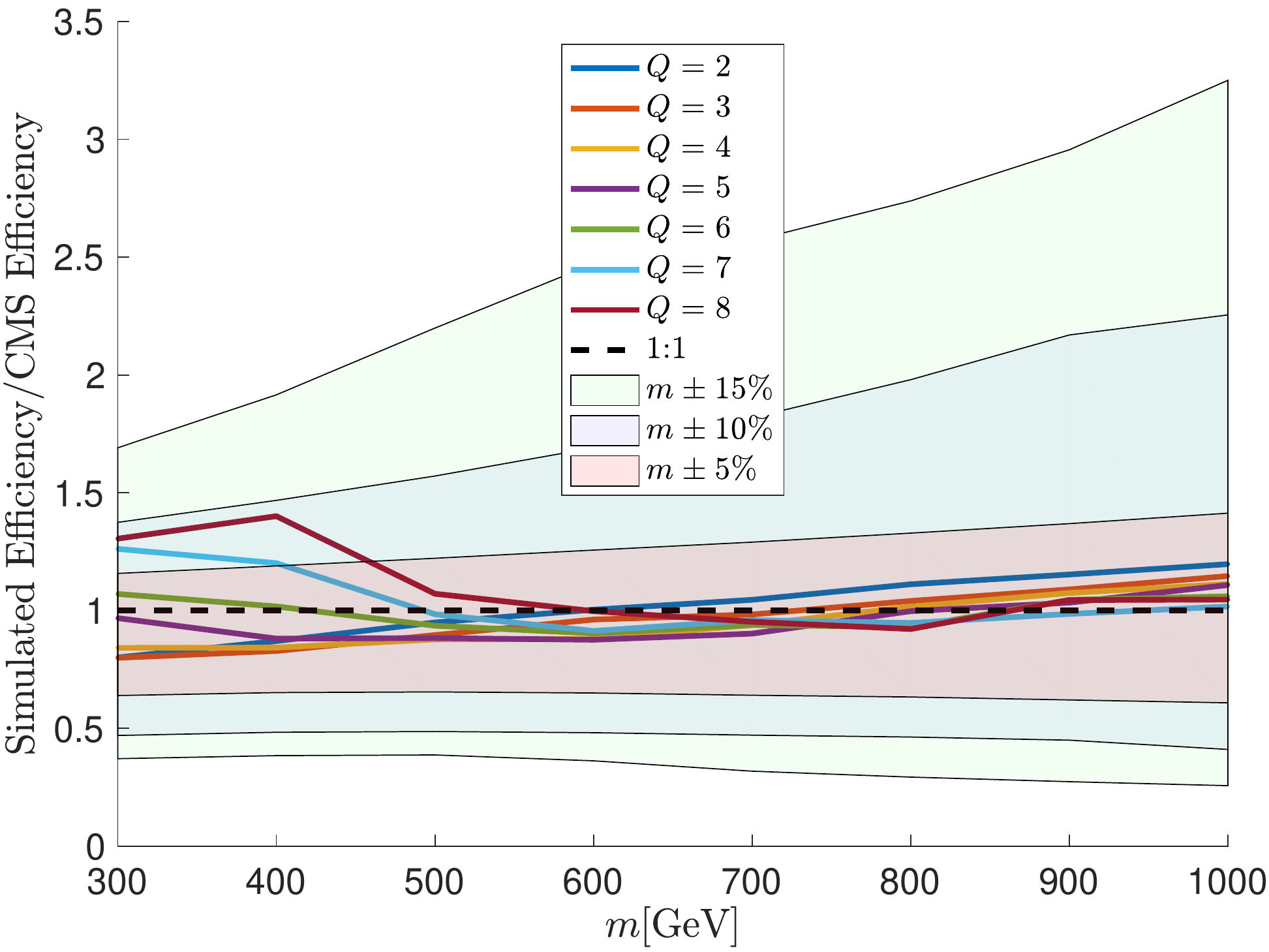}\label{full_ratio_yoke_shower_full_8}}
	\subfigure[]{
		\includegraphics[scale=0.37]{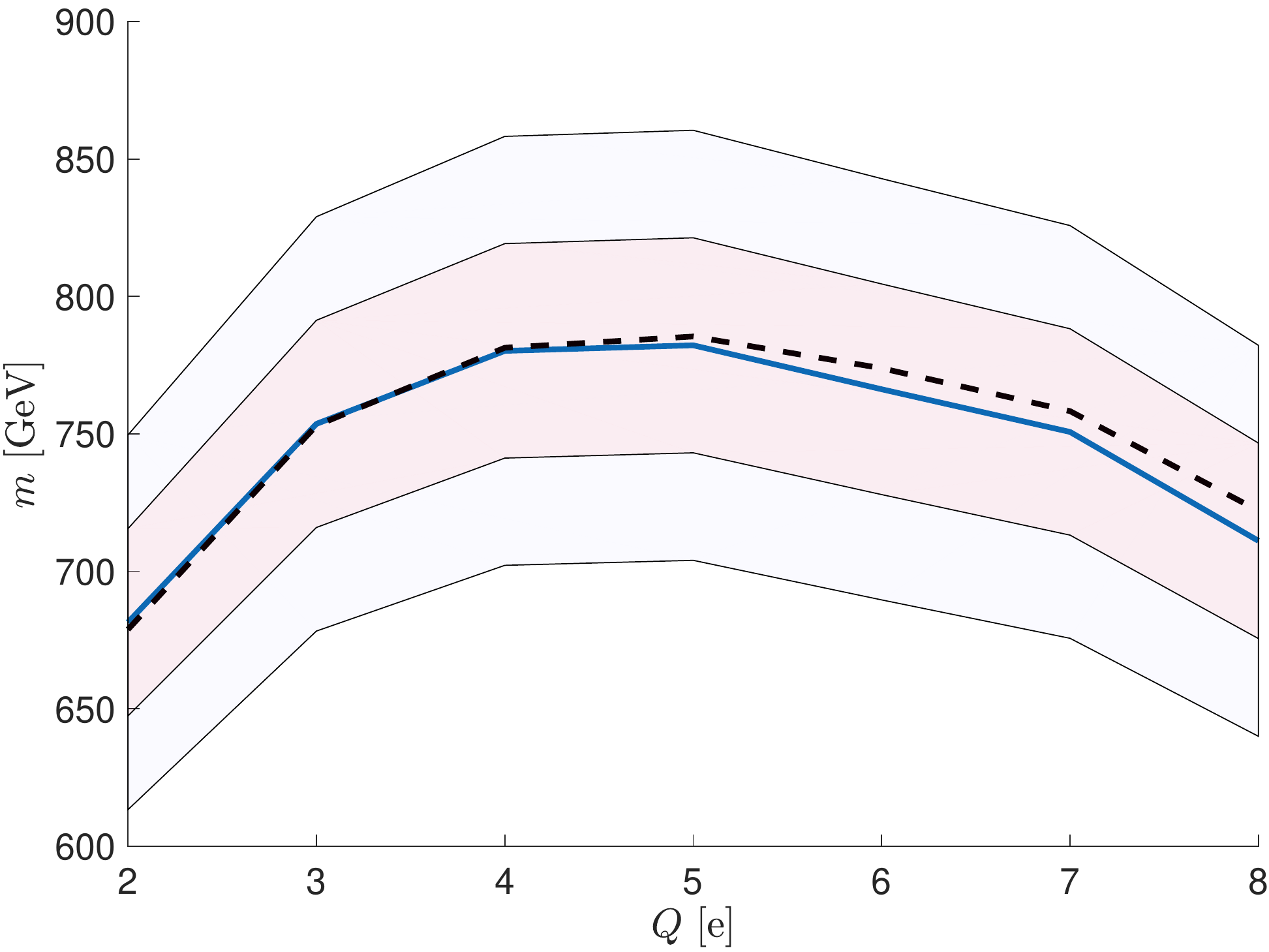}\label{masses_compare}}
	\label{efficiency_final_compare}
	\caption{Simplified efficiency calculation validation.
		(a) The ratio between our resulting efficiencies and the respective CMS efficiencies for $\sqrt{s}=8$~TeV~\cite{StableCMS7and8},~\cite{thesis}. Indicated as well are the efficiency deviation bands corresponding to less than 5\% (\textit{red}), 10\% (\textit{light blue}) and 15\% (\textit{light green}) deviation in the mass bound.
		(b) Reproduced mass bounds for lepton-like particles, following the production mechanism used by CMS. \textit{Dashed} -- the bounds published by CMS~\cite{StableCMS7and8}, using a full detector simulation. \textit{Solid} -- our results using the simplified efficiency calculation. Indicated as well are the 5\% (\textit{red}) and 10\% (\textit{light blue}) mass deviation bands, around the our final mass bounds plot. }
\end{figure}

We compare the overall efficiencies, obtained by our simplified calculation, to the total efficiencies given in~\cite{thesis, StableCMS7and8}. For this purpose, we follow the production prescription in the original analysis by CMS, and generate lepton-like particles by \ac{DY} processes with CTEQ6L1 \acp{PDF}~\cite{CTEQ}. The ratio of the efficiencies is presented in Fig.~\ref{full_ratio_yoke_shower_full_8} for 8~TeV, and a relatively good agreement is established. We find that our efficiency and the results by CMS are less than 40\% apart, for all charges for masses larger than 300~GeV. 

As the cross sections for pair-produced \acp{MCHSP} drop sharply with their mass, the final mass bounds are only weakly sensitive to the exact upper limits on the effective cross section. Therefore, inaccuracies in the efficiency estimation would result in much smaller deviations in the mass bounds. The mass bounds resulting from our efficiency calculation are expected to differ from the corresponding bounds calculated with the full detector simulation by less than 10\% for smaller masses, and by much less than 5\% for masses larger than 500 GeV. Indeed, as shown in Fig.~\ref{masses_compare}, we were able to reproduce the mass bounds for lepton-like particles with excellent accuracy.

When comparing the efficiencies at the muon trigger level with the values given in~\cite{thesis}, we find that other than for $m=100$~GeV, we overestimate the intermediate efficiency by $5\%-40\%$.  There are additional effects, not included in our calculation, that might reduce the number of events passing the muon trigger selection. One such effect is the track reconstruction and matching. Heavy particles with large charges experience large ionization energy loss, and as a result are expected to be less compatible with a global muon pattern. Second, the trigger response and the gaps in the \ac{RPC} and \ac{CSC} coverage may increase the distance a candidate must travel to have a sufficient number of hits. Moreover, we do not consider background effects, both from pileup and from hard particles produced in the interaction, that could affect reconstruction. It may also be that we somewhat underestimate the material budget. However, the final selection filters out particles that are too fast, which are favored by the muon trigger. As a result, the overestimation of the muon trigger efficiency could be compensated, and the total efficiency is therefore in agreement with CMS. Even had these effects not canceled out, the final error for the mass bounds would still be smaller than 15\% for masses larger than 500 GeV.

\section{Bound State Signal at the LHC}
\label{sec:partnerium}
Our second goal is to obtain mass bounds on \acp{CTTP} from their signatures as partnerium bound states. In this section, we will discuss the salient features of the partnerium resonance, and introduce our recast procedure, which will be centered around diphoton channel. 

The partnerium is unstable due to the annihilation of its constituents, and can be detected as a resonance, with invariant-mass peak at $M \approx 2m_{\rm partner}\,$. 
A $J=0$ or $J=2$ partnerium state, made of \ac{EM}-charged constituents, can always decay through annihilation into $\gamma\gamma,\gamma Z$ and $ZZ$. In the case of the color-triplet \acp{CTTP}, it may also decay into a pair of gluons. A $J=1$ partnerium, consisting of fermions, can annihilate into $W^{+}W^{-}$~\cite{Leptonium}, or to any \ac{SM} fermion - anti-fermion pair, through $s$-channel $\gamma/Z$ exchange~\cite{Colored}. Moreover, if the constituent is a top partner, its large coupling to the Higgs implies significant annihilation rates into Higgs pairs and longitudinally polarized \ac{EW} gauge bosons (for $J=0$ or $2$ partnerium made of scalars), or to $hZ$ (for $J=1$ fermion bound states)~\cite{CTTP}. Out of these search channels, the diphoton signal is by far the most sensitive~\cite{CTTP,Colored}, especially for the large electric charges we consider. We will thus solely focus on this final state.

Several authors  have recast \acs{LHC} resonance searches to obtain bounds on \acp{CTTP}. Mass bounds for scalar and fermion \acp{CTTP} of charges $-1/3,2/3,-4/3,5/3$ can be inferred from the plots presented in ref.~\cite{CTTP}. In addition, the authors of~\cite{colored_scalars} obtained bounds for colored scalars with charges~$-7/3,8/3,-10/3$ and of different $SU(2)_\text{weak}$ representations. However, these analyses attributed the dominant partnerium production, binding and decay mechanisms to \ac{QCD}. This is not necessarily the case for partners with larger charges, as we will see. Ref.~\cite{Leptonium} contains the only available resonance analysis for charges 1-8, but is limited to colorless fermions bound in a ``Leptonium"~\cite{Leptonium}. As the leptonium diphoton signal is highly sensitive to the photon \ac{PDF}, we will also see that a more accurate \ac{PDF} choice can lead to significantly different conclusions. Thus, similarly to the open-production case, the existing analyses of partnerium-like signatures are insufficient for constraining the parameter space of \acp{MCHSP}. We therefore recast a diphoton resonance search, to obtain bounds on the masses of \acp{CTTP} and to update the corresponding bounds for lepton-like particles. 

Our recast is based on the latest diphoton search, at $\sqrt{s}=13$~TeV and an integrated luminosity of $35.9\text{ fb}^{-1}$, published by CMS~\cite{diphotoncms13}. As the efficiency of diphoton detection at a given invariant mass is mostly independent of the signal model, we kept it unmodified. We therefore only compute the diphoton production cross section, resulting from a partnerium or a leptonium resonance, accounting for both \ac{QCD} and \ac{EM} effects, and using the more precise LUXqed \ac{PDF} set~\cite{LUX} (see also Section~\ref{sec:opensearches}). The rest of this section is dedicated to the cross section calculation method.

The diphoton resonant production cross section is calculated using the full Breit-Wigner formula~\cite{PDG}. Thus, we are interested in both the production and the decay channels of the intermediate bound state. The partnerium can be produced by photon-fusion and gluon-fusion (projected onto a color-singlet), regardless of the partner's spin. A leptonium, consisting of color-singlet fermions, can be produced via photon-fusion. A fermion-based bound state can also be produced via \ac{DY} processes, mediated by a photon or a $Z$ boson~\cite{Colored}, however it may not decay into a diphoton final state. The allowed decay channels of a diphoton resonance are those of a $J=0,2$ resonance, discussed above. The resulting diphoton cross section would therefore follow 
\begin{align}
\begin{split}
\sigma_{pp \rightarrow B \rightarrow \gamma\gamma}&=8\pi\int_{0}^{1}\left[\frac{1}{64}\mathcal{L}_{gg}\leri{\tau}\Gamma_{B \rightarrow gg}+\mathcal{L}_{\gamma\gamma}\leri{\tau}\Gamma_{B \rightarrow \gamma\gamma}\right]\\
&\times\frac{\Gamma_{B \rightarrow \gamma\gamma}}{(\hat{s}-4m^2)^2+\hat{s}(\Gamma_{B \rightarrow \gamma\gamma}(1+2\tan^2\theta_W+\tan^4\theta_W)+\Gamma_{B \rightarrow gg})^2}\dfrac{d\tau}{\tau}\,,
\end{split}
\end{align}
where  $\tau=\hat{s}/s$, with $\sqrt{\hat{s}}$ being the total partonic \ac{COM} energy, and $\theta_W$ is the weak angle. The parton luminosity for a pair of partons $a$, $b$ is
\begin{align}
\mathcal{L}_{ab}(\tau)=\tau\int_{\tau}^1\dfrac{dx}{x}f_a\leri{x}f_b\leri{\dfrac{\tau}{x}}
\end{align}
where $x$ is the fraction of the proton momentum carried by the parton and $f_a$ is the \ac{PDF} of the parton, which we evaluate  at the factorization scale $m$. For colorless fermions, the diphoton cross section is the same, excluding \ac{QCD} contributions~\cite{Leptonium}. The relevant decay widths for scalar \acp{CTTP} are given by~\cite{Colored,stoponuim}
\begin{align}
&\Gamma_{B \rightarrow \gamma\gamma}=\frac{24\pi\alpha^2Q^4}{M^2}|\Psi(0)|^2\ \ \  \ (\times 2 \text{ for fermions},\times 1/3\text{ for color-singlets}),\\
&\Gamma_{B \rightarrow gg}=\frac{16}{3}\frac{\pi\alpha_s^2}{M^2}|\Psi(0)|^2\ \ \ \ \  \  \ (\times 2 \text{ for fermions}),
\end{align} 
where $M$ is the mass of the resonance, and modification factors for fermions and for color-singlet particles are given in parentheses. The naturalness-enhanced decays of the partnerium were found to be negligible when calculating the total decay width. 

Colored particles of large charges could have a non-negligible contribution to their binding coming from the \ac{EM} force 
\begin{align}
V(r)&=-\frac{C\bar{\alpha}_s+Q^2\alpha}{r}\,,
\end{align}
where C is the Casimir of $SU(3)_{c}$, $C_{3}=4/3$ for a color-triplet and $C_0=0$ for a color-singlet. The wavefunction at the origin is
\begin{align}
\abs{\psi(0)}^2&=\frac{(C\bar{\alpha}_s+Q^2\alpha)^3M^3}{8\pi n}\,,
\end{align}
where $n$ is the radial excitation level. Since the contributions from $n\geq2$ states are negligible, we keep only the ground state contribution~\cite{Colored}. In addition, we only consider the LO effects in the binding potential. The higher order effects have been studied in \cite{Younkin:2009zn,ProbingKats:2012ym,Beneke:2016kvz}. They find a noticeable though not dramatic enhancement of the signal cross section. Therefore, our bounds are somewhat conservative. One should note that in the decay rates and in the wavefunction $M^2\rightarrow\hat{s}$, as $\hat{s}$ is the mass of the resonance~\cite{AnnihilationDecays}.

The decay rates of the partnerium and the leptonium grow significantly with the charge of the constituents. For lepton-like particles, and for \acp{CTTP} with large charges, the bound state annihilation rate approaches a $Q^{10}$-dependence, as a result of the dominant \ac{EM} contributions. Therefore, the diphoton cross section will exhibit high charge sensitivity. 

The signal benchmarks are as described for the open-production channel recast. A resonance treatment is indeed appropriate for all the charges we consider, since $\Gamma/M\lesssim10^{-1}$ for constituents with $Q\lesssim8$. For \acp{CTTP} and lepton-like particles with $Q\leq4$, we have found that the narrow width approximation is more stable numerically. The production cross section for a narrow $\gamma\gamma$ resonance, via the decay of spin-0 partnerium bound state $B$, is given by
\begin{align}
\begin{split}
\sigma_{pp \rightarrow B \rightarrow \gamma\gamma} &= \sigma_{pp \rightarrow B } Br_{B \rightarrow \gamma\gamma}\\
&=\dfrac{\pi^2}{m^3}\left[\dfrac{1}{64}\mathcal{L}_{gg}\leri{\frac{4m^2}{s}}\Gamma_{B \rightarrow gg}+\mathcal{L}_{\gamma\gamma}\leri{\frac{4m^2}{s}}\Gamma_{B \rightarrow \gamma\gamma} \right]\\
&\times\frac{\Gamma_{B \rightarrow \gamma\gamma}}{\Gamma_{B \rightarrow \gamma\gamma}(1+2\tan^2\theta_W+\tan^4\theta_W)+\Gamma_{B \rightarrow gg}},
\end{split}
\end{align}
and in the decay rates and wavefunctions $M^2\rightarrow4m^2$, where $m$ is the mass of the partner.

Following the calculation above, using Mathematica package ManeParse 2.0~\cite{ManeParse} with LUXqed \acp{PDF}~\cite{LUX} and performing numerical integration using Mathematica, we obtain the diphoton cross sections for differently charged \acp{MCHSP}, which can be found in Appendix~\ref{diphoton_plots}. The resulting current and future-projected bounds are discussed in Sections~\ref{sec:currentbounds} and~\ref{sec:projections}.

\acresetall
\section{Current Status -- Recast Bounds}
\label{sec:currentbounds}
\begin{figure}[H]
	\centering
	\subfigure[Colored scalars.]{\includegraphics[scale=0.42]{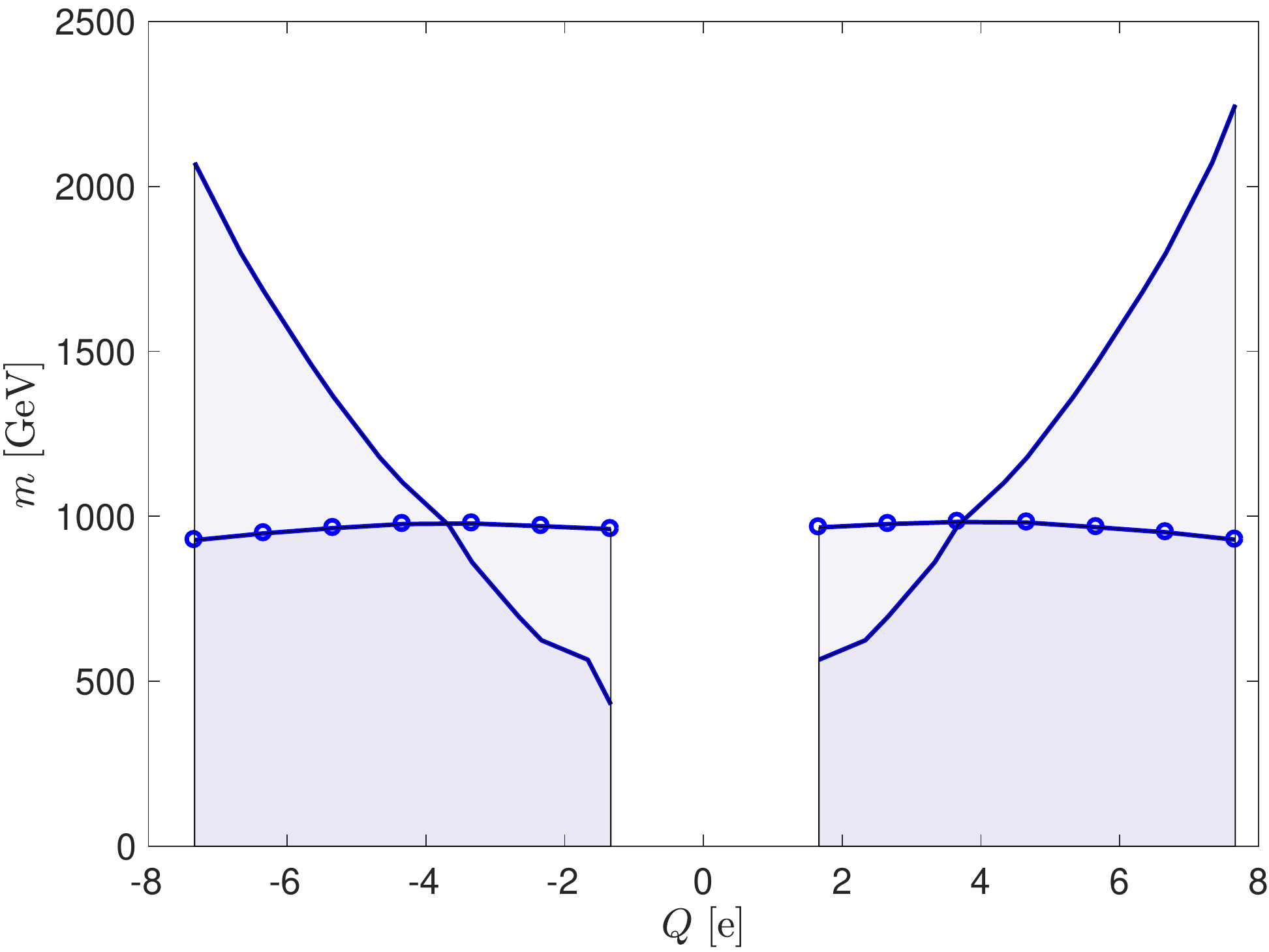}\label{mass_scalars_compare_8_13}}
	\subfigure[Colored fermions.]{\includegraphics[scale=0.42]{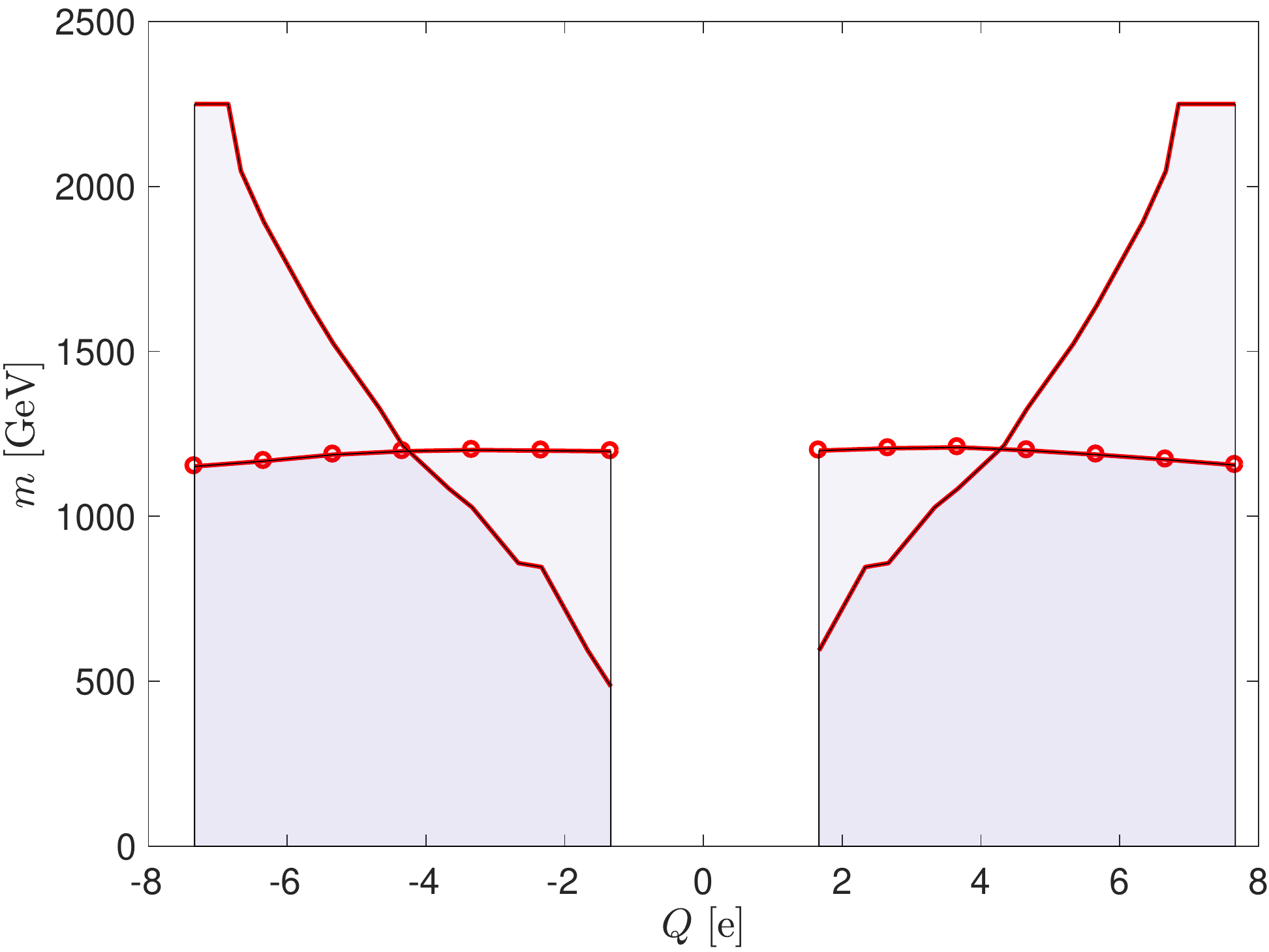}	\label{mass_fermions_compare_8_13}}
	\subfigure[Colorless fermions.]{\includegraphics[scale=0.42]{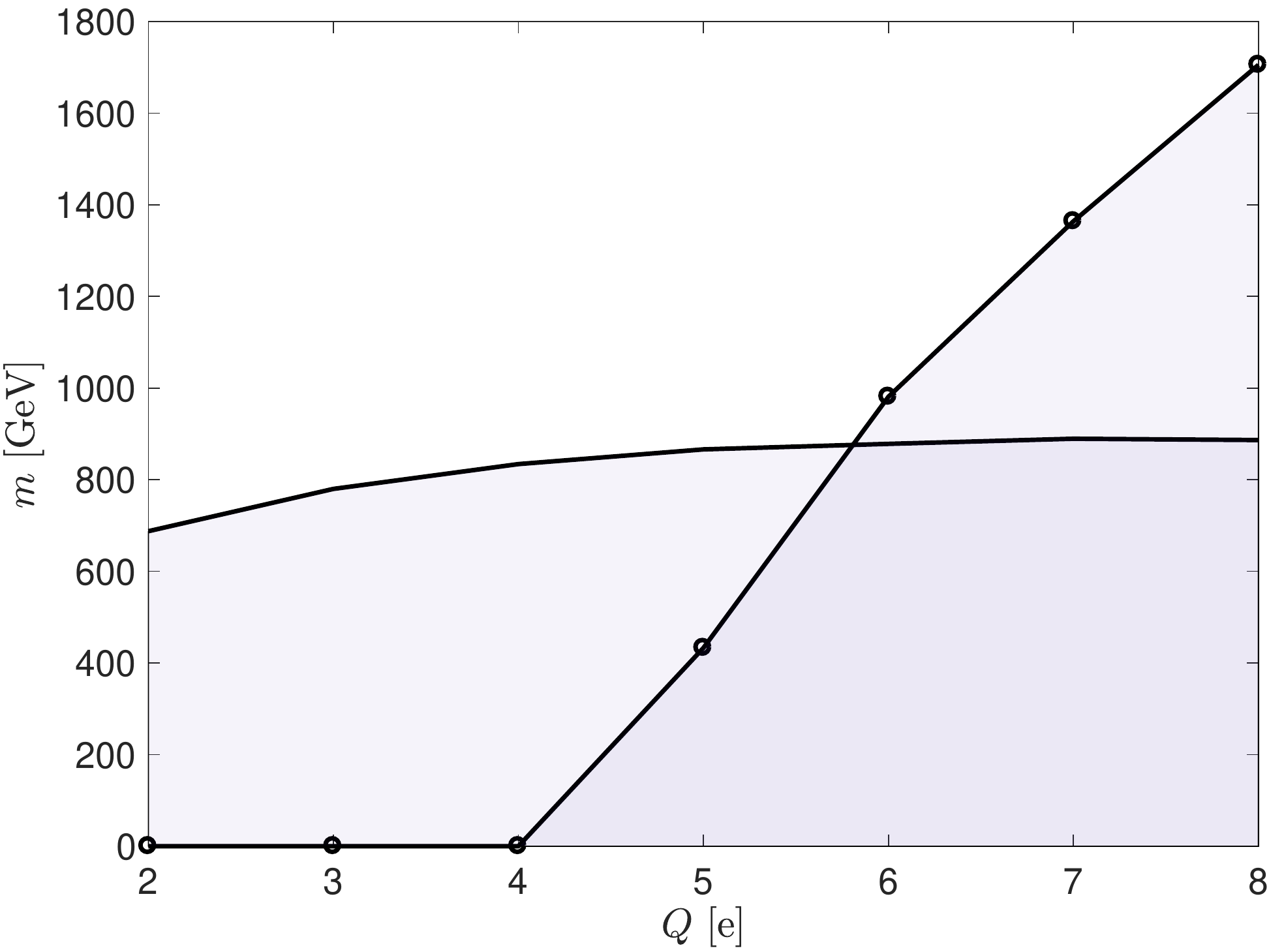}	\label{mass_leptons_compare_8_13}}	
	\caption{Lower mass bounds, as given by the most recent searches.  \textit{Solid} --  a diphoton resonance search at $\sqrt{s}=13$~TeV, $\mathcal{L}=35.9$~$\text{fb}^{-1}$~\cite{diphotoncms13} (closed-production channel).  \textit{Round markers} -- a search for \ac{MCHSP} tracks at $\sqrt{s}=8$~TeV, $\mathcal{L}=18.8$~$\text{fb}^{-1}$~\cite{StableCMS7and8} (open-production channel). \textit{Shaded} -- regions excluded by each channel. More details in text.}
	\label{current_bounds}
\end{figure}

We are now in a position to obtain and compare lower bounds on the masses of \acp{MCHSP} from the (recast) searches for their open-production and closed-production signatures. We begin by describing the current mass bounds, corresponding to the latest observations. Our bounds from the most recently published searches are presented in Table~\ref{results_table} and compared in Figure~\ref{current_bounds}. Conservatively combining the bounds by taking the stricter one for each signal benchmark, we obtain the current mass bounds at a minimal CL of 95\%, highlighted in the table.

To obtain current constraints on \acp{MCHSP} from the open channel, we utilize the most recent search for above-threshold \acp{MCHSP}, conducted by CMS at $\sqrt{s}=8$~TeV~\cite{StableCMS7and8}. The limits on particle masses, in a given signal model, are derived by first obtaining a 95\%-\ac{CL} upper limit on the effective cross section, and then choosing the mass such that the theoretical effective cross section saturates this limit. Following CMS, we apply a hybrid Bayesian-frequentist p-value computation~\cite{hybrid}, with the relevant parameters given in the original analysis. Our resulting upper limit is consistent with that inferred from CMS results. The theoretical effective cross sections are calculated by multiplying the cross sections and the efficiencies, as explained in Sections~\ref{production_cross_section},~\ref{sec:efficiency}, and can be found in Appendix~\ref{effective_cross_sections_plots}.

Analogously to the open channel, we derive mass bounds for \acp{MCHSP} from their bound state signatures as well. For the closed production case, we require the theoretical diphoton production cross section, induced by the bound state resonance, as explained in Section~\ref{sec:partnerium}, to saturate the upper limits at 95\%-\ac{CL}. For the current bound, we employ the CMS limit given in~\cite{diphotoncms13} for $\sqrt{s}=13$~TeV at $\mathcal{L}=35.9~\text{fb}^{-1}$. It should be noted the signal efficiency in~\cite{diphotoncms13} was calculated for gluon-fusion production, and could be slightly different for photon-produced resonances. The experimental bounds on a diphoton resonance in~\cite{diphotoncms13} were given for three resonance-width benchmarks: $\Gamma/M =1.4\cdot 10^{-4}$ (narrow), $\Gamma/M=1.4\cdot 10^{-2}$ (mid-width) and $\Gamma/M=5.6\cdot 10^{-2}$ (wide). Therefore, when available, we use narrow resonance bounds for $\Gamma/M\lesssim 5\cdot 10^{-3}$ ($Q\lesssim5$ for color-triplets, $Q\lesssim 6$ for color-singlets), mid-width resonance bounds for $5\cdot 10^{-3}\lesssim\Gamma/M\lesssim 3\cdot 10^{-2}$ ($5\lesssim Q\lesssim 6$ for color-triplets, $6\lesssim Q\lesssim 7$ for color-singlets) and wide resonance bounds for $\Gamma/M\gtrsim 3\cdot 10^{-2}$ ($6\lesssim Q$ for color-triplets, $7\lesssim Q$ for color-singlets).

The diphoton cross section limit observed in the search was given up to resonance masses of 4500~GeV. However, for colored fermions with $Q>6.9$ the corresponding $\gamma\gamma$ cross section is larger than the observed limit throughout the available mass range. They are thus excluded below $m=2250$~GeV, but their exact mass bound can not be explicitly inferred from this search. 
\renewcommand{\arraystretch}{1.1}
\begin{table}[b!]
	\centering
	\begin{tabular}{|c|ccccccc|c|}
		\hline
		\textbf{Q[e] } & \textbf{5/3 }& \textbf{8/3}& \textbf{11/3} & \textbf{14/3 }  & \textbf{17/3}   & \textbf{20/3}   & \textbf{23/3} & \textbf{channel}  \\ \hline
		\multirow{2}{2.3cm}{\centering color-triplet scalar } & \cellcolor{blue!15}970 & \cellcolor{blue!15}980 & \cellcolor{blue!15}980 & 980 & 970 & 950 & 930 & \textbf{open} \\
		& 570 & 700 & 970  & \cellcolor{blue!15}1180 & \cellcolor{blue!15}1460 & \cellcolor{blue!15}1800 & \cellcolor{blue!15}2250 & \textbf{closed} \\ \hline
		\multirow{2}{2.3cm}{\centering color-triplet fermion } & \cellcolor{red!15}1200 & \cellcolor{red!15}1200   & \cellcolor{red!15}1210  & 1200 & 1190   & 1170   & 1160  & \textbf{open}\\ 
		& 590 & 860 & 1080 & \cellcolor{red!15}1330 & \cellcolor{red!15}1640 & \cellcolor{red!15}2050 & \cellcolor{red!15}2250* &  \textbf{closed} \\\hline
		\textbf{Q[e] }   & \textbf{-4/3} & \textbf{-7/3} & \textbf{-10/3} & \textbf{-13/3}  & \textbf{-16/3}  & \textbf{-19/3 } & \textbf{-22/3} & \textbf{channel}  \\ \hline
		\multirow{2}{2.3cm}{\centering color-triplet scalar} & \cellcolor{blue!15}960 & \cellcolor{blue!15}970  & \cellcolor{blue!15}980  & 980 & 960 & 950  & 930& \textbf{open} \\
		& 430 & 620 & 860  & \cellcolor{blue!15}1100 & \cellcolor{blue!15}1360 & \cellcolor{blue!15}1680 & \cellcolor{blue!15}2070 & \textbf{closed}\\ \hline
		\multirow{2}{2.3cm}{\centering color-triplet fermion }  & \cellcolor{red!15}1200 & \cellcolor{red!15}1200 & \cellcolor{red!15}1200  & 1200 & 1190 & 1170 & 1150 & \textbf{open} \\ 
		& 480 & 850 & 1030 & \cellcolor{red!15}1210 & \cellcolor{red!15}1520 & \cellcolor{red!15}1890 & \cellcolor{red!15}2250*& \textbf{closed}\\ \hline
		\textbf{Q[e]}   & \textbf{2}      & \textbf{3}      & \textbf{4}       & \textbf{5}      & \textbf{6 }     & \textbf{7}      & \textbf{8} & \textbf{channel}      \\ \hline
		\multirow{2}{2.3cm}{\centering color-singlet fermion }  &\cellcolor{black!15} 690&\cellcolor{black!15}	780&	\cellcolor{black!15}840&	\cellcolor{black!15}870&	890&	890&	890& \textbf{open}\\
		& - & - & -  & 570  &\cellcolor{black!15} 980  & \cellcolor{black!15}1380 & \cellcolor{black!15}1710 & \textbf{closed}\\ \hline
	\end{tabular}
	\caption{Current lower bounds on the masses of \acp{MCHSP}. The bounds were obtained from the diphoton resonance signatures at $\sqrt{s}=13$~TeV, $\mathcal{L}=35.9$~$\text{fb}^{-1}$ (closed-production channel) and from the \acp{MCHSP} signatures at $\sqrt{s}=8$~TeV, $\mathcal{L}=18.8$~$\text{fb}^{-1}$ (open-production channel). The colored cells are the corresponding combined bounds, given by naively taking the stricter bound of the two searches. \textit{Blue} -- scalar \acp{CTTP}, \textit{red} -- fermion \acp{CTTP} and \textit{black} -- lepton-like particles. Mass bounds are given in~GeV. *Fermion \acp{CTTP} with $Q=23/3,-22/3$, are excluded below $2250$~GeV, however the exact bound could not be inferred from the search. More details in text. }
	\label{results_table}
\end{table}

\begin{figure}[b!]
	\centering
	\includegraphics[scale=0.44]{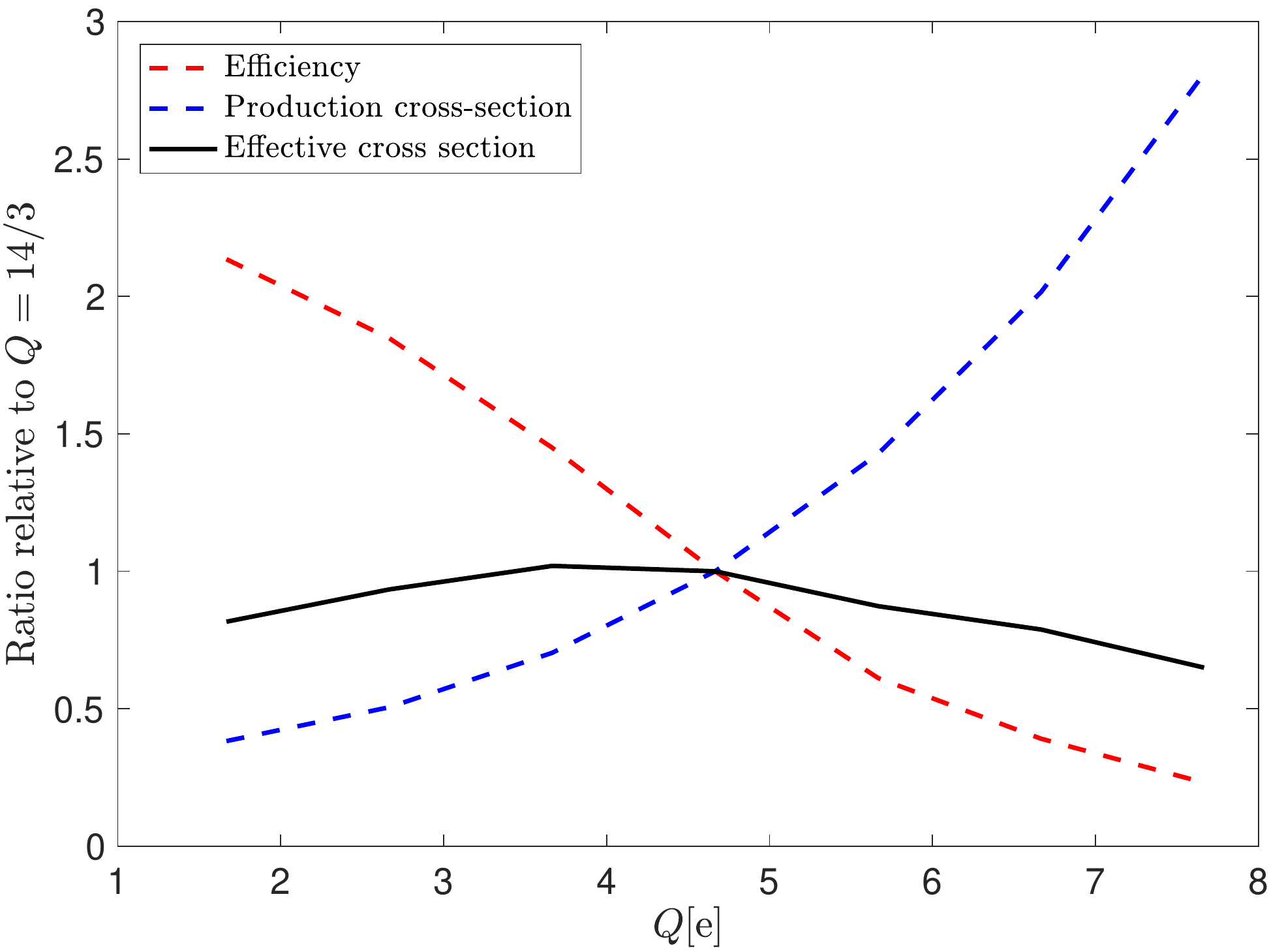}
	\caption{Detection efficiency, production cross section and the resulting effective cross section $\sigma \cdot \epsilon$ for a color-triplet scalar of $m=1000$~GeV, at $\sqrt{s}=8$~TeV. All are presented relative to their value for a color-triplet scalar of $Q=14/3$.}
	\label{flatness_explained}
\end{figure}

\subsection[Bounds from Open Signatures of \acsp{MCHSP}]{Bounds from Open Signatures of \acsp{MCHSP}}
We find that scalar and fermion \acp{CTTP} are excluded below masses of roughly 1~TeV and 1.2~TeV, respectively. Interestingly, the bounds are almost charge independent both for scalar and fermion \acp{CTTP}. As can be seen in Fig.~\ref{flatness_explained}, this is a result of a coincidental balance between the production cross sections and the efficiencies at which color-triplet \acp{MCHSP} could be directly observed. On the one hand, the search becomes less efficient as the charge of the particle increases. For smaller masses, this is mainly a result of the $p_T/Q$ selection, while for larger masses, the timing requirement, imposed by the muon trigger, becomes more important, due to the particle's large ionization energy loss. On the other hand, the cross sections grow with the charge of the particle. The production rate consists of the Q-independent \acs{QCD} processes, the $Q^2$-dependent $g\gamma$-fusion and \acs{EW}-mediated \ac{DY} processes, and the $Q^4$-dependent photon-fusion. As we have shown in Sec.~\ref{production_cross_section}, each subprocess becomes dominant at a different mass scale, resulting in a rather strong charge-dependence for the production rates of heavy partners. The bounds on the masses of lepton-like particles are slightly more charge dependent. We find colorless fermions to be excluded below a mass of$~690$~GeV for $Q=2$, and below$~890$ GeV for $Q=8$. This is a result of the larger charge dependence of the production cross section of lepton-like particles, in the absence of the charge-independent \acs{QCD} production. Due to hadronization, the bounds in the open channel are asymmetric for positively and negatively charged color-triplets.

\subsection{Bounds from Closed Signatures of \acp{MCHSP}}

The diphoton data excludes color-triplet \acp{MCHSP} of charges larger than $\sim4$ ($\sim7$) at masses below 1~TeV (2~TeV). Due to the smaller production and decay rates of bound states consisting of color-singlets, the bounds placed on lepton-like particles are somewhat weaker. Lepton-like particles of charges larger than 5 (8) are excluded below masses of 0.5~TeV (1.7~TeV). The charge dependence of the mass bounds coming from the closed-production signatures is understandably large, due to the dominant \acs{EM} effects contributing to production, binding and decay, as explained in Section~\ref{sec:partnerium}. These result in a significant charge dependence of the diphoton resonant cross section, that can be as much as $Q^{10}$-dependent for lepton-like particles. In addition, the efficiency for the diphoton search is not directly related to the bounded constituents charges. The bounds are symmetric for negative and positive charges, as the diphoton cross section in the Sec.~\ref{sec:partnerium} is an even function of the $Q$.

\subsection{Combined Bounds}
Combining the searches in the open and the closed channels provides powerful constraints on \acp{MCHSP} models. As shown above, the current limits derived from the direct search for \acp{MCHSP} are stronger for charges smaller than~$\sim 4$ for scalar and fermion color-triplets, and for charges smaller than $\sim 6$ for colorless fermions, while for larger charges the diphoton exclusion bounds dominate. Therefore, we benefit from considering both searches, even by naively setting the bound at the larger of the two. Upon further statistical analysis, one should be able to combine the searches as the two channels must be explained simultaneously for stable particles, and thus obtain even stronger mass bounds at 95\% CL.

\subsection{The Leptonic Case -- Comparison to the Literature}
Since lepton-like particles have been studied in the past, we may now compare our new bounds for lepton-like particles to those found in the literature. As we will see, the bounds we have obtained are in disagreement with the existing results. These differences are mainly a result of our new cross section calculations, which are more exhaustive and reliable, compared to previous analyses. 

\begin{figure}[t!]
	\centering
	\includegraphics[scale=0.44]{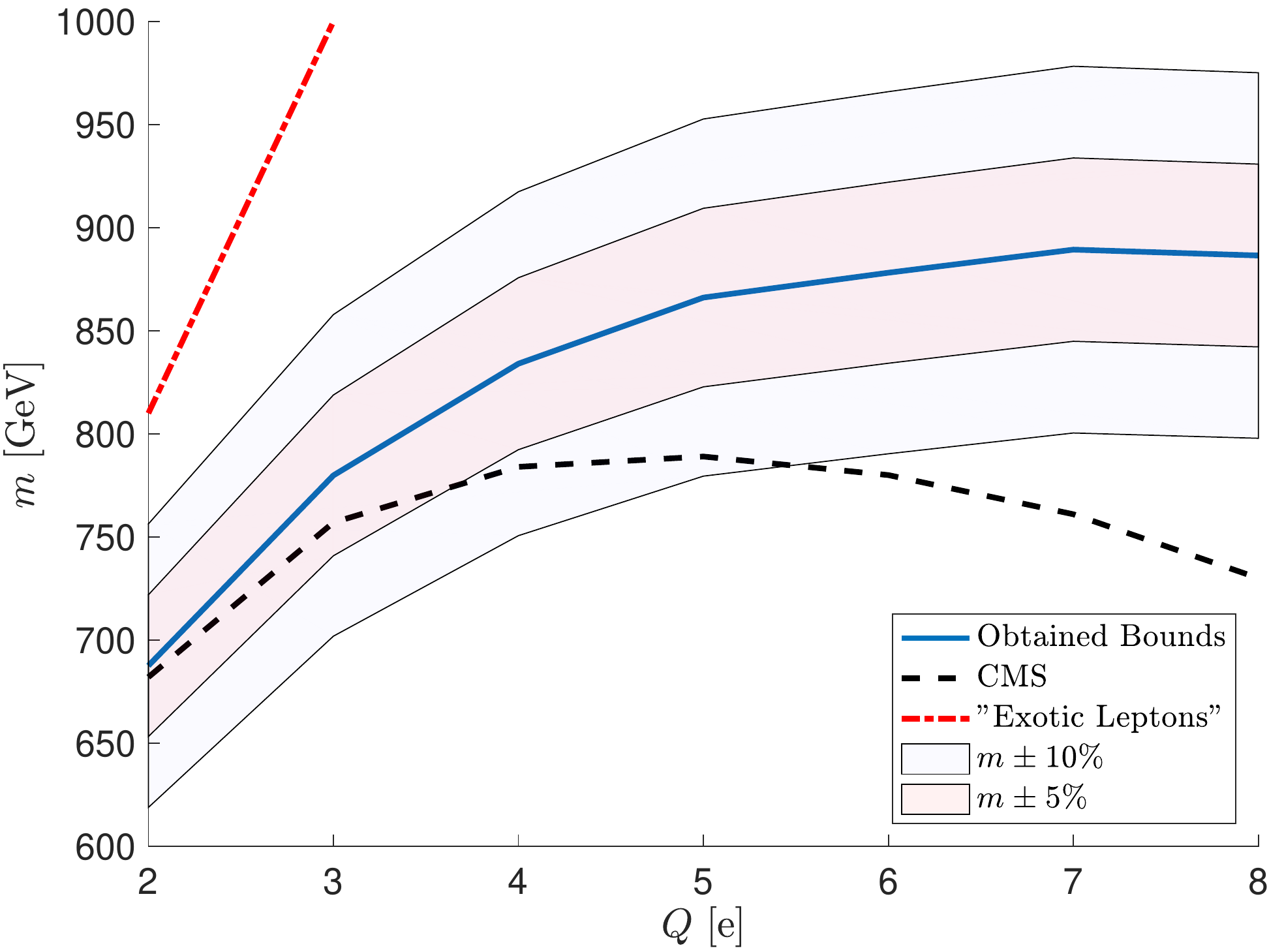}
	\caption{Comparing the lower mass bounds on multiply-charged lepton-like particles, coming from the different analyses of the open-production signature. \textit{Dashed} -- results published by CMS~\cite{StableCMS7and8}.
		\textit{Dash-dotted red} -- bounds for $Q=2,3$ given in~\cite{Leptonium}. 
		\textit{Solid blue} -- mass bounds calculated in this study with 5\% (\textit{Red}) and 10\% (\textit{Light blue}) deviation bands.
	}
	\label{mass_bounds_compare_different_leptons}
\end{figure}

As shown in Fig.~\ref{mass_bounds_compare_different_leptons}, the mass bounds we have obtained from the open-production signature are stricter than those published by CMS~\cite{StableCMS7and8}. While the analysis by CMS considered \ac{DY}-production exclusively, we also include photon-fusion production. Similarly to~\cite{Leptonium}, we find that photo-production processes significantly enhance the cross sections for particles with large charges, and therefore the bounds have strengthened.

\label{pdf_discussion}

The choice of the \acs{PDF} plays an essential role in calculating the production cross sections, and is particularly important when considering photo-production processes. This can be inferred by comparing our mass bounds, obtained using LUXqed \acsp{PDF} set, to the bounds presented in~\cite{Leptonium}, derived using NNPDF2.3QED~\cite{NNPDF}, as both analyses considered the same production processes. As can be seen in Fig.~\ref{mass_bounds_compare_different_leptons}, the mass bounds for colorless fermions, derived from our analysis of the open-production channel, are much weaker than the bounds set by the corresponding analysis in~\cite{Leptonium}. The same trend emerges when comparing the closed-production signature analyses, and we find our bounds to be less stringent than those previously obtained in~\cite{Leptonium}. The origin of these differences can be traced to the choice of the photon \acs{PDF}. As discussed in~\cite{LUX} (see also~\cite{Aad:2016zzw}), the way the photon \acs{PDF} is obtained in the NNPDFx.yQED sets is afflicted by large uncertainties. For the $\gamma\gamma$ parton luminosity at invariant masses of 1-3 TeV, as relevant to our analysis, the resultant uncertainty can be more than an order of magnitude. The precise extraction of the photon~\acs{PDF} via the method of~\cite{LUX,Manohar:2017eqh}, using $ep$ data, implies, via the resulting LUXqed \acs{PDF} set, a photon luminosity which is as much as a factor of 60 lower than that obtained for central values of the NNPDF2.3QED set. As a result, the cross section calculations in ref.~\cite{Leptonium}, which are based on those central values, substantially overestimate the contributions coming from photon fusion (as well as other photon-induced components) to the cross section. Consequently, the bounds in~\cite{Leptonium} need to be corrected down to those derived and presented here.	

\section{Future Scenarios -- Discovery and Exclusion}\label{sec:projections}

\begin{figure}[H]
	\centering
	\subfigure[Colored scalars.]{	\includegraphics[scale=0.42]{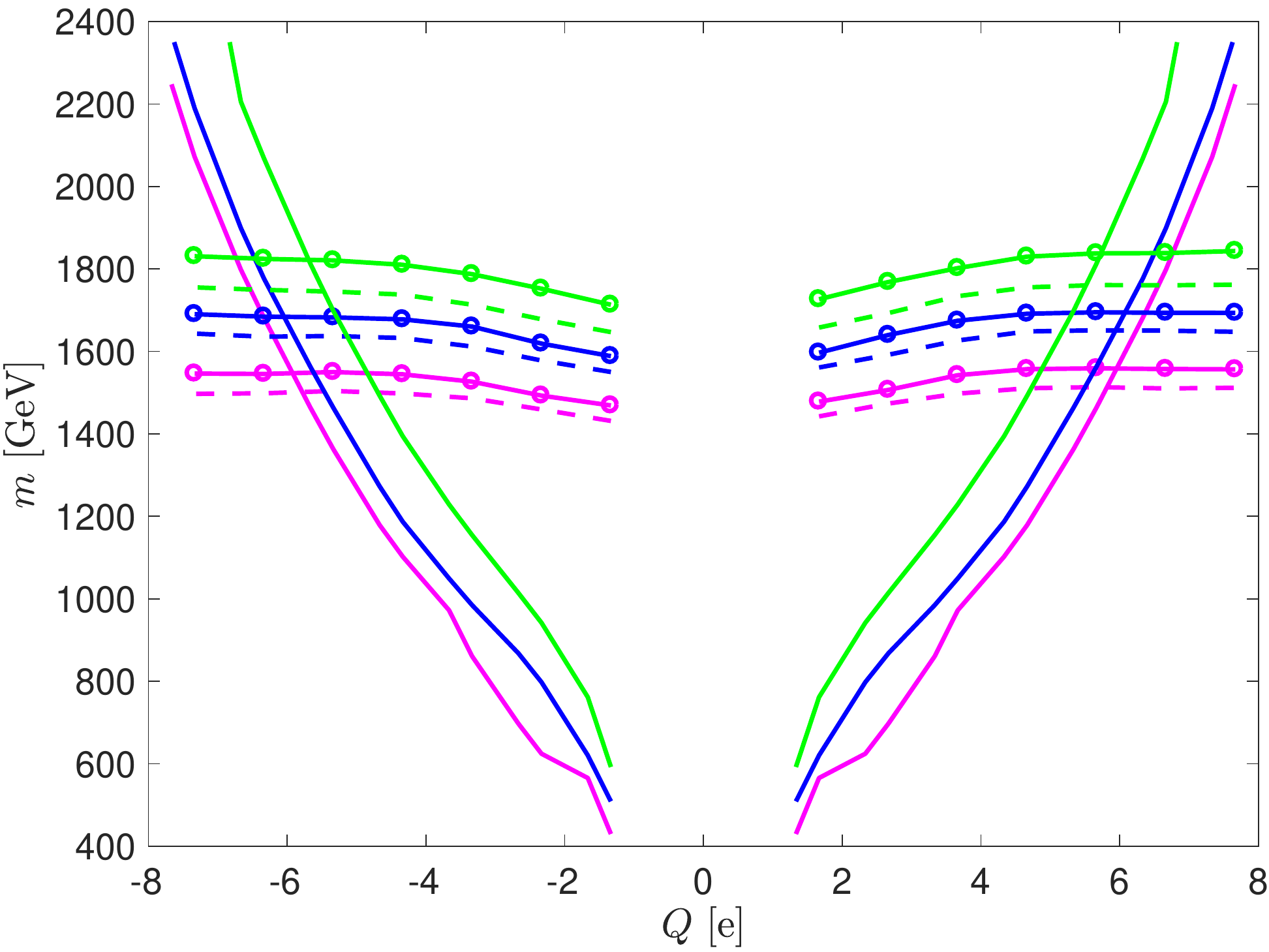}\label{mass_scalars_compare_13_proj}}
	\subfigure[Colored fermions.]	{\includegraphics[scale=0.42]{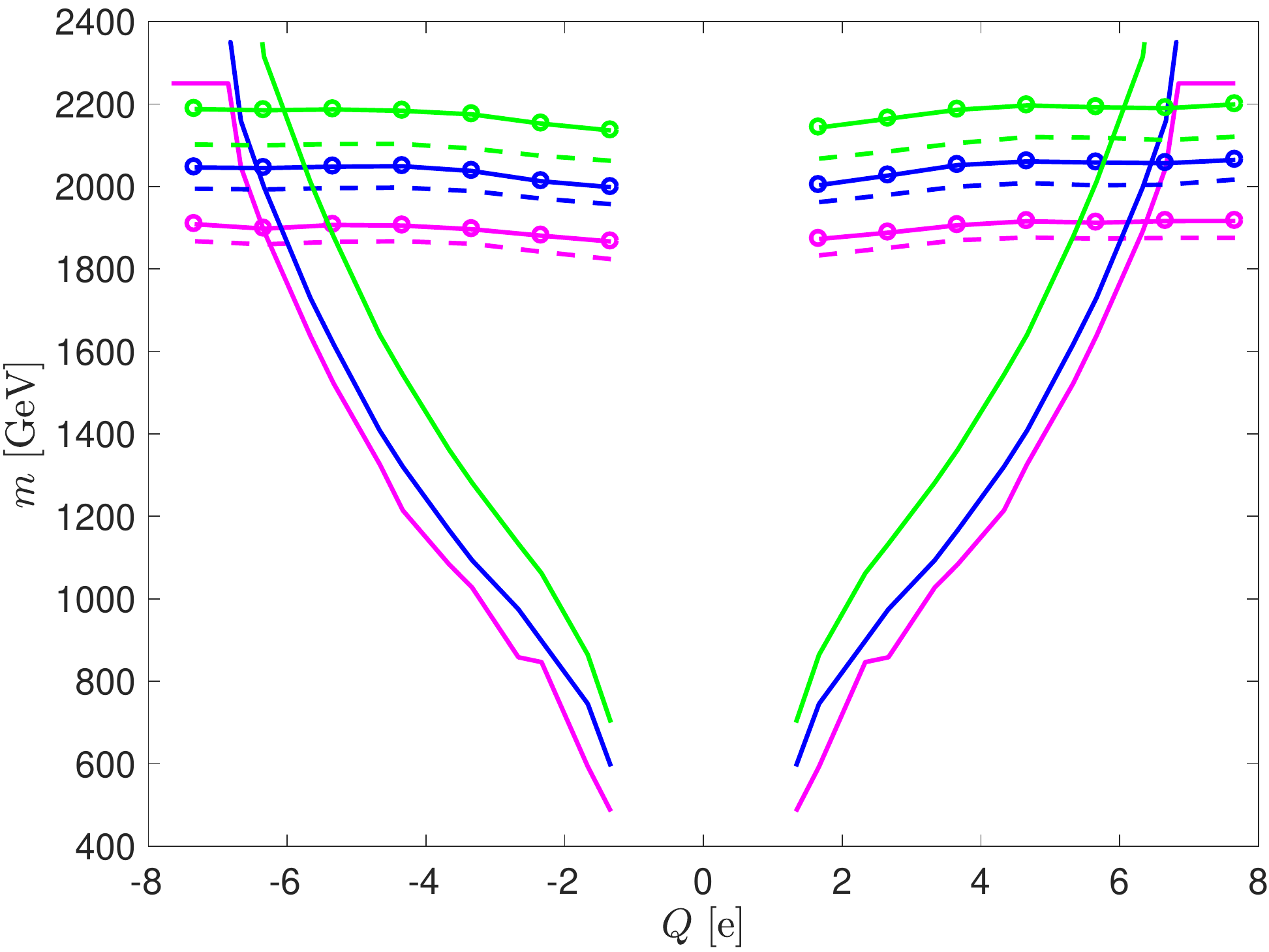}\label{mass_fermions_compare_13_proj}}
	\subfigure[Colorless fermions.]{	\includegraphics[scale=0.42]{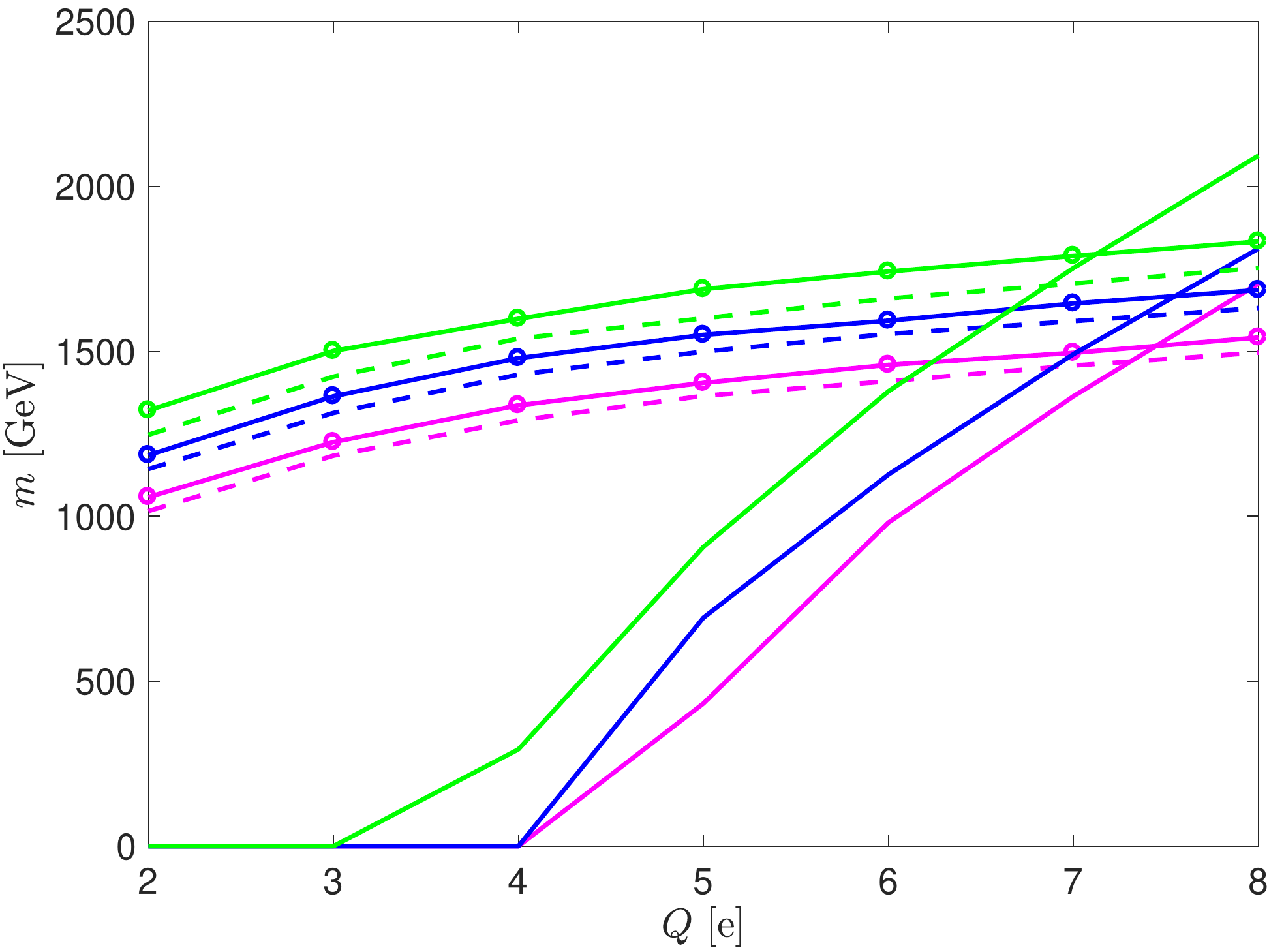}\label{mass_leptons_compare_13_proj}}
	\caption{Expected lower mass bounds at $\sqrt{s}=13$~TeV, $\mathcal{L}=35.9\text{~fb}^{-1}$ (\textit{magenta}),~$100\text{~fb}^{-1}$ (\textit{blue}), and~$300\text{~fb}^{-1}$ (\textit{green}). \textit{Solid} -- diphoton resonance searches (closed-production channel). \textit{Round markers} -- searches for \ac{MCHSP} tracks with luminosity-scaling (open-production channel).\textit{ Dashed} -- searches for \ac{MCHSP} tracks with luminosity and pileup scaling (open-production channel).}
	\label{projected_bounds}
\end{figure}

In order to obtain the prospective mass bounds from \acs{LHC} searches at $\sqrt{s}=13$~TeV, we consider integrated luminosities of $36$~$\text{fb}^{-1}$, $100 \text{fb}^{-1}$(current -- July 2018) and $300$~$\text{fb}^{-1}$ (future). Our projected mass bounds from the two kinds of searches are presented in Figure~\ref{projected_bounds}.

For the closed-production signatures, projected bounds for integrated luminosities of $100$~$\text{fb}^{-1}$ and $300$~$\text{fb}^{-1}$, are calculated using the expected upper limits for ATLAS searches of a photo-produced $J=0$ resonance, as given in~\cite{DiphotonProjection}.

Although the \acs{LHC} has been running in \ac{COM} energy of 13~TeV since 2015, \acp{MCHSP} search results have yet to be updated. Therefore, for the open-production searches, we calculate the expected effective cross section upper limit at 95\%-\ac{CL}, under the background hypothesis. The expected number of background events is calculated by scaling the corresponding $\sqrt{s}=8$~TeV estimate~\cite{StableCMS7and8} in two ways -- by the luminosity ratio and by the luminosity ratio times the pileup ratio. The latter is more conservative, and perhaps more realistic, as some of the selections and the backgrounds involved may depend not only on the luminosity, but also on the amount of pileup in each run. 

Following our analysis, we expect the mass bounds from the open-production searches to improve dramatically with \ac{COM} energy. For $\sqrt{s}=13$~TeV, the bounds could reach about 1-1.5~TeV for lepton-like particles, 1.5~TeV for scalar \acp{CTTP}, and just under 2~TeV for fermion \acp{CTTP}, even when only considering an integrated luminosity of 36~$\text{fb}^{-1}$. We therefore believe that a dedicated experimental search for \acp{MCHSP}, accounting for the additional properties of colored particles, such as nuclear energy loss and charge change, is very much in need.

We find that the interplay between the searches for \ac{MCHSP} tracks and the searches for diphoton resonances leads to an effective way to probe the parameter space of these models. We will now present how the searches in the open and the closed channels could be combined to better study \acp{MCHSP} in the future.

In the case where no excess is observed in both channels, one can combine their results to set upper limits that are significantly stricter than the ones obtained by each search individually. Comparing the two channels assuming the same energy and luminosity, we find that open-production searches are expected to become stronger, and dominate up to charges of about~$\sim6$ for \acp{CTTP}, and $\sim7$ for lepton-like particles. Therefore, these searches are also more likely to carry a potential for discovery. However, in the case of a discovery in the open channel, its analysis might not be able to determine the charge of the observed \ac{MCHSP}, as we have already established. In addition, the measured kinematics of the particle is different from the truth-level kinematics, due to its unknown charge and ionization energy loss, and will thus be difficult to interpret with good accuracy. On the other hand, given its strong charge-dependence, the diphoton search, although typically less sensitive, can be very useful in breaking the charge degeneracy, or at least in narrowing down the range of allowed charges. The situation could be reversed for very large charges, and the diphoton search could become the discovery channel. In the transition region, correlated excesses in both channels, even if insignificant for each one, may be sufficiently significant to point to a discovery of an \ac{MCHSP} when combined.

\begin{figure}[t!]
	\centering
	\includegraphics[scale=0.7]{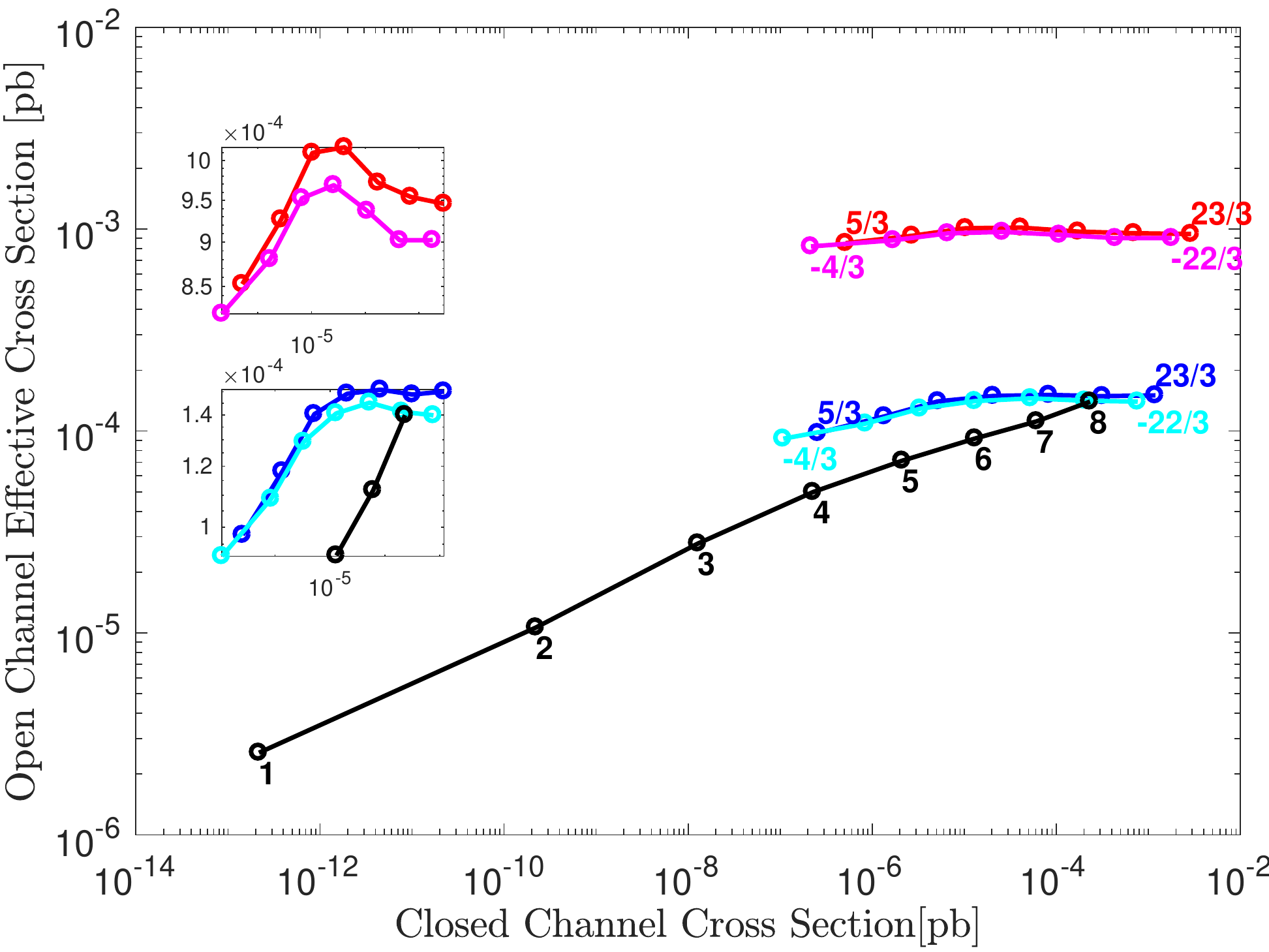}
	\caption{The combined signatures of a hypothetical \ac{MCHSP} with $m=1500$~GeV, for different choices of its quantum numbers. In case of a discovery in both channels, combining the observables measured in the two searches could be used to determine the quantum numbers of the newly discovered particle. The lines correspond to different spin-color combinations studied in this work. \textit{Black}~--~color-singlet fermions. \textit{Blue}~--~color-triplet scalars with positive charges. \textit{Cyan}~--~color-triplet scalars with negative charges. \textit{Red}~--~color-triplet spin-1/2 fermions with positive charges. \textit{Magenta}~--~color-triplet spin-1/2 fermions with negative charges. \textit{Round markers} indicate charges spaced by one unit, colored labels indicate the charges. The two subplots on the top-left are magnified views. \textit{Top box}~--~negatively-charged and positively-charged color-triplet fermions. \textit{Bottom box}~--~negatively-charged and positively-charged color-triplet scalars. }
	\label{discovery_conc}
\end{figure}

In case of a discovery in both channels, not only would one be able to claim an observation of an \ac{MCHSP} with higher significance, but also to better study its properties, as we will now demonstrate. First, the mass of the particle could be determined from the diphoton resonance peak. Given the measured mass, one could calculate the theoretical effective cross section, relevant for the open search, and the theoretical diphoton cross section, relevant for the closed search, for \acp{MCHSP} of different spins, charges and color representations. As demonstrated for $m=1500$~GeV in Fig.~\ref{discovery_conc}, the measurements in both channels would mark a specific point, which could then be related to a specific choice of the particle's quantum numbers. This is true for most of the parameter space, except for the crossing point between a highly charged lepton-like particle and a colored scalar, corresponding to two different choices of quantum numbers. Although measurement uncertainties could make the model distinction less sharp, the appropriate parameter space would be substantially narrowed given the combination of the two measurements. 

\acresetall

\section{Conclusions and Outlook}
\label{sec:conclusions}
We have studied the \acs{LHC} phenomenology of  \acp{MCHSP}. Such particles, that are stable on collider scales and carry exotic electric charges, exist in various extensions of the \acs{SM}. We introduced the signatures of color-triplet \acp{MCHSP}, referred to as \acp{CTTP}, which were proposed as a solution to the hierarchy problem~\cite{CTTP}. In addition, we reanalyzed the signatures of colorless fermion \acp{MCHSP}, referred to as lepton-like particles. We considered both the ``closed" channel -- where the \ac{MCHSP} and its anti-particle form a bound state (partnerium/leptonium), detectable as a diphoton resonance, and the ``open" channel --  where each of the \acp{MCHSP} propagates approximately independently, detectable in designated searches. For this purpose, we have recast existing analyses, including \acs{QCD} effects and an updated treatment of \acs{EM} effects. 

For \acp{MCHSP} with relatively small charges, the open-production searches are more important, albeit with only little sensitivity to the charge of the particle. This is in contrast to the diphoton channel, which is more sensitive to \acp{MCHSP} with large charges, and exhibits a strong charge-dependence. 
Thus, a combined search is useful both for the exclusion and for the discovery of \acp{MCHSP}. We have obtained bounds on \acp{MCHSP} from both production channels, and combined them by taking the more stringent bound for each signal model. We find lower bounds on \ac{CTTP} masses, that are nearly constant at about 1 TeV for charges $|Q| \leq 4$, then raising to $2.3$ TeV at $|Q| = 8$. This behavior is due to the closed (diphoton resonance) signature becoming more constraining than open pair production for $|Q| \geq 4$. The bounds on lepton-like particles display an analogous behavior, beginning at about $0.8$ TeV and starting to rise at $|Q|=6$, to about $1.7$ TeV at $|Q| = 8$. The bounds we obtained for lepton-like particles are significantly weaker than those given in~\cite{Leptonium}, but are stronger than the bounds given in~\cite{StableCMS7and8}. The differences stem from our cross section calculation, which accounts for photo-production processes using LUXqed \acsp{PDF} set, which is more precise for the photon \acs{PDF}.

In addition, we have presented two future scenarios: exclusion and discovery. In the exclusion scenario, where no signal is observed, we have projected the bounds to 13 TeV, three integrated luminosities and with or without the pileup scaling. In all cases we find that the bounds become stricter. We therefore strongly encourage a dedicated experimental analysis for \acp{MCHSP}, which includes colored particles, and which should combine open production and
diphoton resonance signals\footnote{After completing and posting the manuscript, the results of a Run-II ATLAS search for open-production lepton-like particles~\cite{Aaboud:2018kbe} became publicly available. This is the first LHC analysis corresponding to the $\sqrt{s}=13$~TeV, $\mathcal{L}=36\text{~fb}^{-1}$ data in the context of $|Q|>2$ \acp{MCHSP}. Similarly to the run-I CMS analysis~\cite{StableCMS7and8} discussed above, this new ATLAS analysis did not account for photo-production processes. A rough estimate of these effects can be given by recalculating the theoretical production cross sections, as described in our analysis, and comparing them to the cross section upper limits observed by ATLAS to obtain mass bounds. This leads to mass bounds of 1.02 TeV ($Q=2$), 1.36 TeV ($Q=5$) and 1.32 TeV ($Q=7$), which are in good agreement with our future-projected bounds for the same energy and luminosity. A more precise treatment requires a dedicated efficiency computation, considering the relevant aspects of the ATLAS detector and signal selections, and should be addressed through a reanalysis by ATLAS.}. In the event of a discovery, we have shown how combining the measurements at both channels will allow to determine the mass, spin, color, and charge of the observed particle.

In light of our findings, let us briefly comment on the future of open-production searches of \acp{MCHSP}. In order to reduce the impact of pileup, both ATLAS and CMS are considering installing a new timing sub-detector, that is capable of measuring \ac{TOF} at 30 ps resolution~\cite{timingDetectors}. These timing detectors might improve the discovery reach for \acp{MCHSP}, by providing an additional, more accurate, discriminator for slow particles. Moreover, they may be able to measure the \ac{TOF} of a particle prior to its interactions with the material in the calorimeters and in the \ac{MS}, which are the main cause of ionization energy loss, thus improving detection efficiencies. We leave a dedicated study of the implications of incorporating the information collected by the timing detectors in searches for \acp{MCHSP} for future work.

\acknowledgments
We thank Shikma Bressler for her insights and expertise that greatly assisted the research, and for her comments on the manuscript. We also thank Marumi Kado and Yevgeny Kats for the useful discussions. We are grateful to Luke Arpino, Tom Steudtner, Matthias Schlaffer, Sanmay Ganguly and the WIS LCG team for their kind help with different aspects of the project. The work of GP is supported by grants from the BSF, ERC, ISF, the Minerva Foundation, and the Segre Research Award. GP, SJ and SK acknowledge the support through a Weizmann-UK Making Connections grant, including a studentship jointly funded with the School of Mathematical and Physical Sciences at the University of Sussex. The work of SJ was supported by UK STFC Consolidated Grant ST/P000819/1. SJ and SK thank the Weizmann Institute, and IS the University of Sussex, for the hospitality during visits while working on this project. IS and SK are grateful for STSM support from European Cooperation in Science and Technology (COST) action CA15108 ``Connecting Insights in Fundamental Physics".
\appendix

\section{Open-Production Signatures }\label{cross_section_plots}

\subsection{Cross Sections}
\begin{figure}[H]
	\centering
	\subfigure[Positvely-charged colored scalars.]{\includegraphics[scale=0.37]{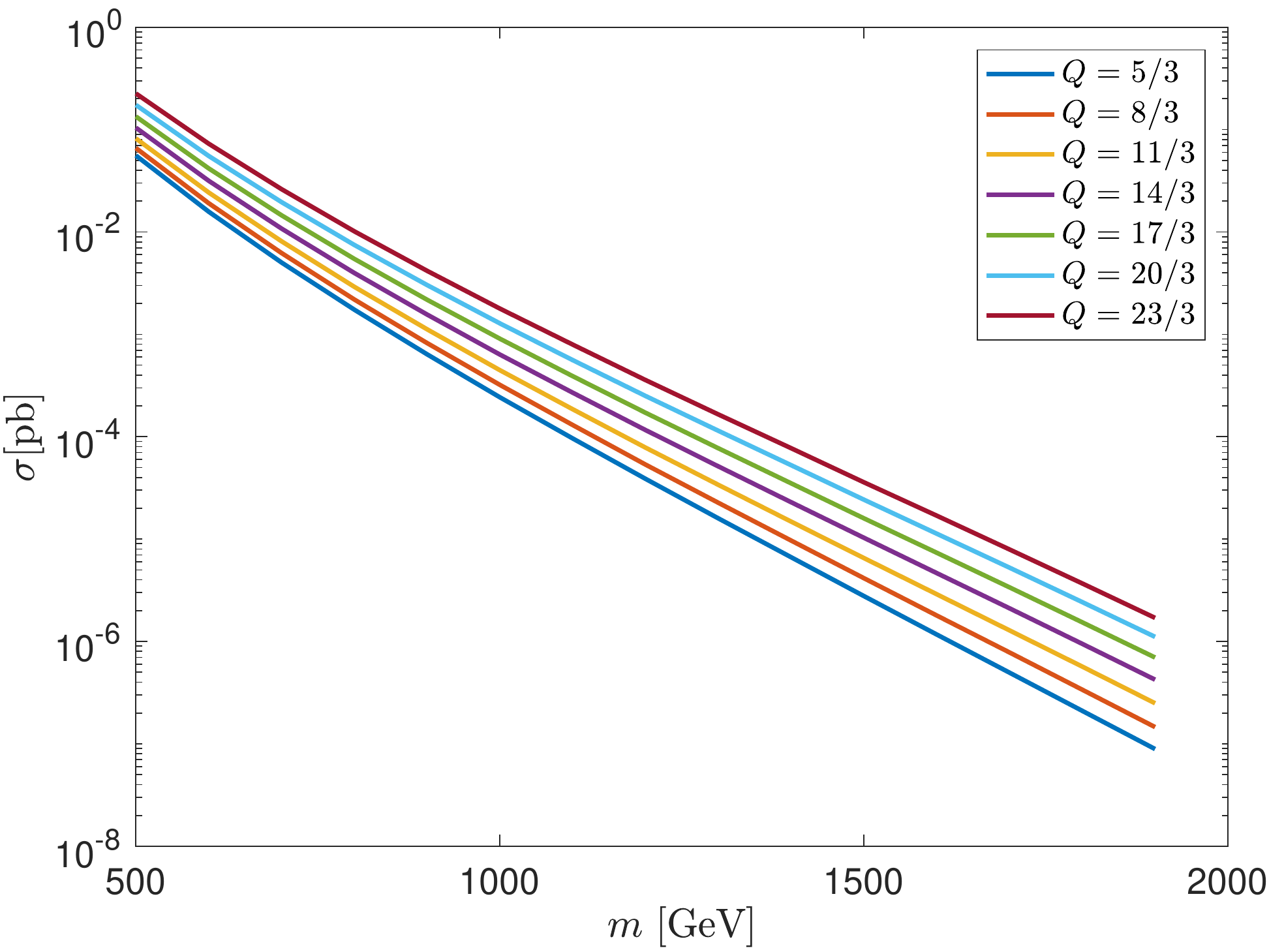}}
	\subfigure[Negatively-charged colored scalars.]{\includegraphics[scale=0.37]{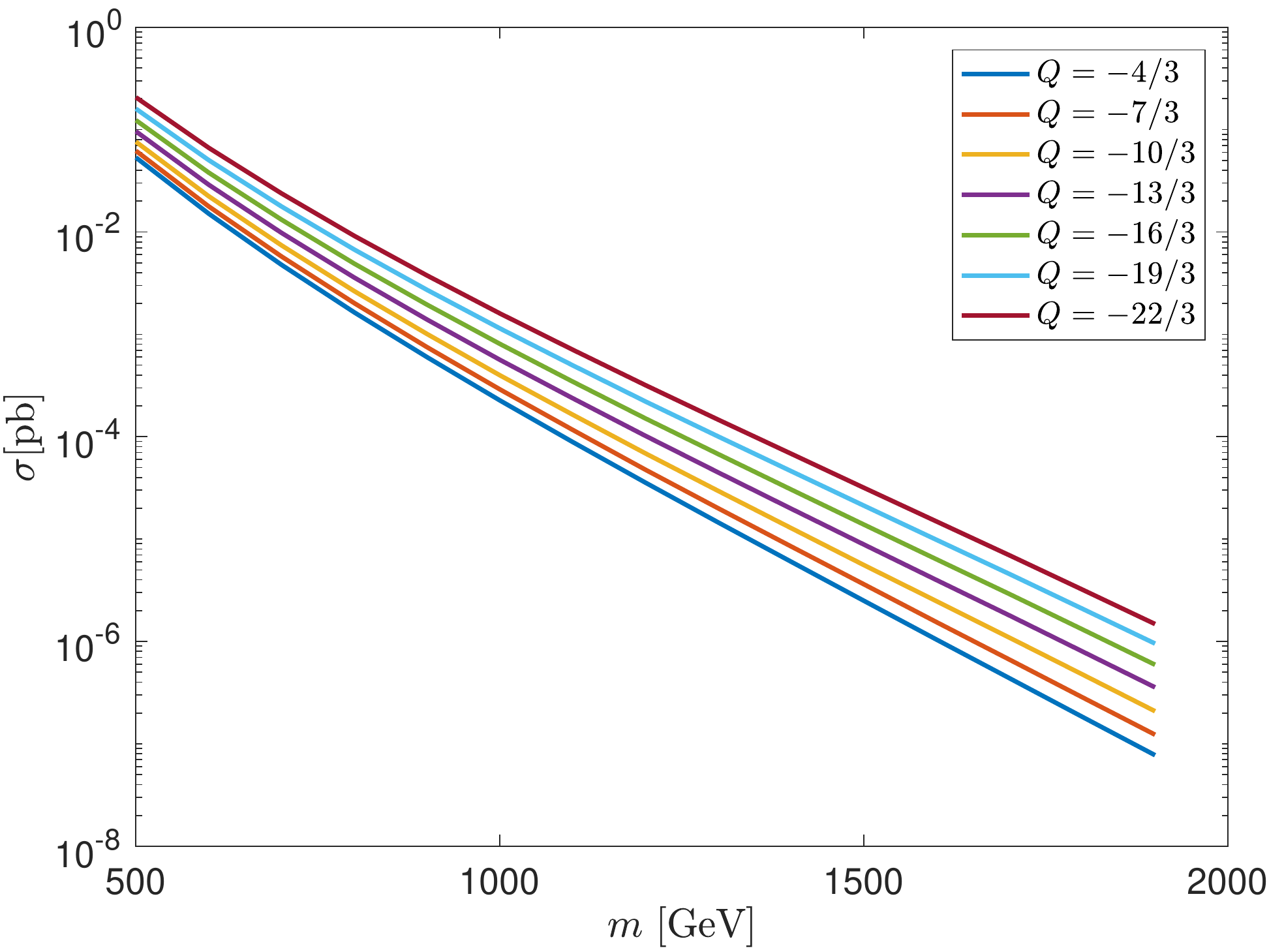}}
	\subfigure[Positvely-charged colored fermions.]{\includegraphics[scale=0.37]{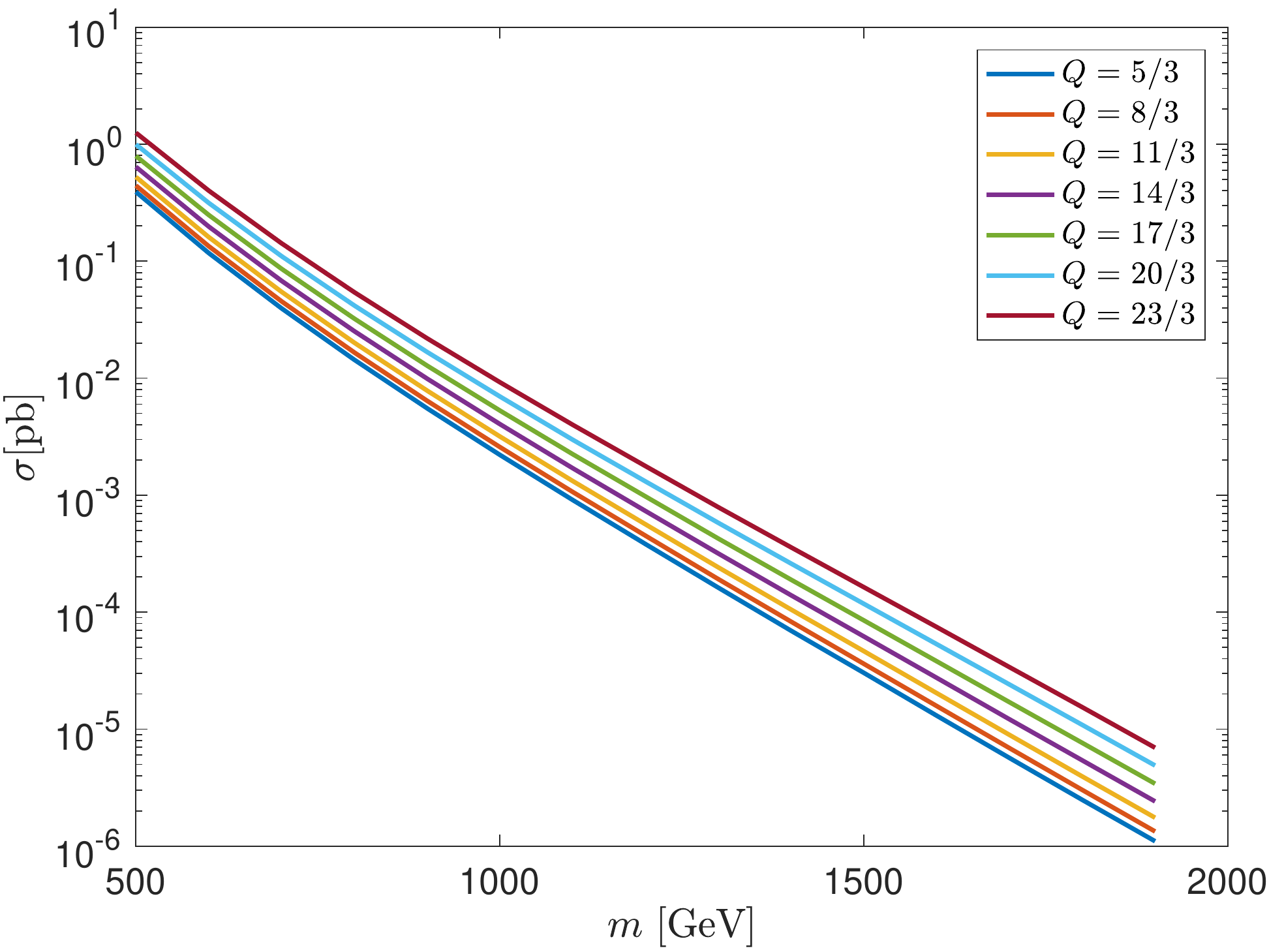}}
	\subfigure[Negatively-charged colored fermions.]{\includegraphics[scale=0.37]{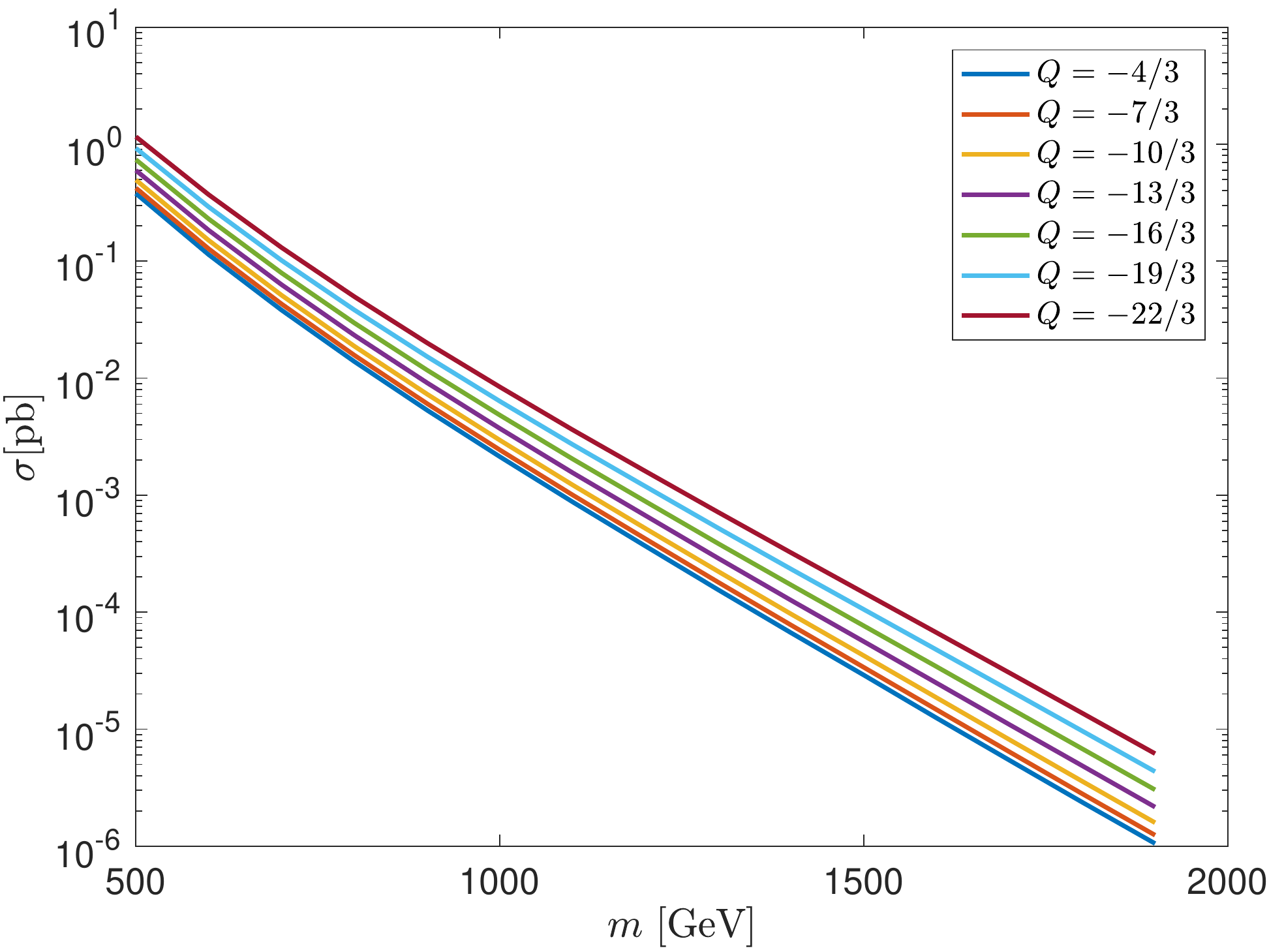}}
	\subfigure[Colorless fermions.]{	\includegraphics[scale=0.37]{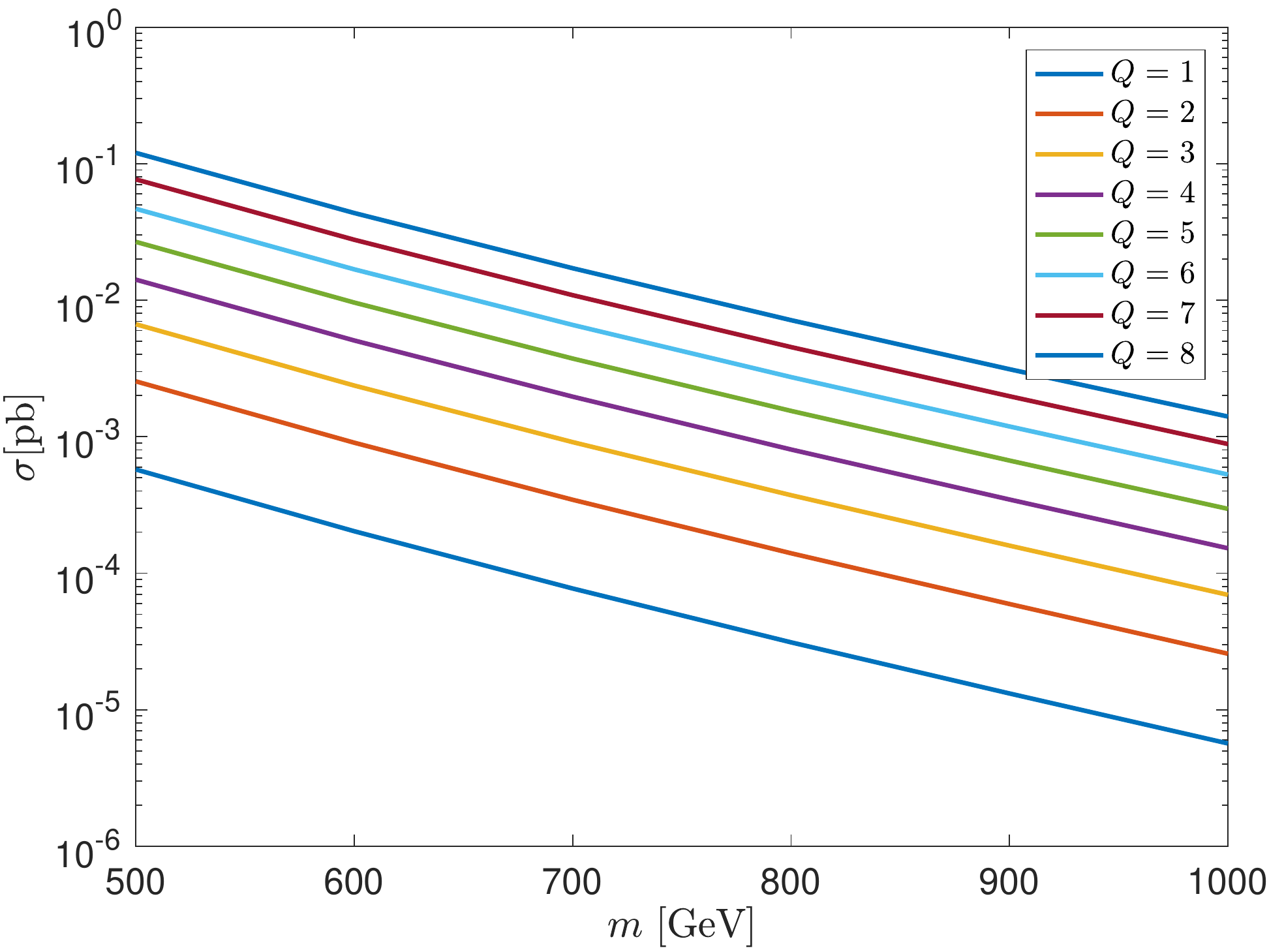}}
	\caption{Open-production cross sections at $\sqrt{s}=8$~TeV.}
	\label{cross_section_8}
\end{figure}

\begin{figure}[H]
	\centering
	\subfigure[Positvely-charged colored scalars.]{\includegraphics[scale=0.37]{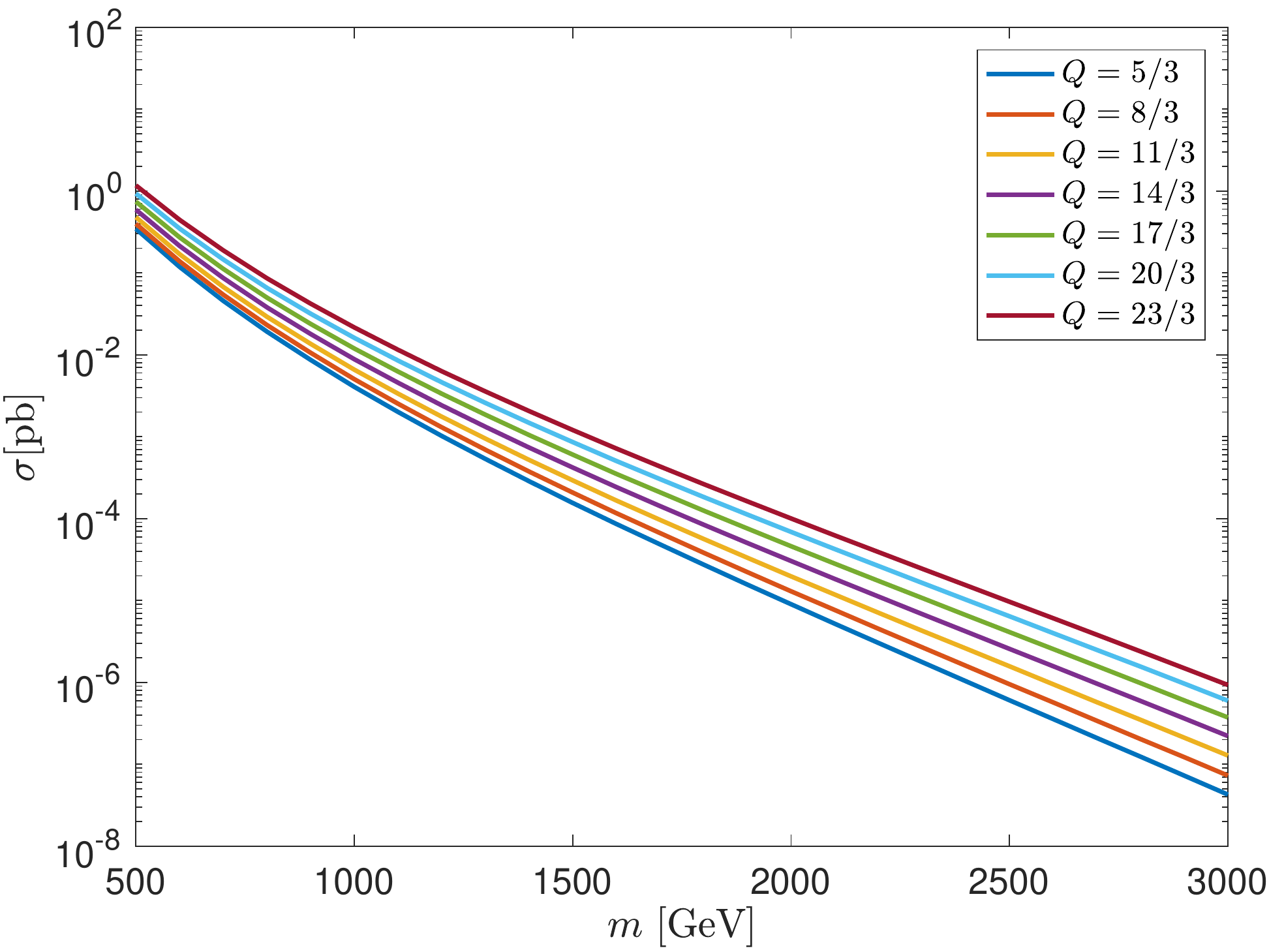}}
	\subfigure[Negatively-charged colored scalars.]{\includegraphics[scale=0.37]{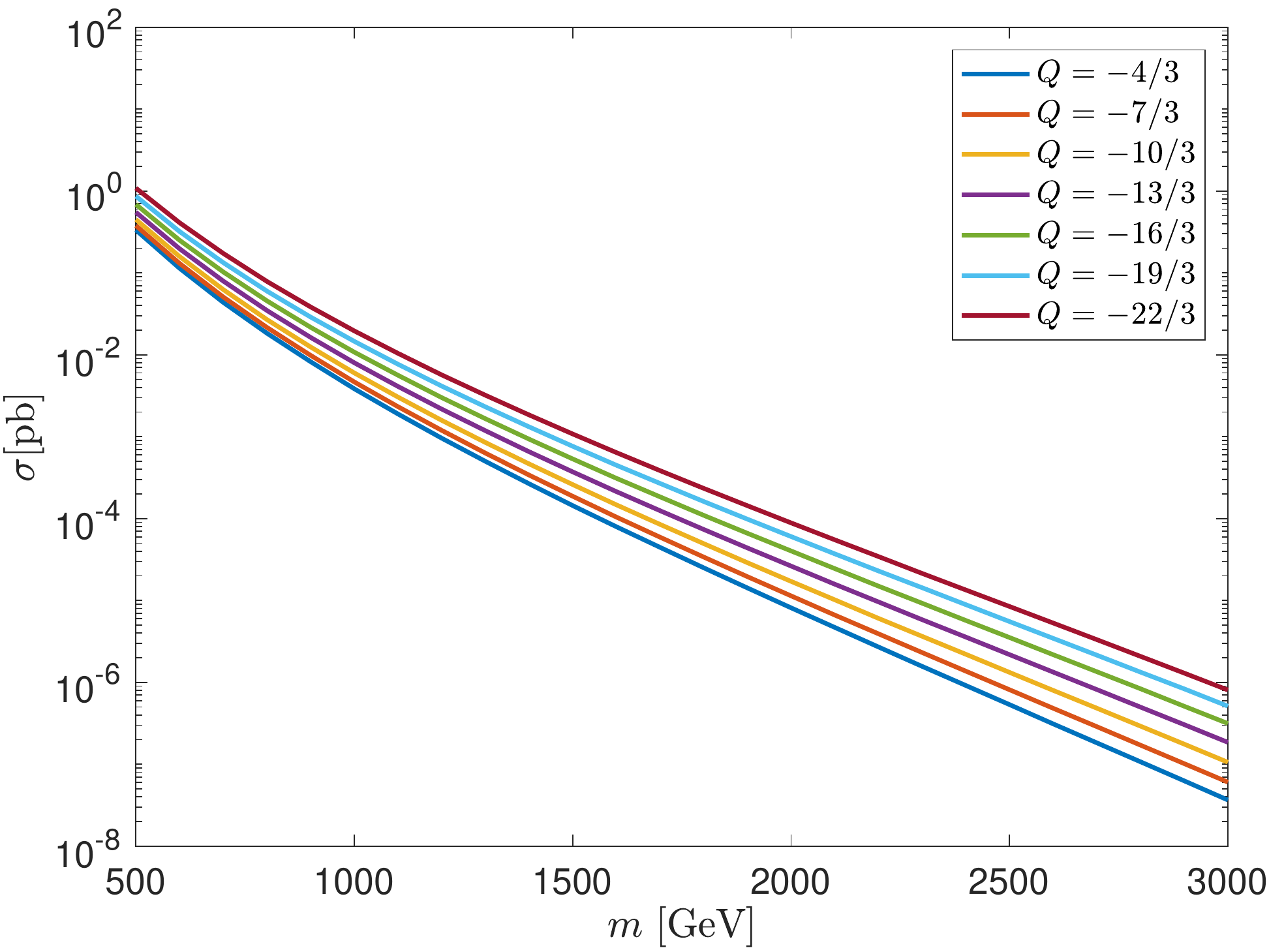}}
	\subfigure[Positvely-charged colored fermions.]{\includegraphics[scale=0.37]{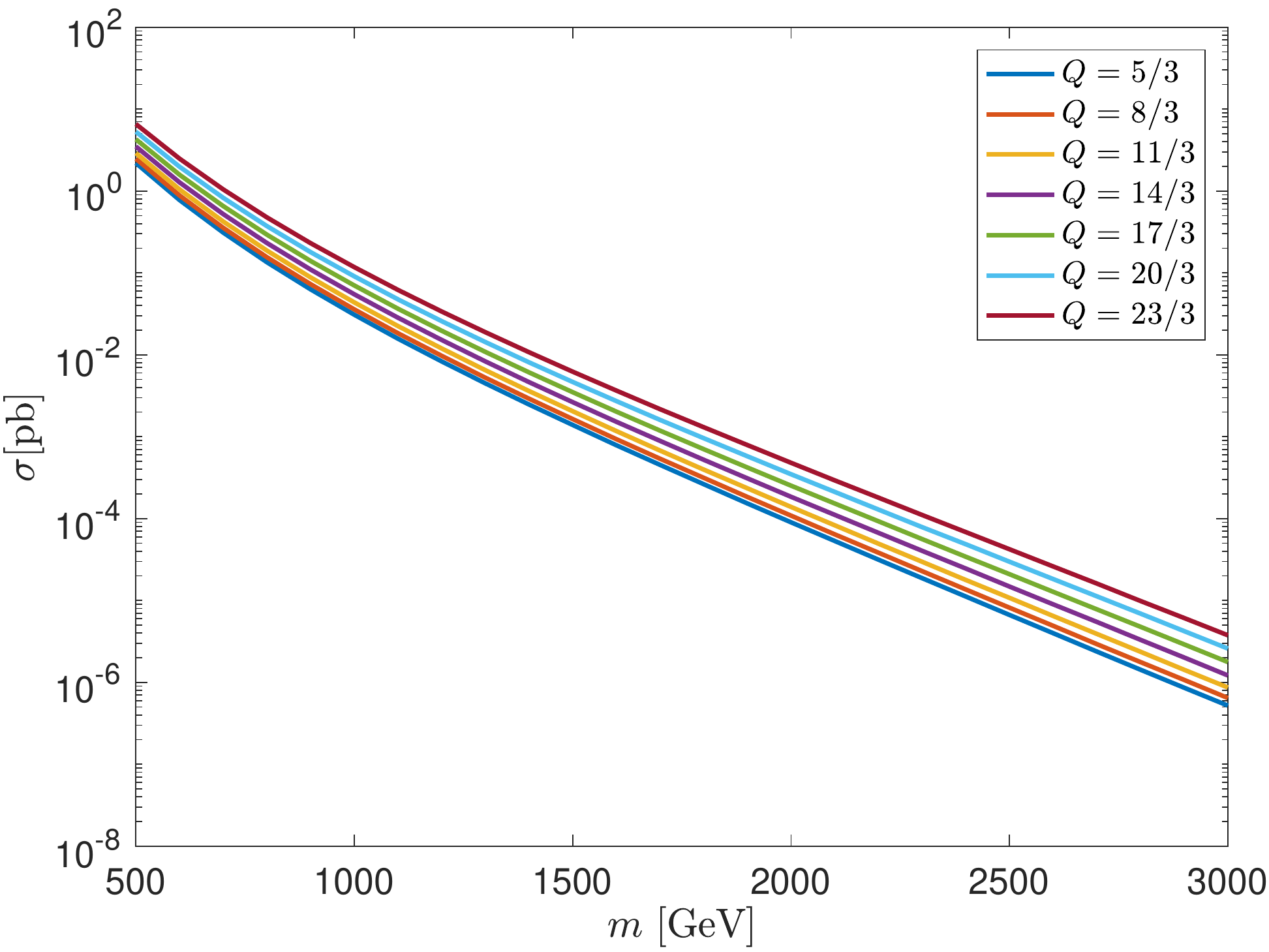}}
	\subfigure[Negatively-charged colored fermions.]{\includegraphics[scale=0.37]{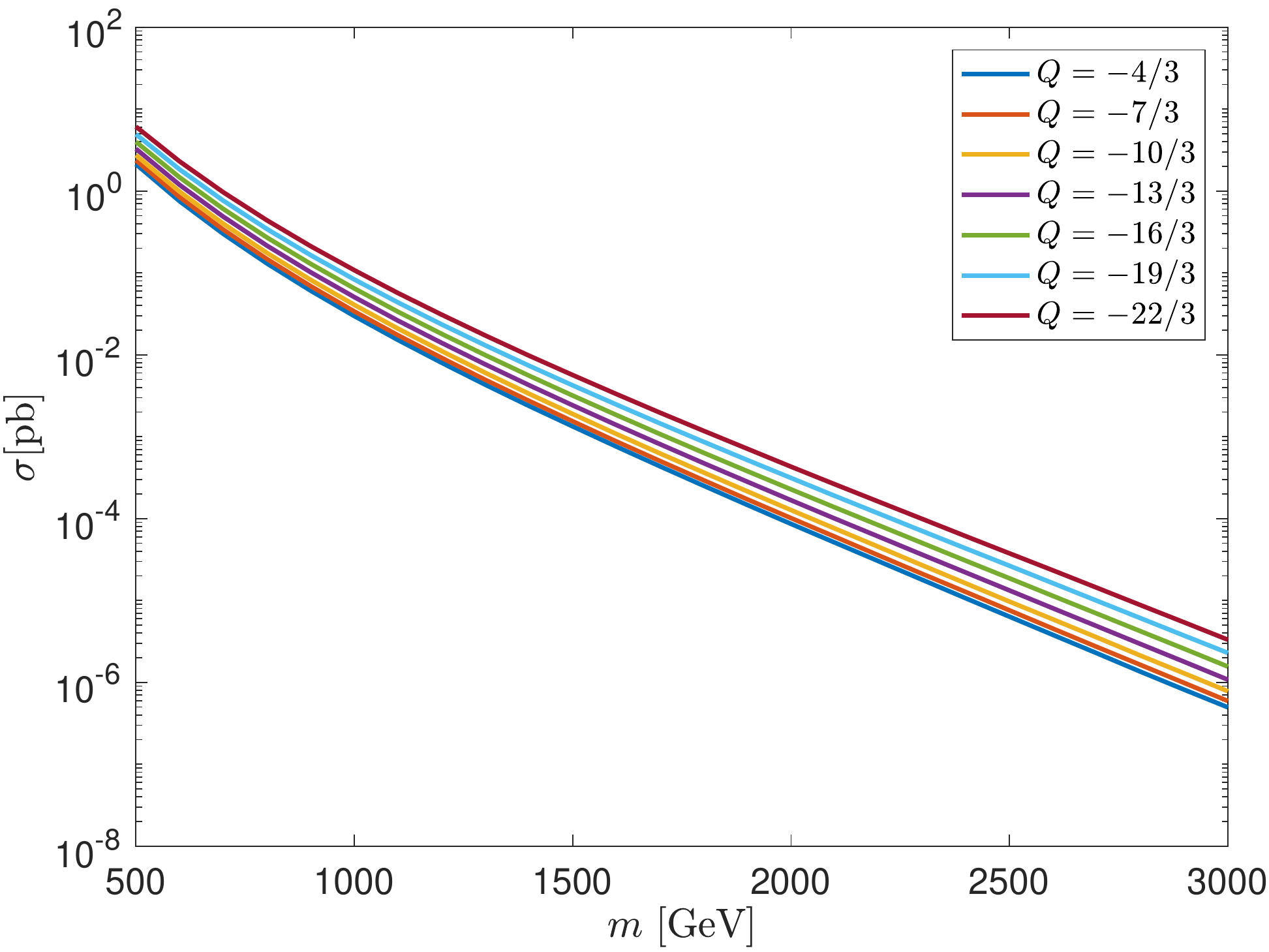}}
	\subfigure[Colorless fermions.]{	\includegraphics[scale=0.37]{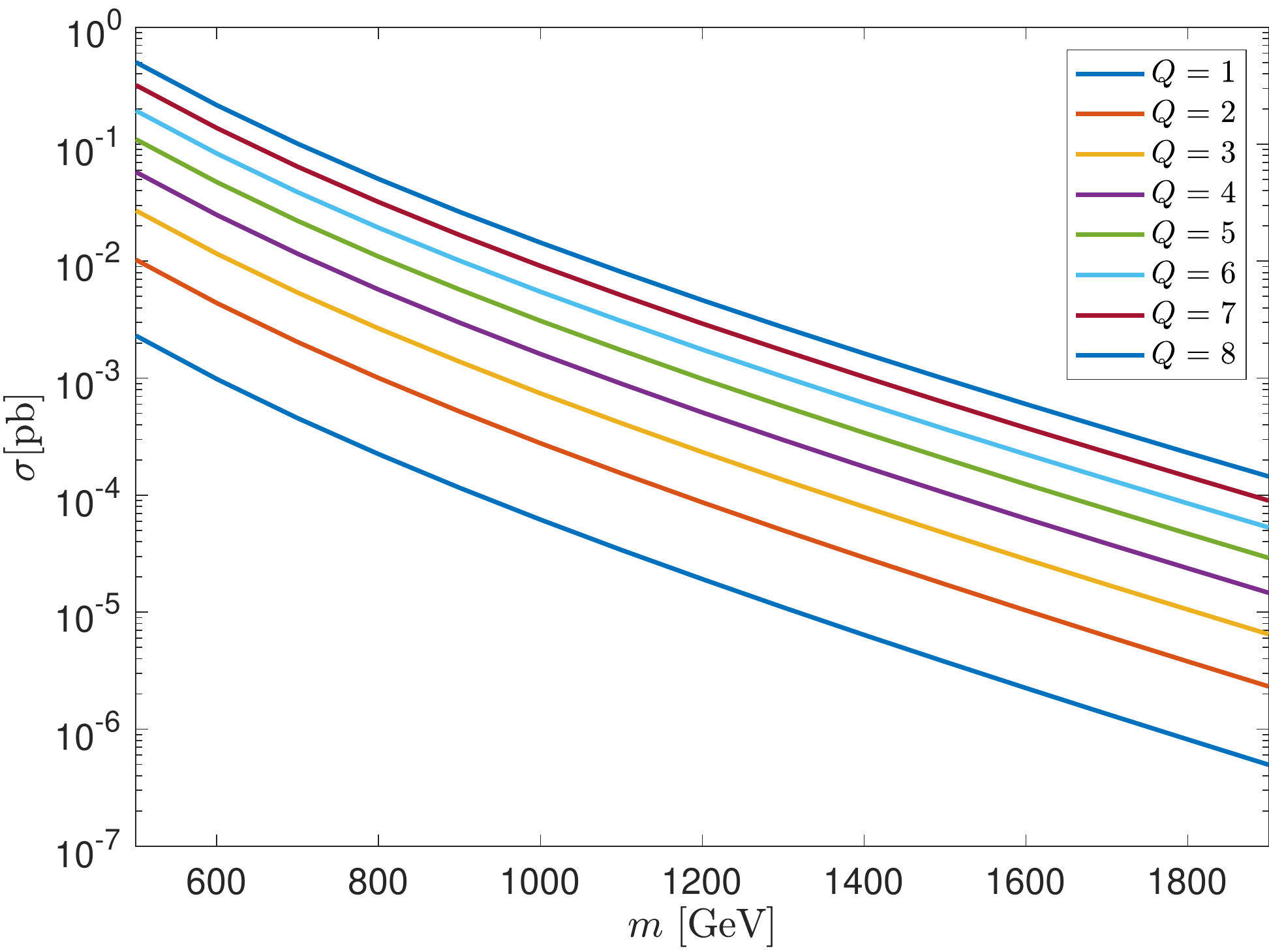}}
	\caption{Open-production cross sections at $\sqrt{s}=13$~TeV. }
	\label{cross_section_13}
\end{figure}

The cross sections for above-threshold pair-production of \acp{MCHSP} are presented for $\sqrt{s}=8$~TeV and $\sqrt{s}=13$~TeV in Figs.~\ref{cross_section_8} and~\ref{cross_section_13}.

\subsection{Simplified Efficiency Calculation}

\subsubsection[Time of Flight Calculation]{\ac{TOF} Calculation}\label{TOF}
\begin{figure}
	\centering
	\subfigure[]{\includegraphics[scale=0.27]{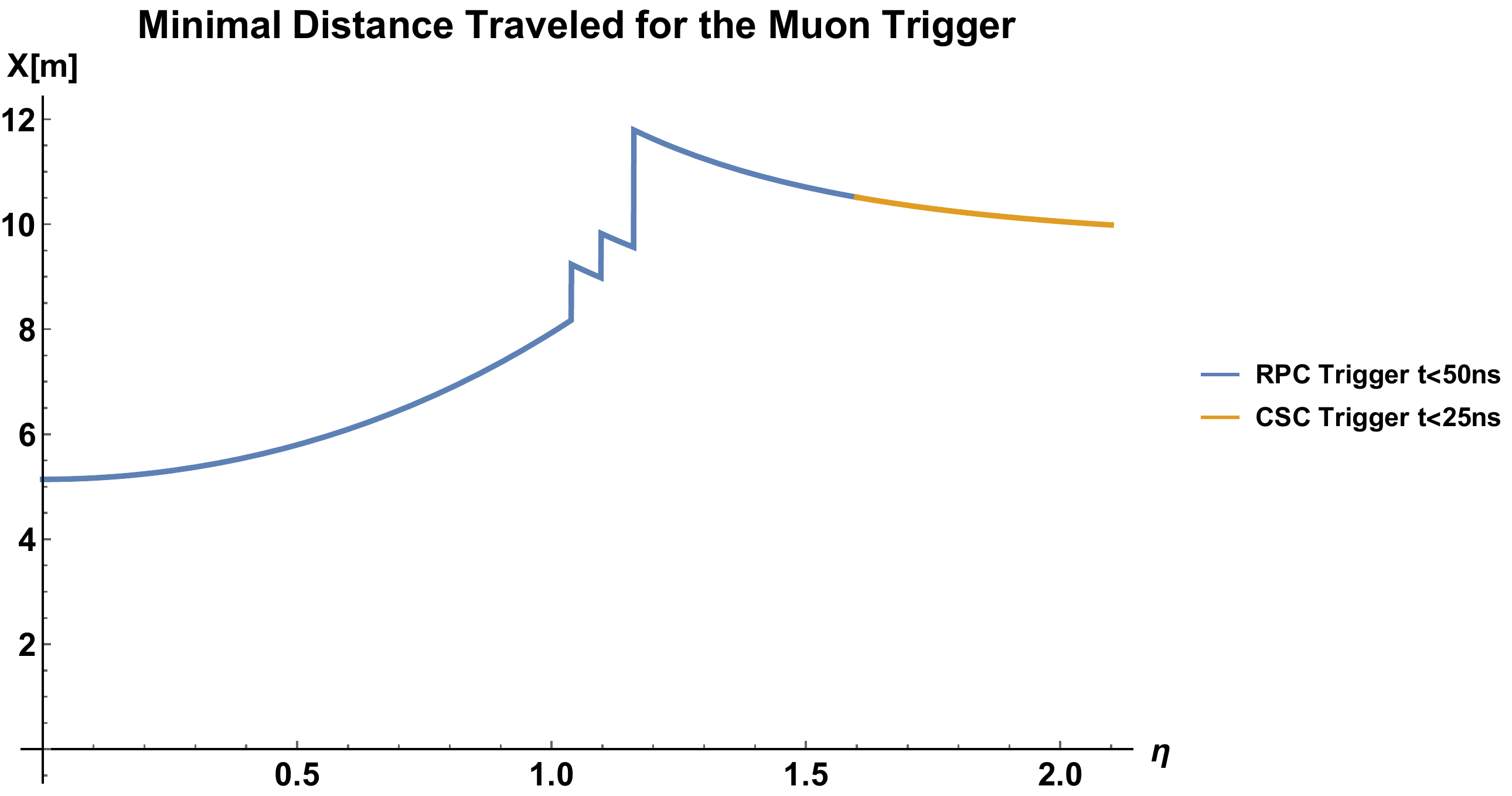}\label{Trigger_Distance}}
	\subfigure[]{\includegraphics[scale=0.27]{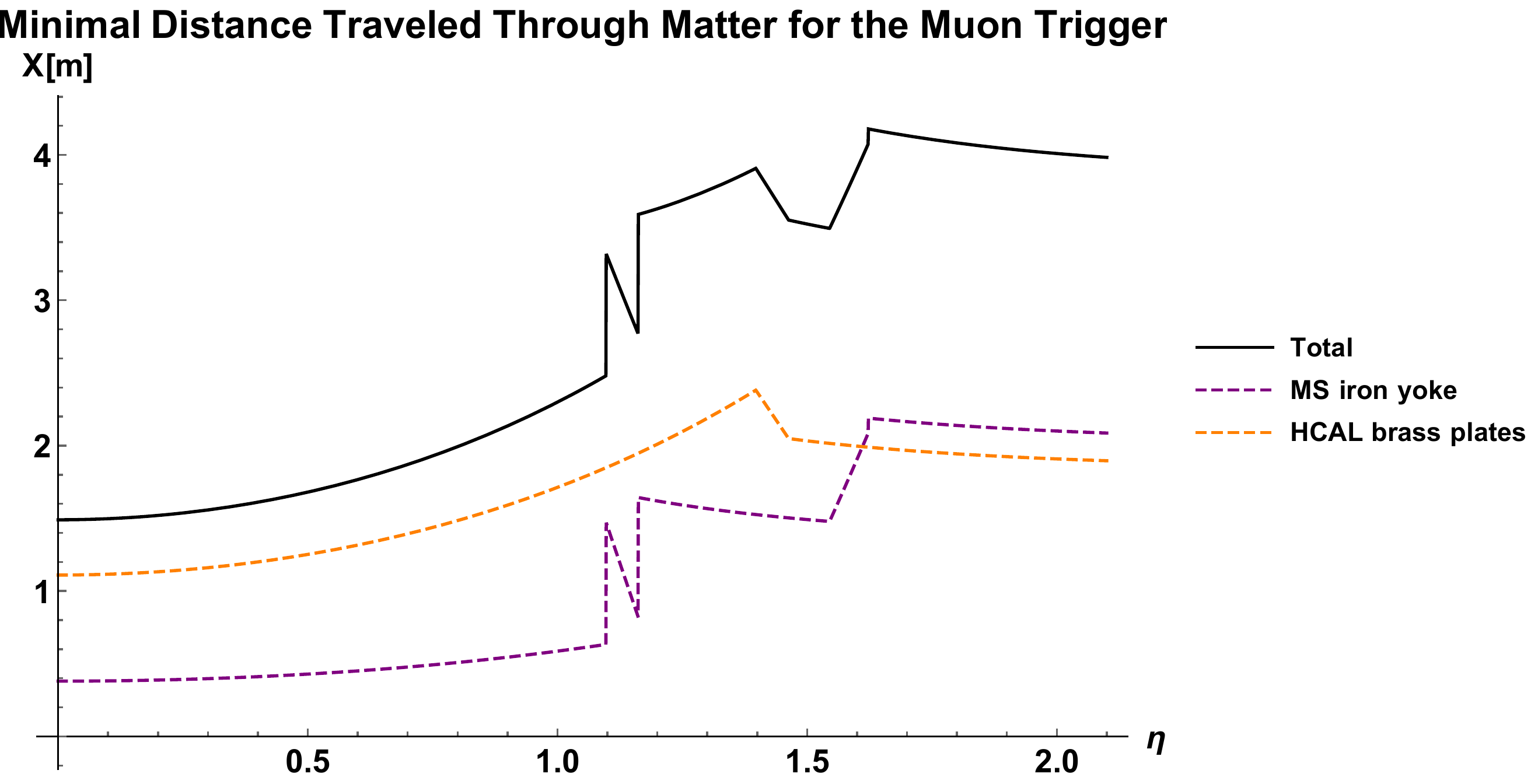}\label{Matter_Distance}}
	\caption{
		(a) Minimal distance traveled within the muon trigger time window for high momentum tracks as a function of $\eta$.
		(b) Distance traveled in matter, relevant for ionization energy loss, within the muon trigger time window as a function of $\eta$. Both (a) and (b) are based on the layout given in~\cite{RPCpicture}.} 
\end{figure}

To determine whether a candidate particle is accepted by the muon trigger,  we calculate its corresponding \ac{TOF} by
\begin{align}
\label{tof_eq} c\cdot t_\text{TOF}&=\dfrac{\gamma_0}{\sqrt{\gamma_0^2-1}}\cdot x^0_\text{HCAL}+\int_0^{x^f_\text{HCAL}-x^0_\text{HCAL}}\dfrac{\gamma_\text{Brass}}{\sqrt{\gamma_\text{Brass}^2-1}}dx +\\ \nonumber
&\dfrac{\gamma_\text{Brass}(x^f_\text{HCAL})}{\sqrt{\gamma_\text{Brass}(x^f_\text{HCAL})^2-1}}\cdot(x_\text{trigger}-x^f_\text{HCAL}-\Delta x_\text{IY}) +
\int_0^{\Delta x_\text{IY}}\dfrac{\gamma_\text{Iron}}{\sqrt{\gamma_\text{Iron}^2-1}}dx\,,
\end{align}
where $x_\text{trigger}$ is the minimal distance a particle must travel, within the trigger time window, in order to be triggered as a muon. As explained in Section~\ref{sec:efficiency}, $x_\text{trigger}$ is $\eta$-dependent and it is presented in Figure.~\ref{Trigger_Distance}. $ x^0_\text{HCAL}$, $x^f_\text{HCAL}$ are, respectively --  the distance a particle would travel to the entrance and to the exit of the \ac{HCAL}. The minimal distance a triggering particle would travel in the brass absorber of the \ac{HCAL}, $x^f_\text{HCAL}-x^0_\text{HCAL}$, and in the iron absorber of the iron yoke, $\Delta x_\text{IY}$, are also $\eta$-dependent and are shown in Fig.~\ref{Matter_Distance}. $\gamma(x)$ is the Lorentz factor $\gamma=1/\sqrt{1-\beta^2}$, and it is calculated by numerically solving
\begin{align}
\dfrac{d\gamma_\text{Brass}}{dx}(x)&=\dfrac{Q^2}{m}\dfrac{dE}{dx}_\text{Brass}(\gamma) ,& \gamma_\text{Brass}(0)&=\gamma_0\\ 
\dfrac{d\gamma_\text{Iron}}{dx}(x)&=\dfrac{Q^2}{m}\dfrac{dE}{dx}_\text{Iron}(\gamma) ,& \gamma_\text{Iron}(0)&=\gamma_\text{Brass}(x^f_\text{HCAL}-x^0_\text{HCAL}) \label{gamma_calc}\,,
\end{align}
where $\gamma_0$ is $\gamma$ at production, $Q$ is the charge of the particle and $m$ is the mass of the particle. $dE/dx$ is the energy loss function in the appropriate material for $Q=1$, and is taken from~\cite{brass_energy_loss} (brass) and~\cite{iron_energy_loss} (iron).

\subsubsection{Straight Tracks Approximation}
\label{curved}
We treat candidates as moving in straight lines, since the bending due to the magnetic field is negligible for particles passing the $p_T/Q\geq40$ selection. A particle tracing a curved track of radius $R$ would travel a distance $l$ in the $r-\theta$ plane before propagating $\Delta r$ in the radial direction, where
\begin{align}
l&=R\arcsin{\frac{\Delta r}{R}}\,,\\
R&=\frac{p_T}{Q\cdot B\cdot 0.303}\,.
\end{align}
The magnetic field in the CMS detector is about 2 T in the \ac{MS} and 3.8 T in the \ac{ID}~\cite{magnetic_field_cms}. Assuming a maximal 4 T magnetic field, the $p_T$ cut allows minimal $R$ of
\begin{align}
R_\text{min}=\frac{40}{4\cdot 0.303}\approx 33.00\,.
\end{align}
Consider the maximal possible $\Delta r$ distance, which is from the interaction point to the furthest RPC at $\Delta r_\text{max}\approx7$~m, 
\begin{align}
\frac{l}{\Delta r}\Big |_\text{max}&=\dfrac{\arcsin{\frac{\Delta r}{R}}\Big|_\text{max}}{\dfrac{\Delta r}{R}\Big|_\text{max}}\approx 1.0077\,,
\end{align}
which is indeed a negligible correction to the distance traveled in a straight track.

\subsubsection{Global Muon Offline Selection}
\label{global_muon}

In the analysis by CMS, the fraction of particles passing the global-muon selection, relative to the total number of particles produced, is given by
\begin{align}
\epsilon^\text{CMS}_{\text{particles}_\text{global-muon}}&=\epsilon^\text{CMS}_\text{online}\cdot\epsilon^\text{CMS}_{\text{offline}_\text{global-muon}}\nonumber\\
&=\frac{\text{Events}^\text{CMS}(\text{muon-trigger}\cup E_T^\text{miss})}{\text{Events}}\cdot\epsilon^\text{CMS}_{\text{offline}_\text{global-muon}}\,,
\end{align}
where $\epsilon^\text{CMS}_\text{online}$ is the fraction of events passing the online selection, relative to the total number of events. $\epsilon^\text{CMS}_{\text{offline}_\text{global-muon}}$ is the fraction of particles passing the global-muon criterion, out of the particles passing the online selection. $\text{\textit{Events}}$ is the total number of events and $\text{Events}^\text{CMS}(\text{selection})$ is the number of events passing a selection.
We claim that the particle-level global-muon efficiency can be written as 
\begin{align}
\epsilon^\text{CMS}_{\text{particles}_\text{global-muon}}&=\frac{\text{Events}^\text{CMS}(\text{muon-trigger})}{\text{Events}} \cdot\frac{\epsilon^\text{CMS}_\text{online}\cdot\epsilon^\text{CMS}_{\text{offline}_\text{global-muon}}}{\epsilon^\text{CMS}_{\text{events}_\text{muon-trigger}}}\nonumber\\
&\equiv\alpha^\text{CMS}(m,q)\cdot\epsilon^\text{CMS}_{\text{events}_\text{muon-trigger}}=f(m,q)\cdot\epsilon^\text{CMS}_{\text{particles}_\text{muon-trigger}}\,,
\end{align}
where $\epsilon^\text{CMS}_{\text{events}_\text{muon-trigger}}$ is the fraction of events passing the muon-trigger selection, relative to the total number of events. $\epsilon^\text{CMS}_{\text{particles}_\text{muon-trigger}}$ is the fraction of particles satisfying the muon-trigger requirements, relative to the total number of particles produced, and we hypothesize $f(m,q)\approx 1$. 

In our simplified efficiency calculation, we accept only particles that individually satisfy the muon trigger requirements, and omit the global muon selection. So
\begin{align}
\epsilon^\text{sim}_{\text{particles}_\text{global-muon}}&=\frac{\text{Particles}^\text{sim}(\text{muon-trigger})}{\text{Particles}}\nonumber\\
&=\frac{\text{Events}^\text{sim}(\text{muon-trigger})}{\text{Events}}\cdot\frac{\epsilon^\text{sim}_{\text{particle}s_\text{muon-trigger}}}{\epsilon^\text{sim}_{\text{events}_\text{muon-trigger}}}
\nonumber\\
&\equiv\alpha^{sim}(m,q)\cdot\epsilon^\text{sim}_{\text{events}_\text{muon-trigger}\,.}
\end{align}

\begin{figure}[H]
	\centering
	\includegraphics[scale=0.42]{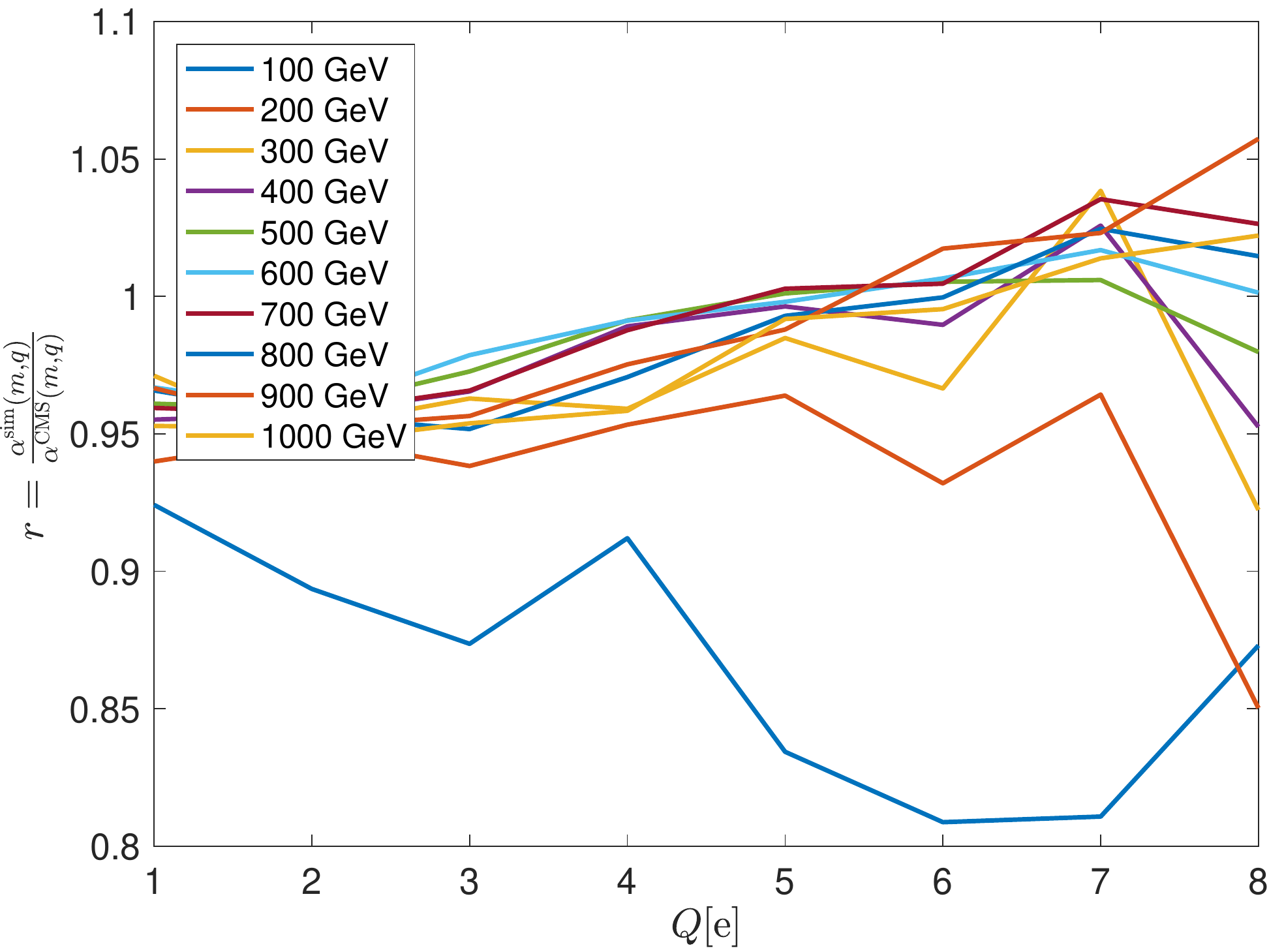}
	\caption{$r=\alpha^\text{sim}(m,q)/\alpha^\text{CMS}(m,q)$, the ratio of multiplicative factors required to convert the muon trigger event efficiency into the global-muon offline particle efficiency for our procedure, and for CMS.}
	\label{muon_trigger_pr_ratio}
\end{figure}

To check the validity of our assumption, independently of our muon-trigger simulation, we calculate the ratio between $\alpha^\text{sim}(m,q)$ and $\alpha^\text{CMS}(m,q)$
\begin{align}
r=\frac{\alpha^\text{sim}(m,q)}{\alpha^\text{CMS}(m,q)}=	\frac{\frac{\epsilon^\text{sim}_{\text{particle}s_\text{muon-trigger}}}{\epsilon^\text{sim}_{\text{events}_\text{muon-trigger}}}}{\frac{\epsilon^\text{CMS}_\text{online}\cdot\epsilon^\text{CMS}_{\text{offline}_\text{global-muon}}}{\epsilon^\text{CMS}_{\text{events}_\text{muon-trigger}}}}\,.
\end{align}
where the $\epsilon^\text{CMS}$ efficiencies are taken from~\cite{thesis}, and $\epsilon^\text{sim}$ efficiencies are obtained from our calculation. Indeed, as seen in Fig.~\ref{muon_trigger_pr_ratio}, $r\approx 1$ for all masses and charges for $\sqrt{s}=8$~TeV. Therefore, we conclude that accepting only particles passing the muon-trigger requirements to be subject for further selection is a reasonable approximation for $\sqrt{s}=8$~TeV. 

\subsubsection{Efficiency Values}\label{efficiency_values}
Here we list the final efficiencies, resulting from our simplified calculation described in Section~\ref{sec:efficiency}. The values for a $\sqrt{s}=8$~TeV search are given for color-triplet scalars, color-triplet fermions and color-singlet fermions in Tables~\ref{eff_scalars_8},~\ref{eff_fermions_8} and~\ref{eff_leptons_8}, respectively.  The values for a future search at $\sqrt{s}=13$~TeV are given in Tables~\ref{eff_scalars_13},~\ref{eff_fermions_13} and~\ref{eff_leptons_13}, respectively. Masses are in units of GeV.

\begin{table}[H]
	\centering
	\(\begin{array}{|c|cccccccccccccc|}
	\hline
	m/Q& -\frac{22}{3} & -\frac{19}{3} & -\frac{16}{3} & -\frac{13}{3} & -\frac{10}{3}
	& -\frac{7}{3} & -\frac{4}{3} & \frac{5}{3} & \frac{8}{3} & \frac{11}{3} &
	\frac{14}{3} & \frac{17}{3} & \frac{20}{3} & \frac{23}{3} \\ \hline
	500 & 0.043 & 0.071 & 0.13 & 0.22 & 0.34 & 0.46 & 0.55 & 0.55 & 0.44 & 0.32 & 0.20 &
	0.12 & 0.072 & 0.038 \\
	600 & 0.055 & 0.091 & 0.16 & 0.25 & 0.39 & 0.50 & 0.59 & 0.58 & 0.50 & 0.36 & 0.23 &
	0.14 & 0.084 & 0.049 \\
	700 & 0.065 & 0.11 & 0.17 & 0.28 & 0.41 & 0.54 & 0.61 & 0.60 & 0.52 & 0.39 & 0.26 & 0.17
	& 0.097 & 0.059 \\
	800 & 0.071 & 0.11 & 0.19 & 0.30 & 0.43 & 0.55 & 0.63 & 0.62 & 0.53 & 0.41 & 0.28 & 0.17
	& 0.10 & 0.067 \\
	900 & 0.076 & 0.12 & 0.20 & 0.31 & 0.44 & 0.56 & 0.65 & 0.64 & 0.54 & 0.42 & 0.29 & 0.18
	& 0.11 & 0.069 \\
	1000 & 0.074 & 0.12 & 0.20 & 0.32 & 0.45 & 0.56 & 0.63 & 0.63 & 0.54 & 0.42 & 0.29 &
	0.18 & 0.11 & 0.068 \\
	1100 & 0.073 & 0.12 & 0.20 & 0.31 & 0.45 & 0.57 & 0.65 & 0.64 & 0.56 & 0.43 & 0.30 &
	0.19 & 0.11 & 0.065 \\
	1200 & 0.074 & 0.12 & 0.20 & 0.31 & 0.45 & 0.57 & 0.65 & 0.65 & 0.55 & 0.43 & 0.29 &
	0.18 & 0.11 & 0.066 \\
	1300 & 0.070 & 0.12 & 0.19 & 0.31 & 0.45 & 0.56 & 0.66 & 0.64 & 0.55 & 0.42 & 0.29 &
	0.18 & 0.11 & 0.062 \\
	1400 & 0.067 & 0.11 & 0.19 & 0.31 & 0.45 & 0.57 & 0.66 & 0.65 & 0.55 & 0.43 & 0.28 &
	0.17 & 0.099 & 0.058 \\
	1500 & 0.059 & 0.10 & 0.18 & 0.29 & 0.44 & 0.56 & 0.65 & 0.63 & 0.54 & 0.40 & 0.28 &
	0.16 & 0.090 & 0.055 \\
	1600 & 0.054 & 0.094 & 0.17 & 0.28 & 0.42 & 0.56 & 0.65 & 0.63 & 0.53 & 0.40 & 0.26 &
	0.15 & 0.087 & 0.048 \\
	1700 & 0.047 & 0.087 & 0.16 & 0.27 & 0.41 & 0.55 & 0.64 & 0.63 & 0.52 & 0.39 & 0.25 &
	0.14 & 0.079 & 0.041 \\
	1800 & 0.040 & 0.079 & 0.15 & 0.26 & 0.40 & 0.53 & 0.64 & 0.61 & 0.52 & 0.38 & 0.24 &
	0.13 & 0.071 & 0.040 \\
	1900 & 0.039 & 0.072 & 0.13 & 0.25 & 0.39 & 0.52 & 0.63 & 0.61 & 0.50 & 0.37 & 0.23 &
	0.12 & 0.062 & 0.035 \\ \hline
	\end{array} \)
	\caption{Efficiencies for color-triplet scalars at $\sqrt{s}=8$~TeV.}
		\label{eff_scalars_8}
\end{table}

\begin{table}[H]
	\centering
	\(
	\begin{array}{|c|cccccccccccccc|}
	\hline
	m/Q & -\frac{22}{3} & -\frac{19}{3} & -\frac{16}{3} & -\frac{13}{3} & -\frac{10}{3} & -\frac{7}{3} & -\frac{4}{3} & \frac{5}{3} & \frac{8}{3} & \frac{11}{3} &
	\frac{14}{3} & \frac{17}{3} & \frac{20}{3} & \frac{23}{3} \\ \hline
	500 & 0.049 & 0.084 & 0.15 & 0.24 & 0.36 & 0.47 & 0.55 & 0.54 & 0.46 & 0.34 & 0.23 & 0.14 & 0.077 & 0.045 \\
	600 & 0.067 & 0.11 & 0.17 & 0.27 & 0.40 & 0.51 & 0.59 & 0.58 & 0.49 & 0.38 & 0.26 & 0.16 & 0.099 & 0.058 \\
	700 & 0.076 & 0.12 & 0.19 & 0.29 & 0.42 & 0.54 & 0.61 & 0.60 & 0.52 & 0.41 & 0.28 & 0.18 & 0.12 & 0.075 \\
	800 & 0.084 & 0.13 & 0.21 & 0.32 & 0.44 & 0.55 & 0.62 & 0.61 & 0.53 & 0.43 & 0.30 & 0.20 & 0.12 & 0.081 \\
	900 & 0.092 & 0.15 & 0.22 & 0.33 & 0.46 & 0.55 & 0.63 & 0.63 & 0.55 & 0.43 & 0.31 & 0.21 & 0.13 & 0.086 \\
	1000 & 0.094 & 0.14 & 0.22 & 0.33 & 0.45 & 0.55 & 0.62 & 0.61 & 0.53 & 0.44 & 0.31 & 0.21 & 0.14 & 0.087 \\
	1100 & 0.094 & 0.14 & 0.23 & 0.33 & 0.46 & 0.55 & 0.63 & 0.62 & 0.55 & 0.44 & 0.32 & 0.21 & 0.14 & 0.085 \\
	1200 & 0.089 & 0.14 & 0.23 & 0.34 & 0.46 & 0.56 & 0.62 & 0.62 & 0.55 & 0.45 & 0.32 & 0.21 & 0.13 & 0.087 \\
	1300 & 0.088 & 0.14 & 0.21 & 0.33 & 0.46 & 0.55 & 0.63 & 0.62 & 0.55 & 0.44 & 0.32 & 0.21 & 0.14 & 0.083 \\
	1400 & 0.085 & 0.13 & 0.22 & 0.33 & 0.46 & 0.55 & 0.62 & 0.60 & 0.54 & 0.43 & 0.31 & 0.21 & 0.12 & 0.081 \\
	1500 & 0.079 & 0.13 & 0.21 & 0.32 & 0.44 & 0.55 & 0.61 & 0.61 & 0.54 & 0.43 & 0.30 & 0.20 & 0.12 & 0.075 \\
	1600 & 0.074 & 0.12 & 0.20 & 0.31 & 0.44 & 0.53 & 0.61 & 0.60 & 0.52 & 0.41 & 0.30 & 0.19 & 0.11 & 0.070 \\
	1700 & 0.069 & 0.11 & 0.19 & 0.30 & 0.43 & 0.53 & 0.59 & 0.59 & 0.51 & 0.40 & 0.28 & 0.18 & 0.11 & 0.062 \\
	1800 & 0.062 & 0.11 & 0.18 & 0.29 & 0.41 & 0.51 & 0.58 & 0.57 & 0.50 & 0.40 & 0.27 & 0.16 & 0.10 & 0.054 \\
	1900 & 0.055 & 0.10 & 0.17 & 0.27 & 0.40 & 0.50 & 0.56 & 0.57 & 0.49 & 0.38 & 0.26 & 0.15 & 0.089 & 0.049 \\ \hline
	\end{array}
	\)
	\caption{Efficiencies for color-triplet fermions at $\sqrt{s}=8$~TeV.}
	\label{eff_fermions_8}
\end{table}

\begin{table}[H]
	\centering
	\(
	\begin{array}{|c|cccccccc|}
	\hline
	\text{m/Q} & 1 & 2 & 3 & 4 & 5 & 6 & 7 & 8 \\ \hline
	500 & 0.61 & 0.53 & 0.42 & 0.28 & 0.18 & 0.11 & 0.061 & 0.039 \\
	600 & 0.65 & 0.58 & 0.47 & 0.33 & 0.21 & 0.13 & 0.080 & 0.049 \\
	700 & 0.67 & 0.60 & 0.50 & 0.36 & 0.23 & 0.15 & 0.095 & 0.061 \\
	800 & 0.69 & 0.62 & 0.52 & 0.39 & 0.26 & 0.16 & 0.11 & 0.067 \\
	900 & 0.69 & 0.62 & 0.53 & 0.40 & 0.27 & 0.17 & 0.11 & 0.071 \\
	1000 & 0.68 & 0.60 & 0.53 & 0.41 & 0.28 & 0.17 & 0.11 & 0.072 \\ \hline
	\end{array}
	\)
	\caption{Efficiencies for color-singlet fermions at $\sqrt{s}=8$~TeV.}
		\label{eff_leptons_8}
\end{table}

\begin{table}[H]
	\centering
		\(
	\begin{array}{|c|cccccccccccccc|}
	\hline
	m/Q & -\frac{22}{3} & -\frac{19}{3} & -\frac{16}{3} & -\frac{13}{3} & -\frac{10}{3} & -\frac{7}{3} & -\frac{4}{3} & \frac{5}{3} & \frac{8}{3} & \frac{11}{3} &
	\frac{14}{3} & \frac{17}{3} & \frac{20}{3} & \frac{23}{3} \\ \hline
	500 & 0.041 & 0.069 & 0.12 & 0.19 & 0.28 & 0.40 & 0.48 & 0.47 & 0.38 & 0.27 & 0.18 & 0.10 & 0.061 & 0.036 \\
	600 & 0.058 & 0.088 & 0.14 & 0.22 & 0.33 & 0.43 & 0.50 & 0.50 & 0.42 & 0.31 & 0.21 & 0.13 & 0.083 & 0.054 \\
	700 & 0.072 & 0.11 & 0.17 & 0.25 & 0.36 & 0.46 & 0.52 & 0.52 & 0.45 & 0.34 & 0.24 & 0.16 & 0.10 & 0.067 \\
	800 & 0.084 & 0.12 & 0.19 & 0.28 & 0.39 & 0.47 & 0.54 & 0.54 & 0.47 & 0.37 & 0.27 & 0.18 & 0.11 & 0.077 \\
	900 & 0.094 & 0.14 & 0.21 & 0.31 & 0.42 & 0.50 & 0.55 & 0.55 & 0.49 & 0.41 & 0.29 & 0.19 & 0.13 & 0.091 \\
	1000 & 0.10 & 0.15 & 0.22 & 0.31 & 0.43 & 0.50 & 0.56 & 0.56 & 0.50 & 0.40 & 0.30 & 0.21 & 0.14 & 0.096 \\
	1100 & 0.11 & 0.15 & 0.23 & 0.34 & 0.45 & 0.53 & 0.58 & 0.57 & 0.51 & 0.43 & 0.31 & 0.22 & 0.14 & 0.10 \\
	1200 & 0.12 & 0.17 & 0.25 & 0.35 & 0.47 & 0.55 & 0.60 & 0.59 & 0.53 & 0.45 & 0.33 & 0.23 & 0.16 & 0.11 \\
	1300 & 0.12 & 0.17 & 0.25 & 0.37 & 0.47 & 0.55 & 0.61 & 0.60 & 0.54 & 0.45 & 0.34 & 0.24 & 0.16 & 0.12 \\
	1400 & 0.13 & 0.18 & 0.26 & 0.38 & 0.49 & 0.56 & 0.62 & 0.61 & 0.56 & 0.47 & 0.36 & 0.24 & 0.17 & 0.12 \\
	1500 & 0.13 & 0.18 & 0.27 & 0.38 & 0.50 & 0.58 & 0.64 & 0.63 & 0.57 & 0.48 & 0.35 & 0.25 & 0.17 & 0.12 \\
	1600 & 0.13 & 0.18 & 0.27 & 0.39 & 0.51 & 0.58 & 0.65 & 0.64 & 0.58 & 0.48 & 0.37 & 0.26 & 0.18 & 0.12 \\
	1700 & 0.14 & 0.19 & 0.27 & 0.39 & 0.51 & 0.60 & 0.66 & 0.64 & 0.59 & 0.50 & 0.37 & 0.26 & 0.18 & 0.12 \\
	1800 & 0.13 & 0.19 & 0.27 & 0.39 & 0.51 & 0.60 & 0.66 & 0.65 & 0.58 & 0.49 & 0.37 & 0.26 & 0.18 & 0.13 \\
	1900 & 0.13 & 0.18 & 0.28 & 0.39 & 0.51 & 0.60 & 0.66 & 0.66 & 0.58 & 0.49 & 0.37 & 0.26 & 0.17 & 0.13 \\
	2000 & 0.13 & 0.18 & 0.27 & 0.39 & 0.52 & 0.60 & 0.66 & 0.65 & 0.59 & 0.49 & 0.37 & 0.25 & 0.18 & 0.12 \\
	2100 & 0.13 & 0.18 & 0.26 & 0.39 & 0.51 & 0.60 & 0.67 & 0.66 & 0.58 & 0.49 & 0.37 & 0.25 & 0.17 & 0.12 \\
	2200 & 0.12 & 0.17 & 0.27 & 0.39 & 0.51 & 0.60 & 0.66 & 0.66 & 0.58 & 0.49 & 0.37 & 0.25 & 0.16 & 0.12 \\
	2300 & 0.11 & 0.17 & 0.26 & 0.38 & 0.51 & 0.60 & 0.66 & 0.65 & 0.58 & 0.49 & 0.35 & 0.24 & 0.16 & 0.11 \\
	2400 & 0.11 & 0.17 & 0.25 & 0.37 & 0.50 & 0.59 & 0.66 & 0.65 & 0.58 & 0.49 & 0.36 & 0.24 & 0.16 & 0.11 \\
	2500 & 0.11 & 0.16 & 0.25 & 0.37 & 0.50 & 0.58 & 0.66 & 0.65 & 0.57 & 0.48 & 0.35 & 0.23 & 0.15 & 0.10 \\
	2600 & 0.10 & 0.15 & 0.23 & 0.36 & 0.49 & 0.58 & 0.65 & 0.64 & 0.57 & 0.48 & 0.35 & 0.23 & 0.15 & 0.099 \\
	2700 & 0.097 & 0.15 & 0.23 & 0.35 & 0.49 & 0.57 & 0.65 & 0.64 & 0.55 & 0.46 & 0.34 & 0.21 & 0.14 & 0.094 \\
	2800 & 0.094 & 0.14 & 0.22 & 0.34 & 0.47 & 0.57 & 0.65 & 0.63 & 0.55 & 0.45 & 0.33 & 0.21 & 0.14 & 0.089 \\
	2900 & 0.089 & 0.14 & 0.22 & 0.34 & 0.46 & 0.56 & 0.64 & 0.63 & 0.54 & 0.45 & 0.32 & 0.20 & 0.13 & 0.085 \\
	3000 & 0.083 & 0.13 & 0.21 & 0.33 & 0.46 & 0.55 & 0.64 & 0.62 & 0.54 & 0.43 & 0.31 & 0.20 & 0.12 & 0.075 \\ \hline
	\end{array}
	\)
	\caption{Efficiencies for color-triplet scalars at $\sqrt{s}=13$~TeV.}
	\label{eff_scalars_13}
\end{table}

\begin{table}[H]
	\centering
	\(
	\begin{array}{|c|cccccccccccccc|}
	\hline
	m/Q & -\frac{22}{3} & -\frac{19}{3} & -\frac{16}{3} & -\frac{13}{3} & -\frac{10}{3} & -\frac{7}{3} & -\frac{4}{3} & \frac{5}{3} & \frac{8}{3} & \frac{11}{3} &
	\frac{14}{3} & \frac{17}{3} & \frac{20}{3} & \frac{23}{3} \\ \hline
	500 & 0.046 & 0.077 & 0.13 & 0.21 & 0.31 & 0.40 & 0.47 & 0.47 & 0.40 & 0.29 & 0.20 & 0.12 & 0.073 & 0.043 \\
	600 & 0.067 & 0.10 & 0.16 & 0.25 & 0.35 & 0.45 & 0.51 & 0.50 & 0.43 & 0.34 & 0.23 & 0.15 & 0.10 & 0.063 \\
	700 & 0.083 & 0.12 & 0.19 & 0.28 & 0.38 & 0.46 & 0.52 & 0.52 & 0.45 & 0.36 & 0.26 & 0.18 & 0.12 & 0.078 \\
	800 & 0.10 & 0.14 & 0.21 & 0.30 & 0.41 & 0.49 & 0.54 & 0.54 & 0.48 & 0.39 & 0.29 & 0.20 & 0.14 & 0.094 \\
	900 & 0.12 & 0.16 & 0.23 & 0.33 & 0.42 & 0.51 & 0.55 & 0.55 & 0.48 & 0.42 & 0.31 & 0.22 & 0.15 & 0.11 \\
	1000 & 0.12 & 0.16 & 0.24 & 0.34 & 0.44 & 0.50 & 0.56 & 0.56 & 0.50 & 0.43 & 0.33 & 0.23 & 0.16 & 0.12 \\
	1100 & 0.14 & 0.17 & 0.25 & 0.36 & 0.45 & 0.51 & 0.58 & 0.57 & 0.52 & 0.45 & 0.34 & 0.24 & 0.18 & 0.13 \\
	1200 & 0.14 & 0.19 & 0.27 & 0.37 & 0.48 & 0.54 & 0.59 & 0.58 & 0.53 & 0.46 & 0.35 & 0.26 & 0.18 & 0.13 \\
	1300 & 0.15 & 0.20 & 0.28 & 0.38 & 0.49 & 0.55 & 0.60 & 0.59 & 0.55 & 0.48 & 0.37 & 0.27 & 0.19 & 0.14 \\
	1400 & 0.15 & 0.21 & 0.29 & 0.39 & 0.49 & 0.56 & 0.61 & 0.60 & 0.55 & 0.48 & 0.38 & 0.27 & 0.20 & 0.14 \\
	1500 & 0.16 & 0.21 & 0.29 & 0.40 & 0.50 & 0.57 & 0.61 & 0.61 & 0.56 & 0.49 & 0.39 & 0.28 & 0.20 & 0.15 \\
	1600 & 0.16 & 0.22 & 0.30 & 0.41 & 0.52 & 0.58 & 0.62 & 0.62 & 0.57 & 0.50 & 0.39 & 0.29 & 0.21 & 0.15 \\
	1700 & 0.16 & 0.22 & 0.31 & 0.41 & 0.52 & 0.59 & 0.63 & 0.63 & 0.58 & 0.51 & 0.39 & 0.29 & 0.21 & 0.16 \\
	1800 & 0.16 & 0.22 & 0.30 & 0.41 & 0.52 & 0.59 & 0.63 & 0.63 & 0.58 & 0.51 & 0.40 & 0.29 & 0.21 & 0.16 \\
	1900 & 0.17 & 0.22 & 0.31 & 0.42 & 0.52 & 0.59 & 0.63 & 0.62 & 0.57 & 0.50 & 0.39 & 0.29 & 0.21 & 0.15 \\
	2000 & 0.16 & 0.22 & 0.31 & 0.42 & 0.52 & 0.58 & 0.64 & 0.63 & 0.58 & 0.52 & 0.40 & 0.29 & 0.21 & 0.16 \\
	2100 & 0.16 & 0.22 & 0.30 & 0.42 & 0.51 & 0.58 & 0.63 & 0.63 & 0.58 & 0.51 & 0.40 & 0.29 & 0.20 & 0.15 \\
	2200 & 0.16 & 0.22 & 0.31 & 0.41 & 0.53 & 0.59 & 0.63 & 0.62 & 0.58 & 0.51 & 0.40 & 0.28 & 0.20 & 0.15 \\
	2300 & 0.15 & 0.21 & 0.29 & 0.41 & 0.52 & 0.58 & 0.63 & 0.62 & 0.57 & 0.50 & 0.39 & 0.28 & 0.20 & 0.15 \\
	2400 & 0.15 & 0.20 & 0.29 & 0.40 & 0.52 & 0.57 & 0.62 & 0.61 & 0.57 & 0.50 & 0.38 & 0.27 & 0.20 & 0.14 \\
	2500 & 0.14 & 0.19 & 0.28 & 0.40 & 0.51 & 0.58 & 0.62 & 0.61 & 0.56 & 0.50 & 0.38 & 0.27 & 0.19 & 0.14 \\
	2600 & 0.14 & 0.20 & 0.28 & 0.39 & 0.50 & 0.57 & 0.62 & 0.60 & 0.56 & 0.49 & 0.37 & 0.26 & 0.19 & 0.13 \\
	2700 & 0.14 & 0.19 & 0.27 & 0.39 & 0.50 & 0.55 & 0.60 & 0.59 & 0.56 & 0.49 & 0.37 & 0.26 & 0.18 & 0.13 \\
	2800 & 0.13 & 0.18 & 0.26 & 0.38 & 0.49 & 0.55 & 0.59 & 0.60 & 0.54 & 0.47 & 0.36 & 0.25 & 0.17 & 0.12 \\
	2900 & 0.12 & 0.17 & 0.26 & 0.37 & 0.48 & 0.54 & 0.58 & 0.58 & 0.54 & 0.47 & 0.36 & 0.25 & 0.17 & 0.11 \\
	3000 & 0.11 & 0.17 & 0.26 & 0.36 & 0.48 & 0.53 & 0.57 & 0.57 & 0.53 & 0.46 & 0.35 & 0.24 & 0.16 & 0.11 \\ \hline
	\end{array}
	\)
\caption{Efficiencies for color-triplet fermions at $\sqrt{s}=13$~TeV.}
		\label{eff_fermions_13}
\end{table}

\begin{table}[H]
	\centering
	\(
	\begin{array}{|c|cccccccc|}
	\hline
	\text{m/Q} & 1 & 2 & 3 & 4 & 5 & 6 & 7 & 8 \\ \hline
	500 & 0.49 & 0.44 & 0.35 & 0.25 & 0.16 & 0.10 & 0.067 & 0.041 \\
	600 & 0.54 & 0.49 & 0.40 & 0.29 & 0.19 & 0.13 & 0.086 & 0.055 \\
	700 & 0.56 & 0.52 & 0.44 & 0.32 & 0.22 & 0.15 & 0.10 & 0.066 \\
	800 & 0.59 & 0.53 & 0.46 & 0.36 & 0.26 & 0.17 & 0.12 & 0.082 \\
	900 & 0.61 & 0.55 & 0.49 & 0.38 & 0.28 & 0.19 & 0.13 & 0.095 \\
	1000 & 0.61 & 0.55 & 0.51 & 0.41 & 0.29 & 0.20 & 0.14 & 0.10 \\
	1100 & 0.64 & 0.57 & 0.53 & 0.42 & 0.31 & 0.21 & 0.15 & 0.11 \\
	1200 & 0.65 & 0.58 & 0.55 & 0.44 & 0.33 & 0.23 & 0.16 & 0.12 \\
	1300 & 0.66 & 0.60 & 0.56 & 0.45 & 0.33 & 0.24 & 0.17 & 0.13 \\
	1400 & 0.67 & 0.60 & 0.57 & 0.46 & 0.34 & 0.24 & 0.18 & 0.13 \\
	1500 & 0.68 & 0.61 & 0.59 & 0.48 & 0.35 & 0.25 & 0.18 & 0.14 \\
	1600 & 0.66&  0.60 & 0.56 &  0.45 & 0.35 & 0.26 & 0.19& 0.15\\
	1700 & 0.66& 0.60 & 0.57 &  0.47 & 0.36 & 0.26& 0.19 & 0.15 \\
	1800 & 0.66 & 0.61 & 0.57 & 0.47 & 0.36 & 0.25 &  0.19  & 0.15 \\
	1900 &  0.67& 0.61  & 0.57 &  0.47 & 0.36 & 0.26 & 0.20 & 0.15 \\ \hline
	\end{array}
	\)
	\caption{Efficiencies for color-singlet fermions at $\sqrt{s}=13$~TeV.}
		\label{eff_leptons_13}
\end{table}

\subsection{ Effective Cross Sections}\label{effective_cross_sections_plots}
\begin{figure}
	\centering
	\subfigure[Positvely-charged colored scalars.]{\includegraphics[scale=0.34]{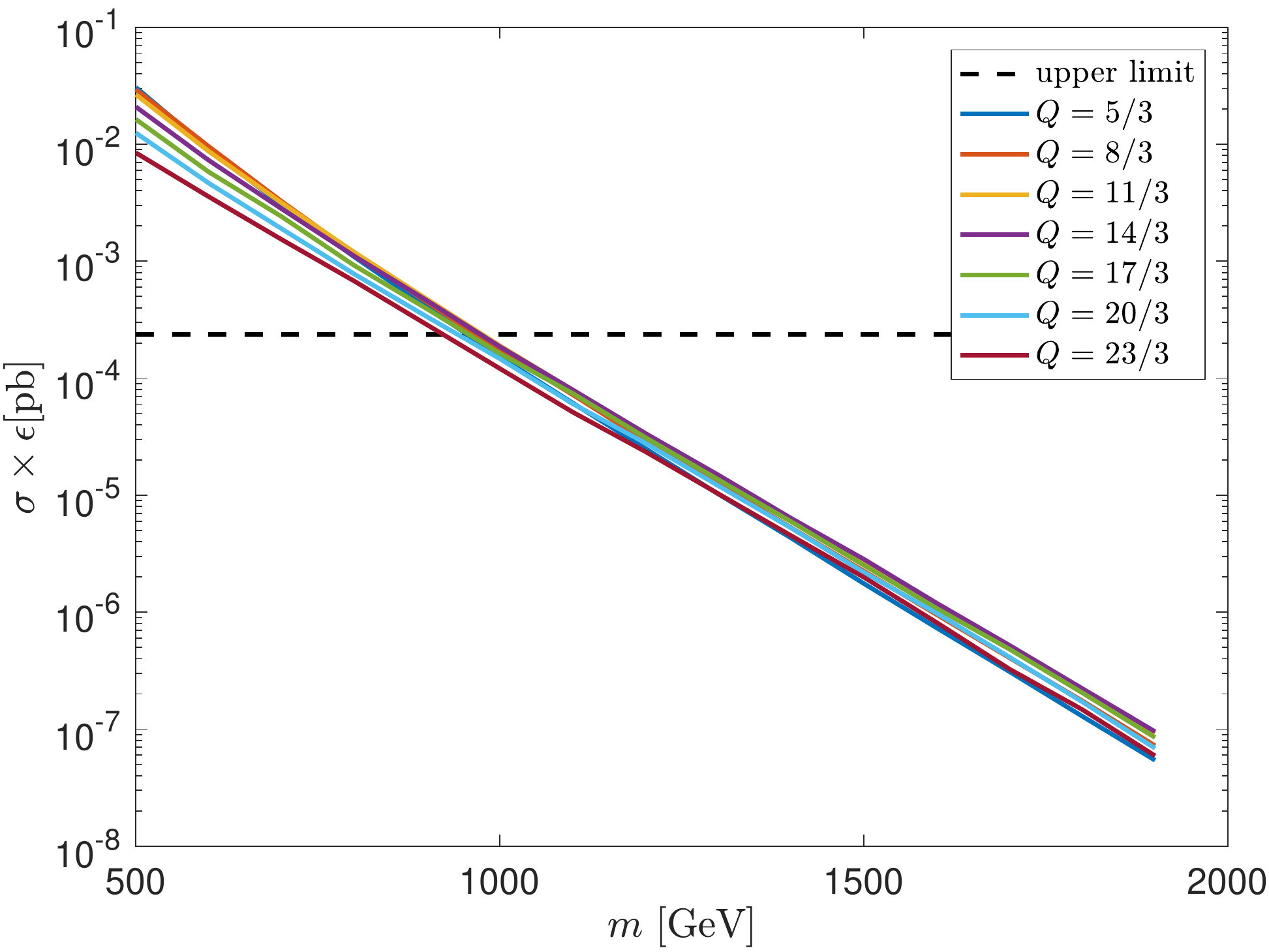}}
	\subfigure[Negatively-charged colored scalars.]{\includegraphics[scale=0.34]{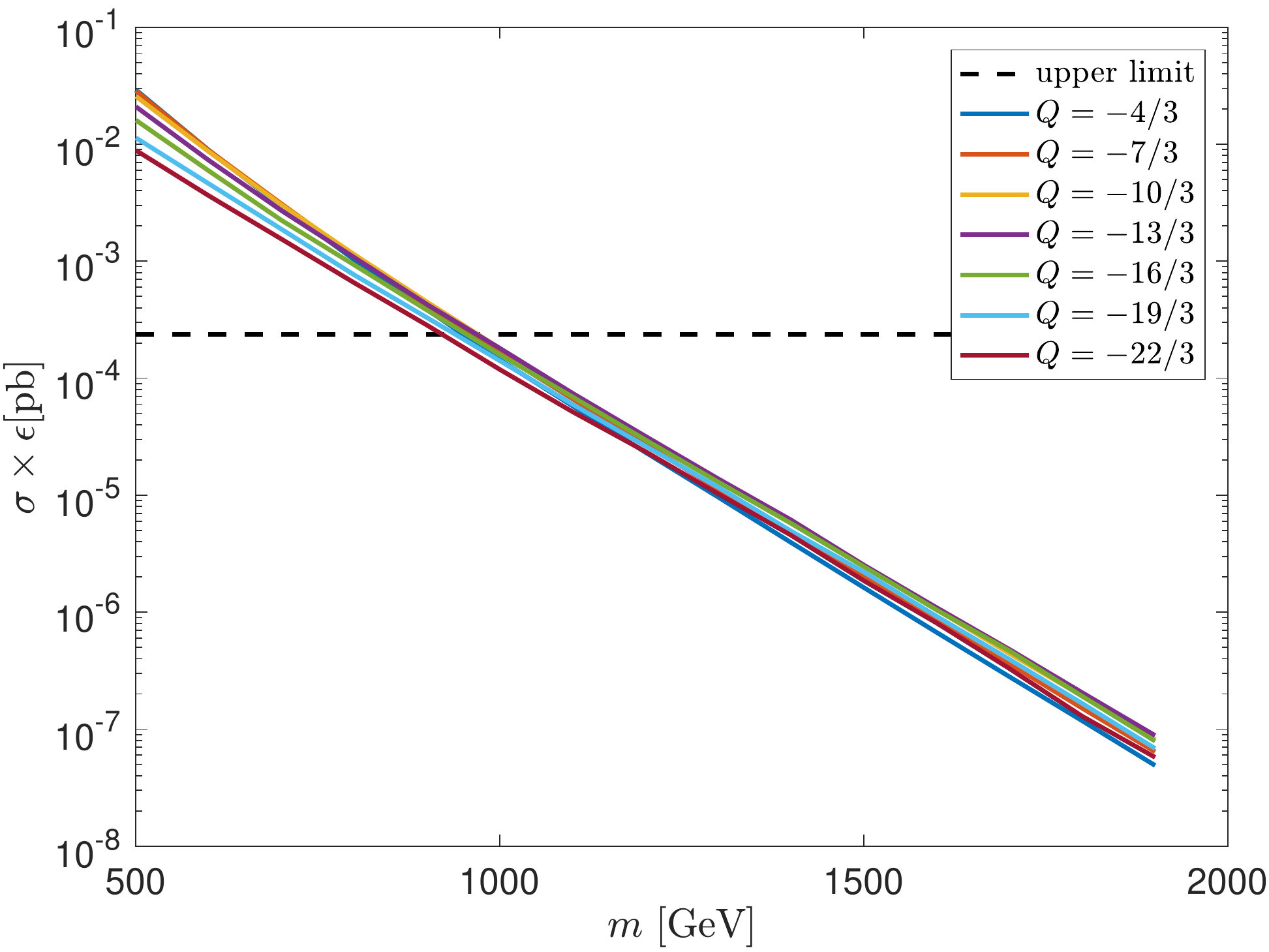}}
	\subfigure[Positvely-charged colored fermions.]{\includegraphics[scale=0.34]{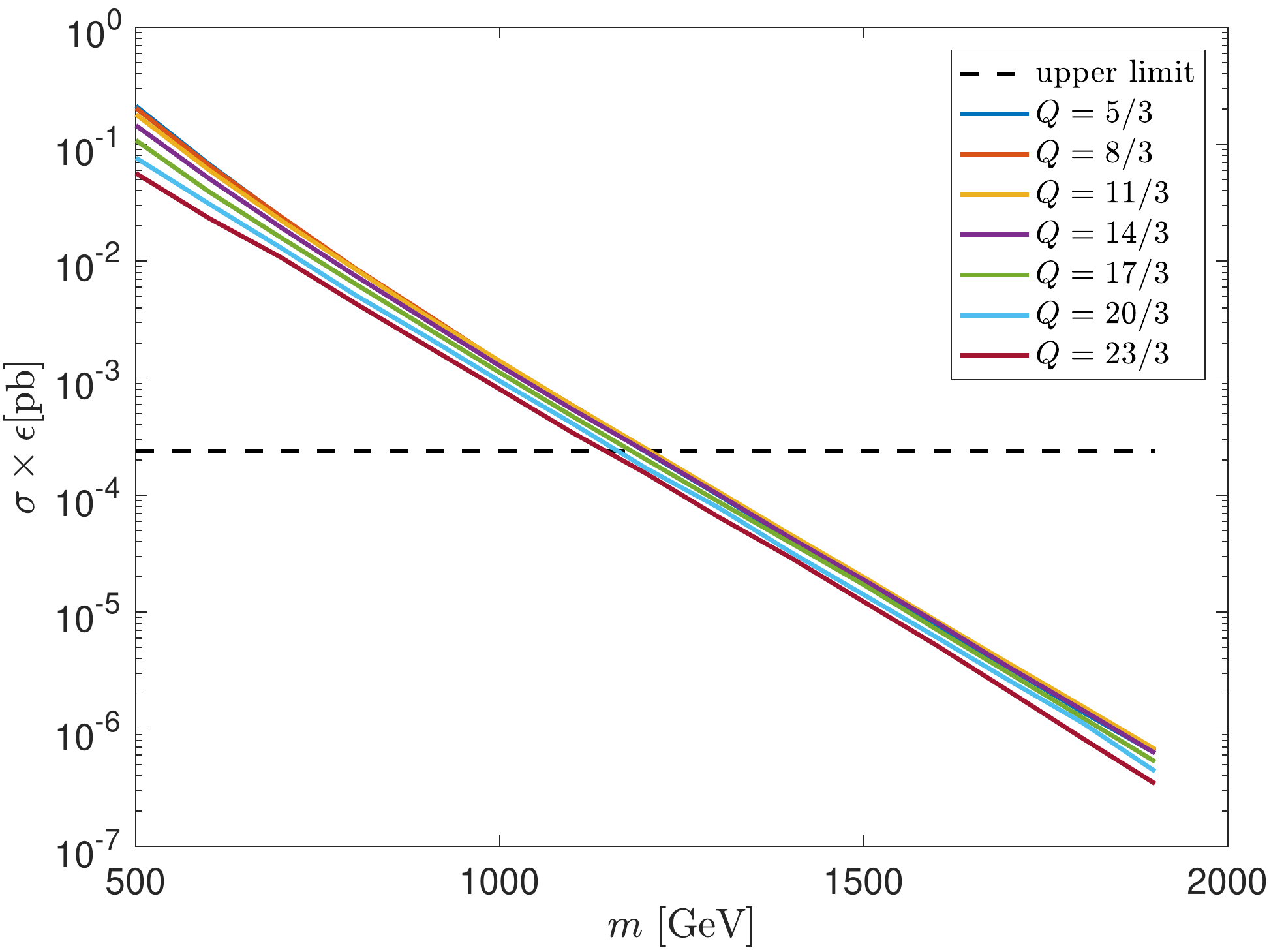}}
	\subfigure[Negatively-charged colored fermions.]{\includegraphics[scale=0.34]{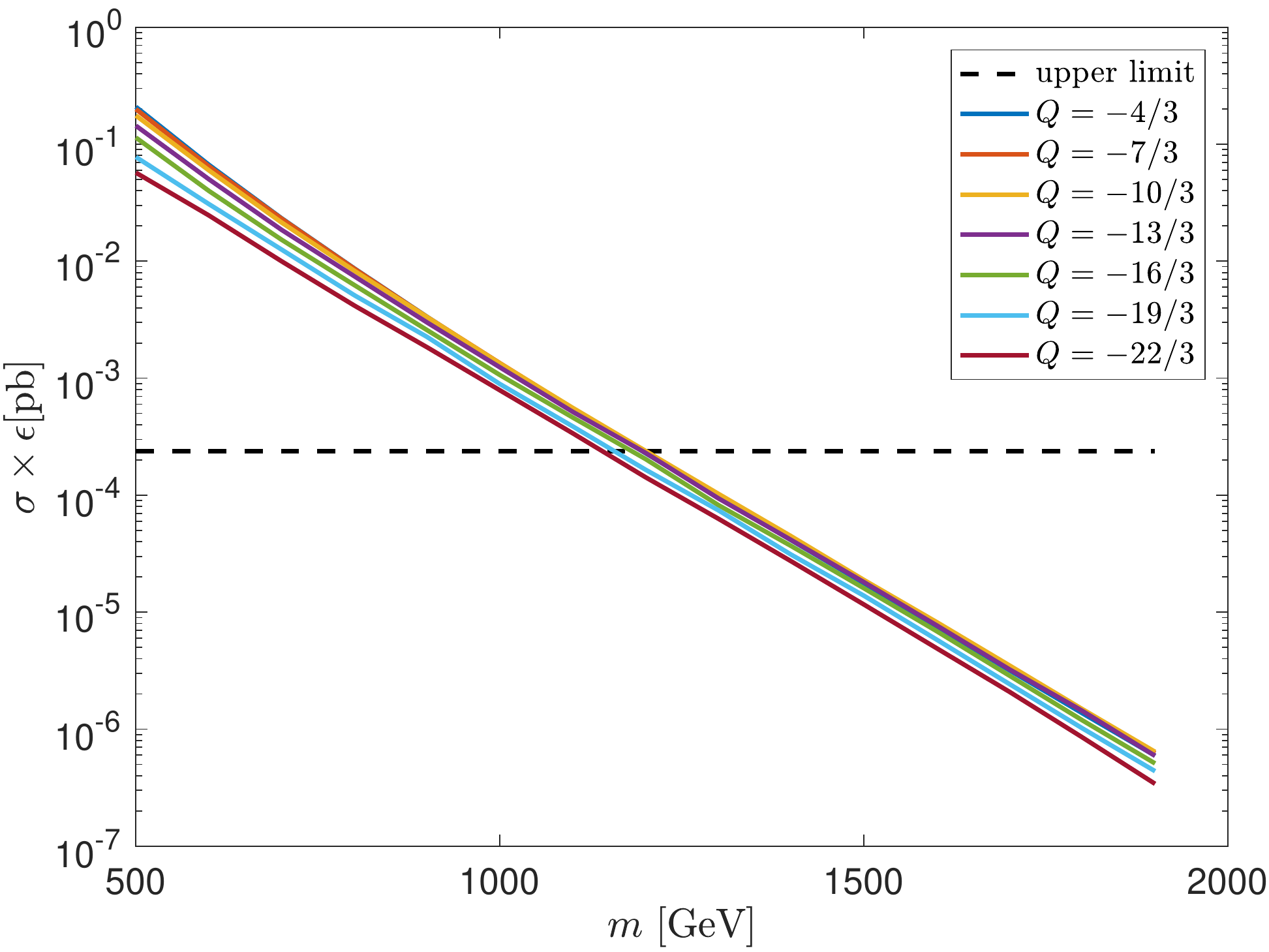}}
	\subfigure[Colorless fermions.]{	\includegraphics[scale=0.34]{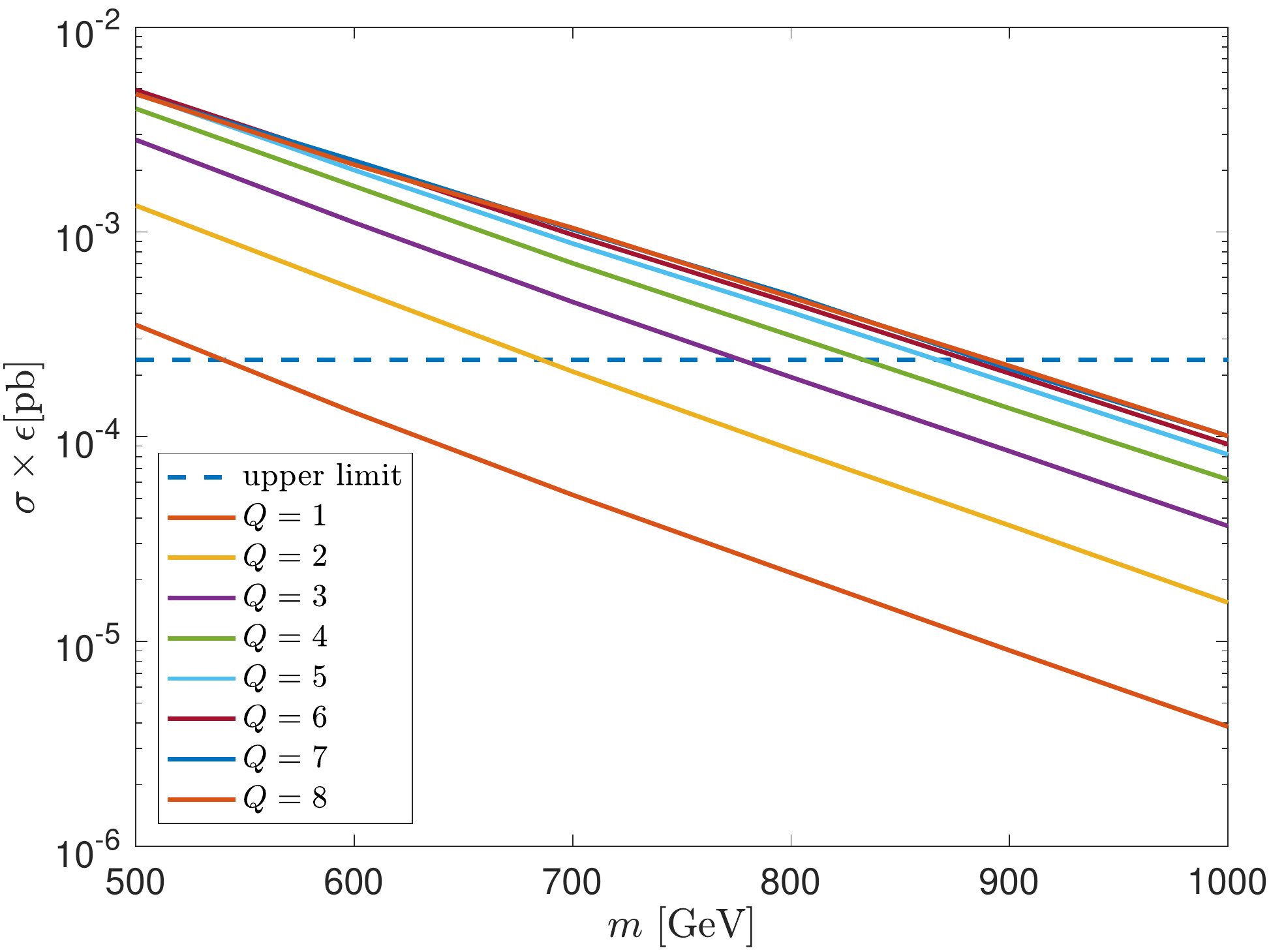}}
	\caption{Open-production channel signatures. Effective cross sections $\sigma\cdot \epsilon$ for CMS search~\cite{StableCMS7and8} at $\sqrt{s}=8$~TeV, together with the observed upper bound. \textit{Solid} -- theoretical effective cross sections, \textit{dashed} -- observed limit.}
	\label{effective_cross_section_8}
\end{figure}

\begin{figure}
	\centering
	\subfigure[Positvely-charged colored scalars.]{\includegraphics[scale=0.34]{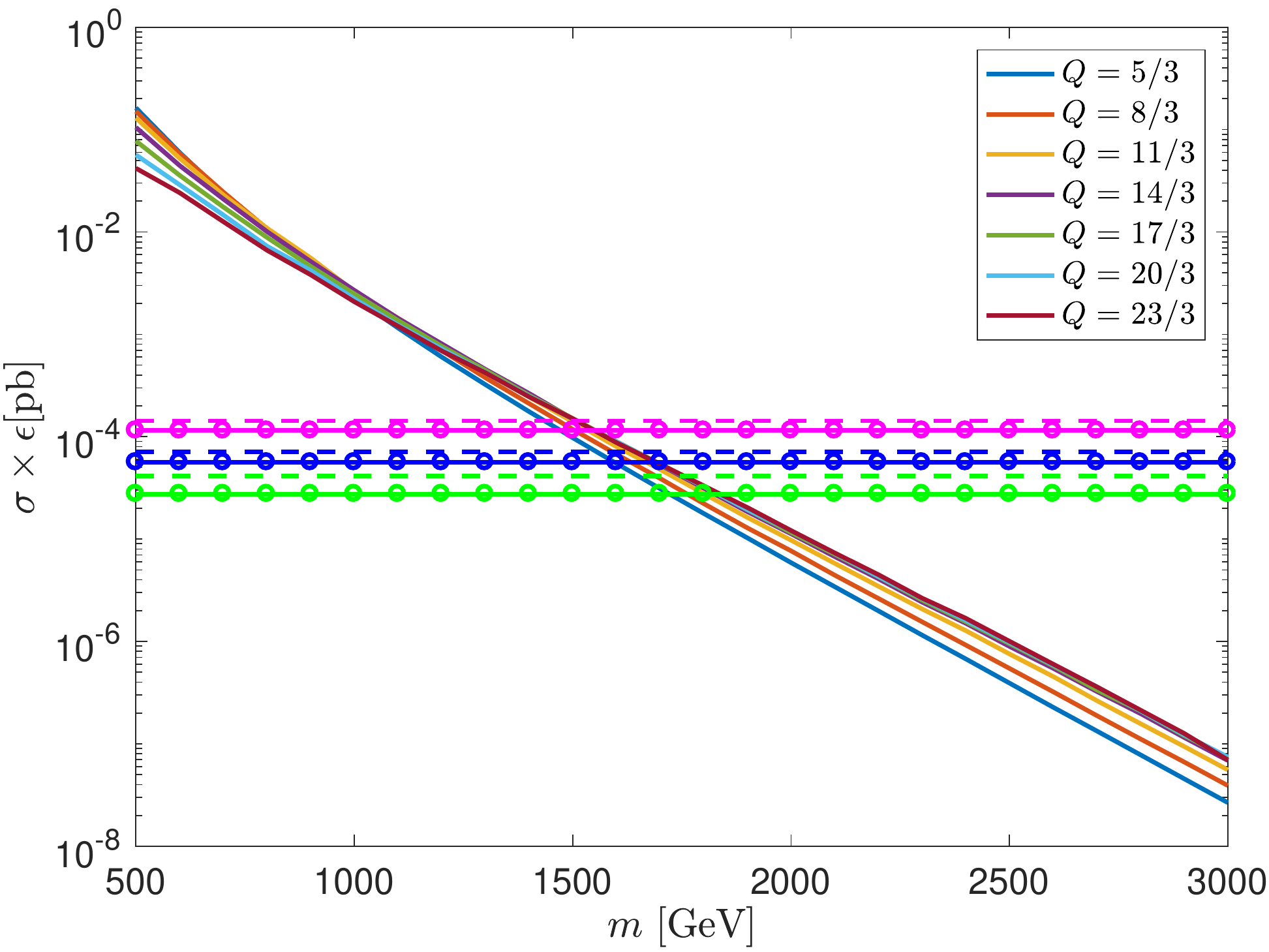}}
	\subfigure[Negatively-charged colored scalars.]{\includegraphics[scale=0.34]{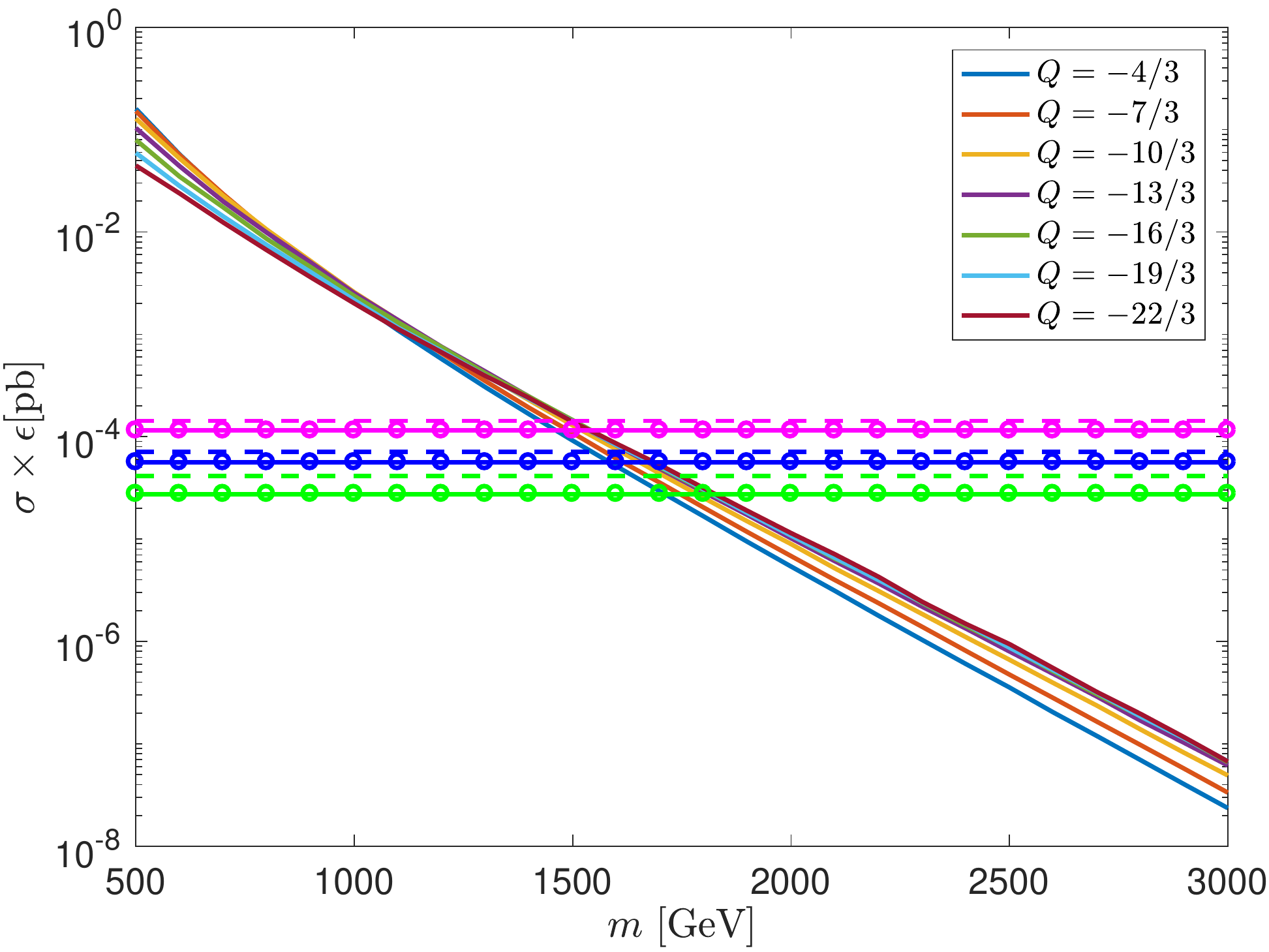}}
	\subfigure[Positvely-charged colored fermions.]{\includegraphics[scale=0.34]{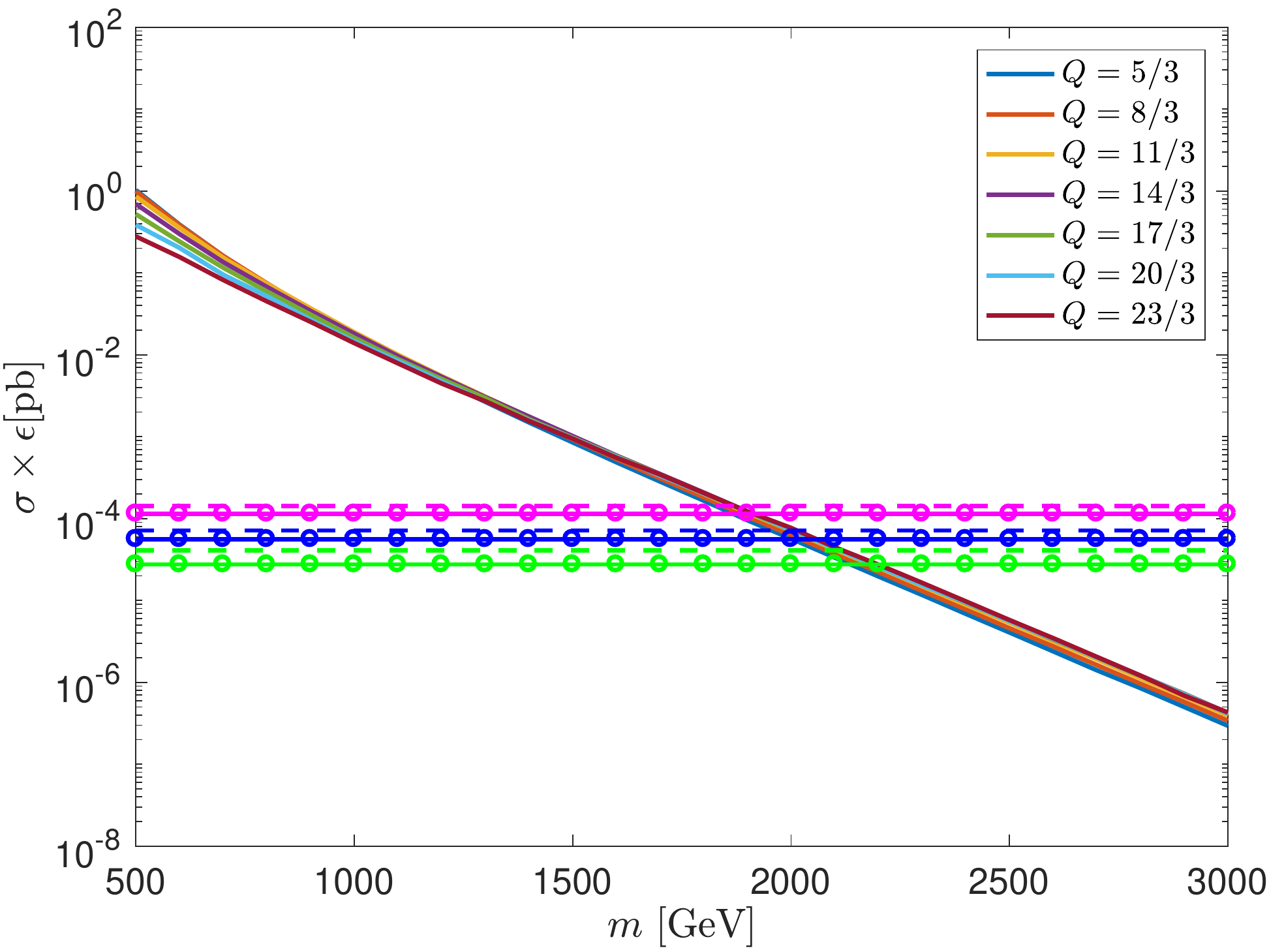}}
	\subfigure[Negatively-charged colored fermions.]{\includegraphics[scale=0.34]{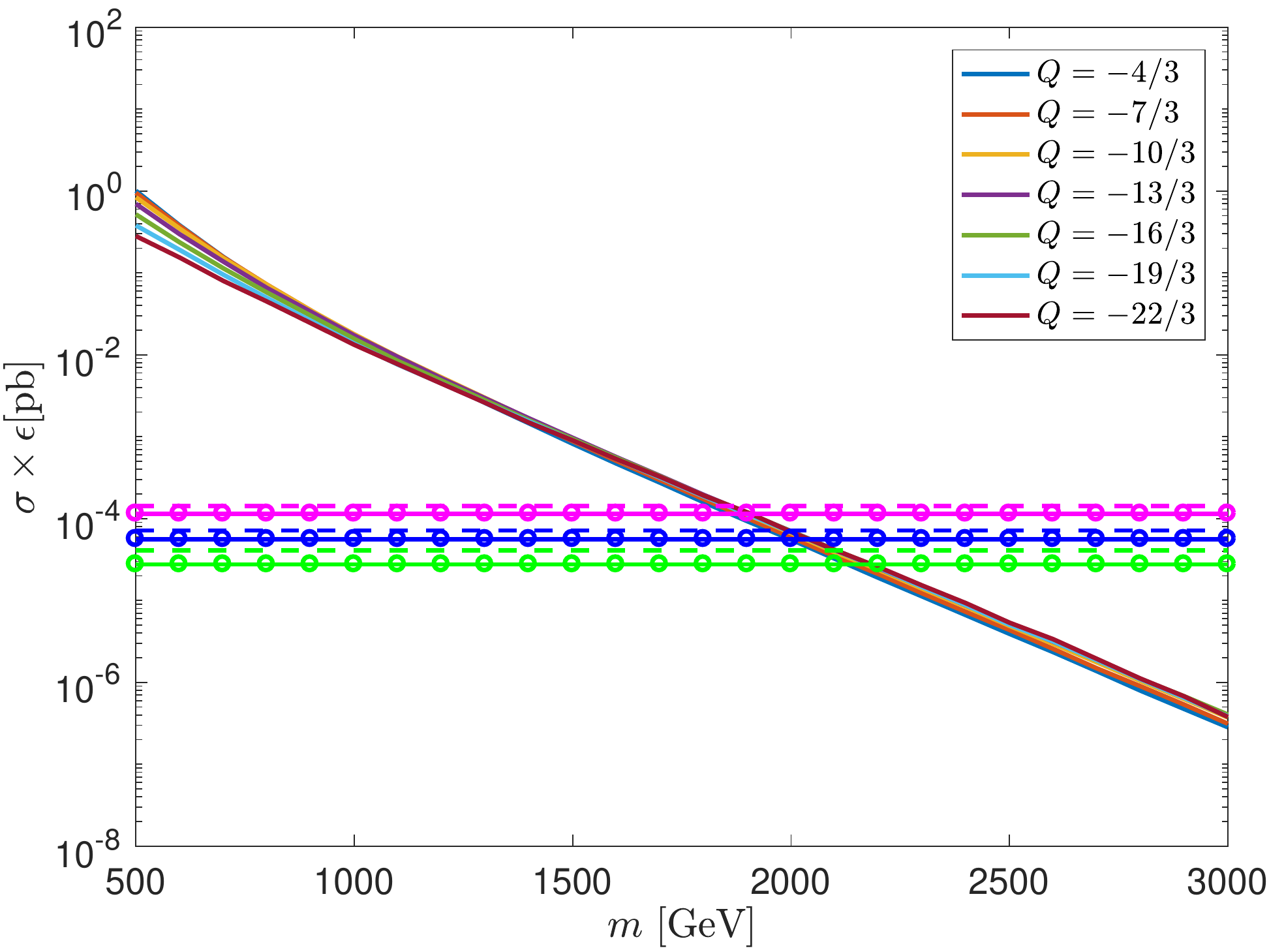}}
	\subfigure[Colorless fermions.]{	\includegraphics[scale=0.34]{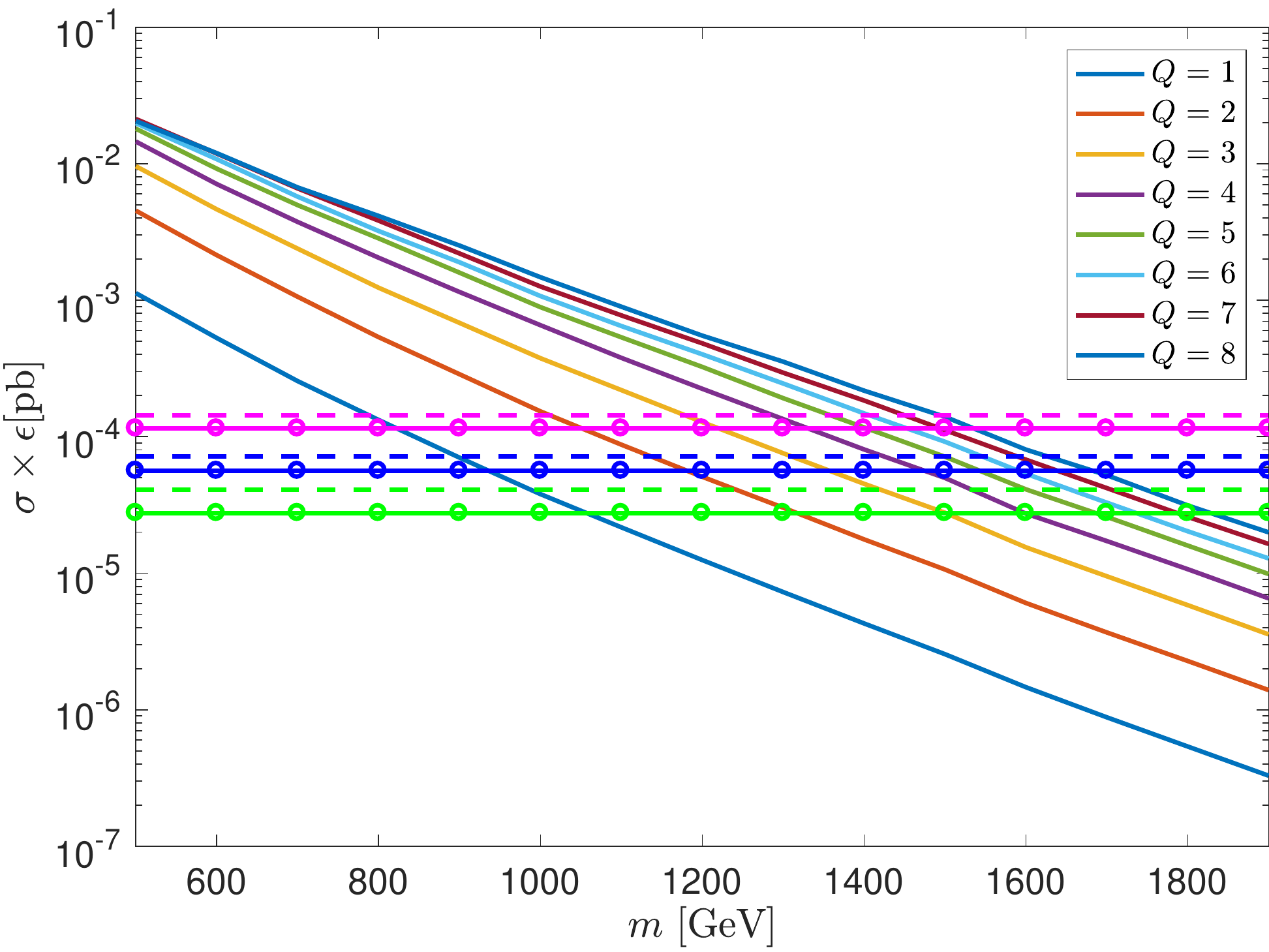}}
	\caption{Open-production channel signatures. Effective cross sections $\sigma\cdot \epsilon$ for future CMS searches at $\sqrt{s}=13$~TeV, together with expected upper bounds. \textit{Solid} -- theoretical effective cross sections. \textit{Round markers} -- luminosity scaling. \textit{Dashed} -- luminosity scaling and pileup scaling. \textit{Magenta} -- $\mathcal{L}=35.9$~$\text{fb}^{-1}$, \textit{blue} --$\mathcal{L}=100$~$\text{fb}^{-1}$ , \textit{green} -- $\mathcal{L}=300$~$\text{fb}^{-1}$. }
	\label{effective_cross_section_13}
\end{figure}

The effective cross sections for \acp{MCHSP}, obtained as a product of the cross sections and the efficiencies corresponding to open-production searches, are presented together with the observed upper limit for $\sqrt{s}=8$~TeV, and the projected upper limits for $\sqrt{s}=13$~TeV, in Figs.~\ref{effective_cross_section_8} and~\ref{effective_cross_section_13}.

\section{Closed-Production Signatures -- Diphoton Cross Sections}\label{diphoton_plots}
\begin{figure}[H]
	\centering
	\subfigure[Positvely-charged colored scalars. ]{\includegraphics[scale=0.34]{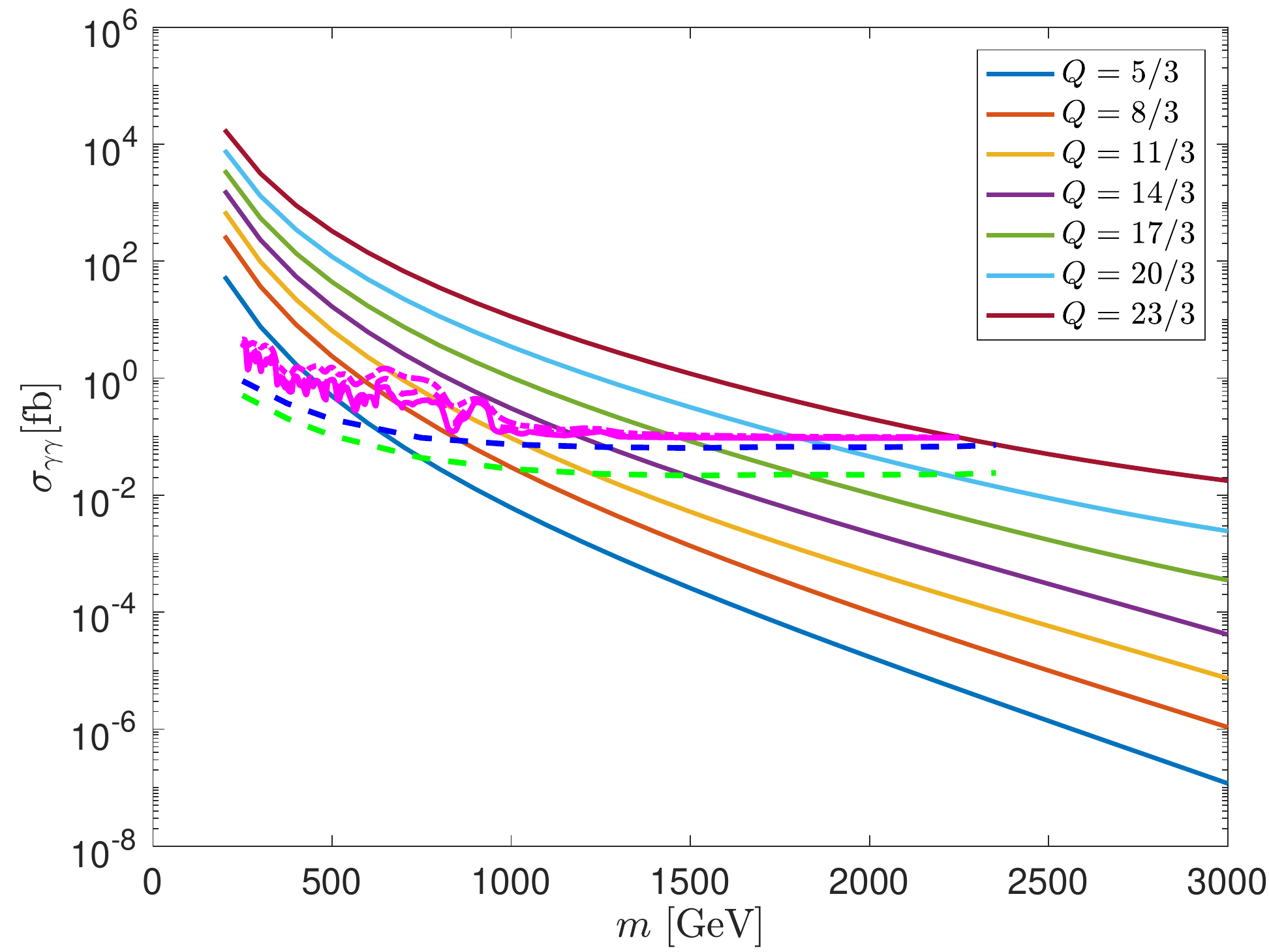}}
	\subfigure[Negatively-charged colored scalars.]{\includegraphics[scale=0.34]{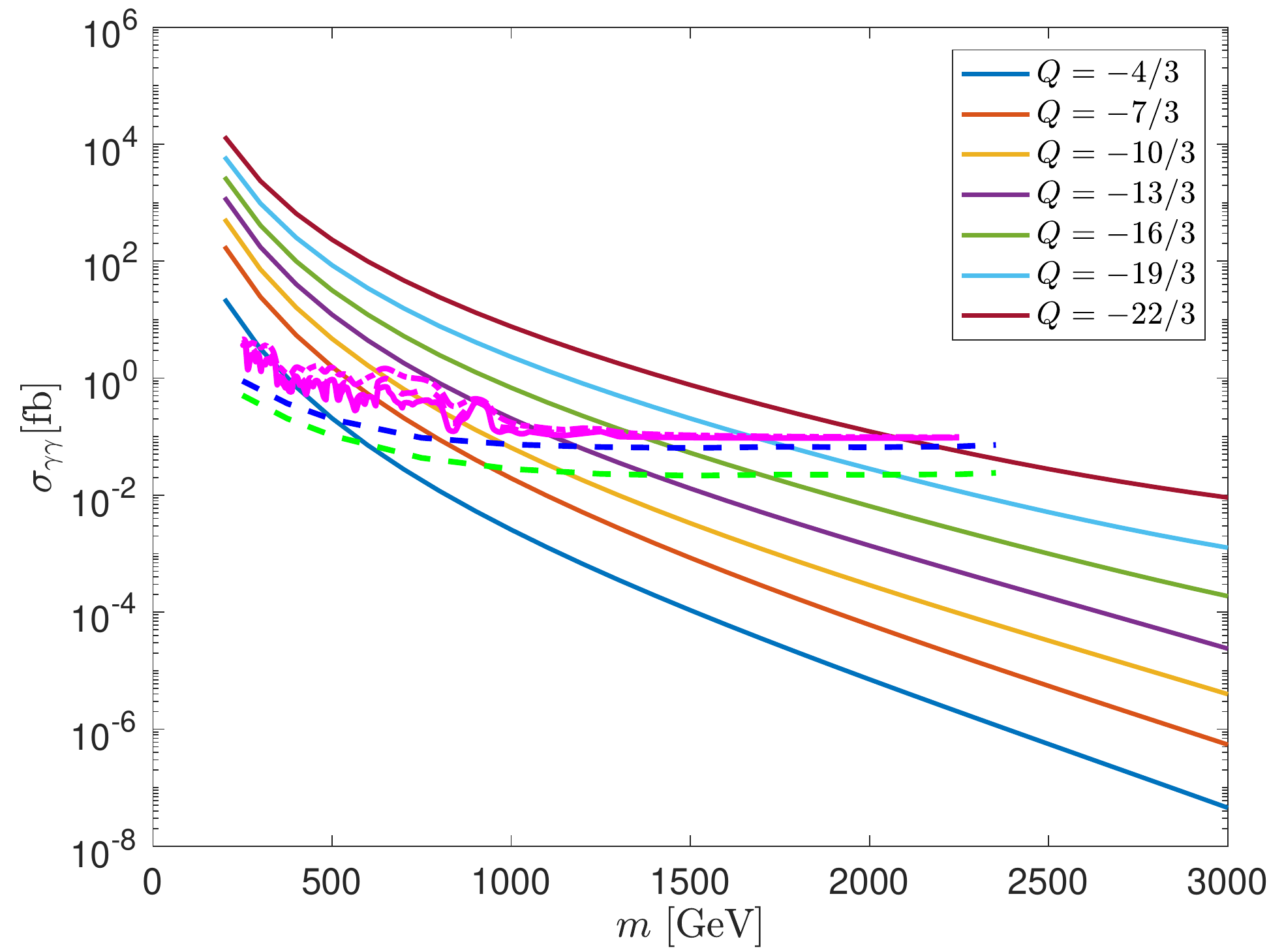}}
	\subfigure[Positvely-charged colored fermions.]{\includegraphics[scale=0.34]{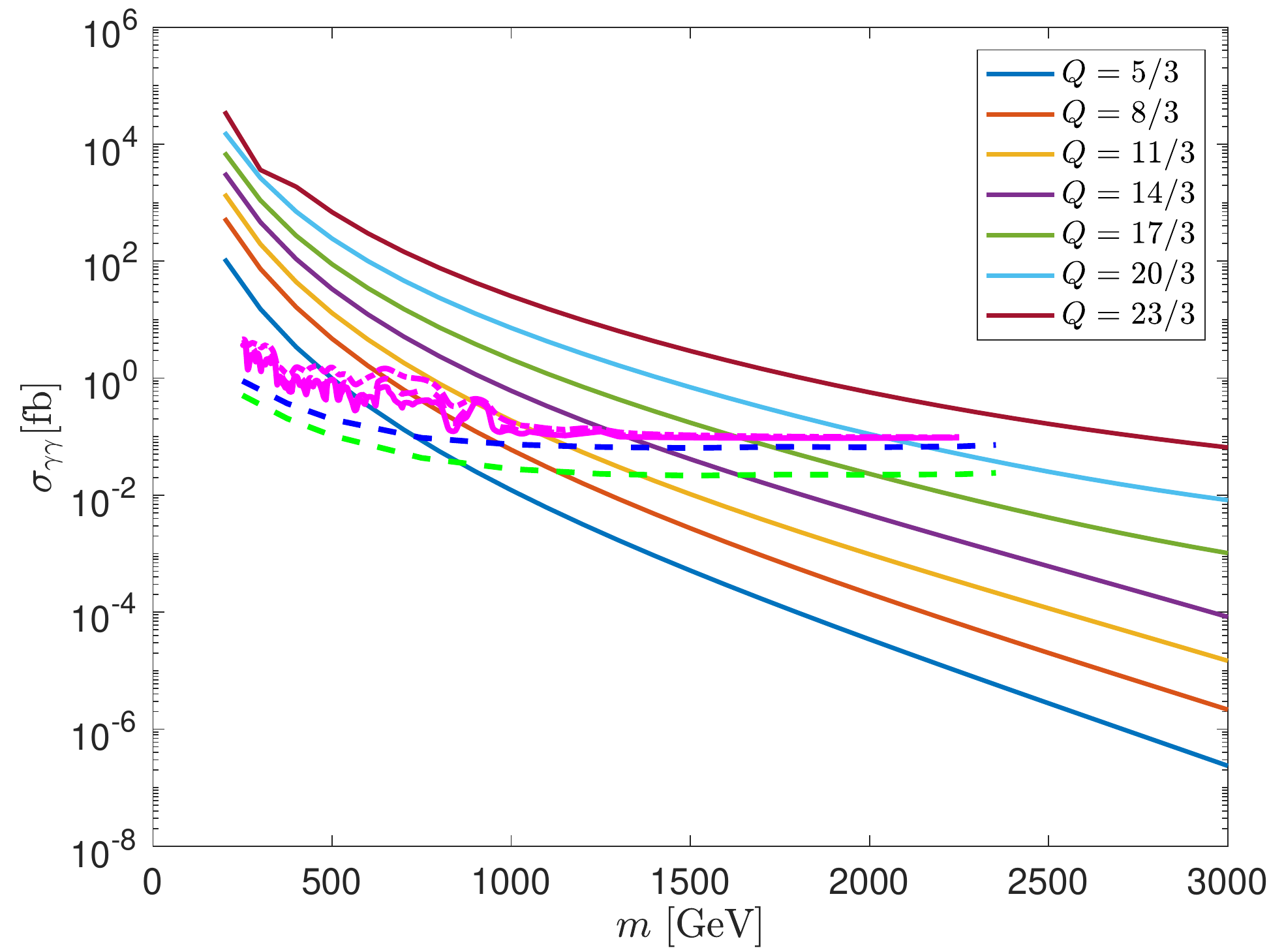}}
	\subfigure[Negatively-charged colored fermions.]{\includegraphics[scale=0.34]{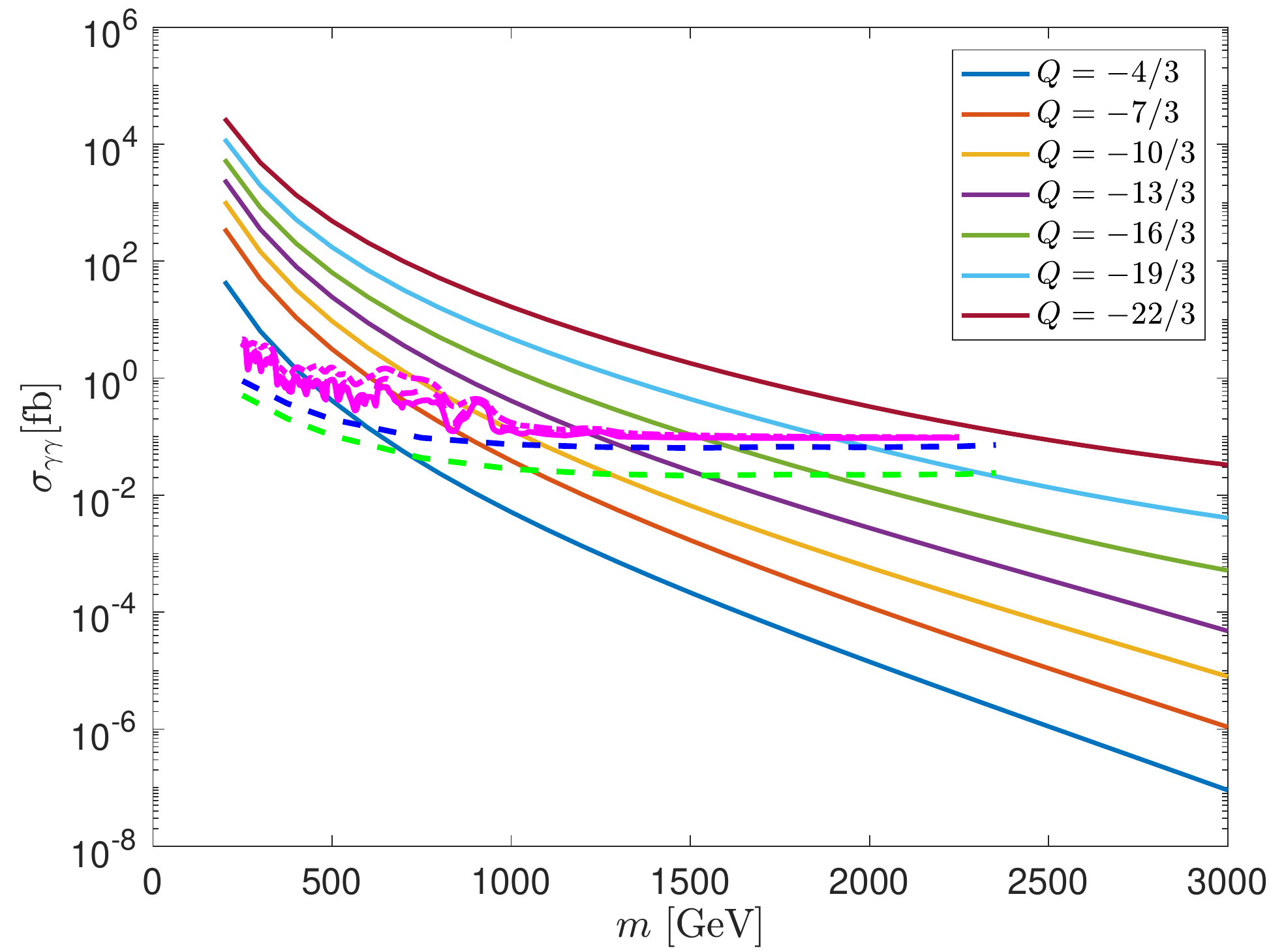}}
	\subfigure[Colorless fermions.]{\includegraphics[scale=0.34]{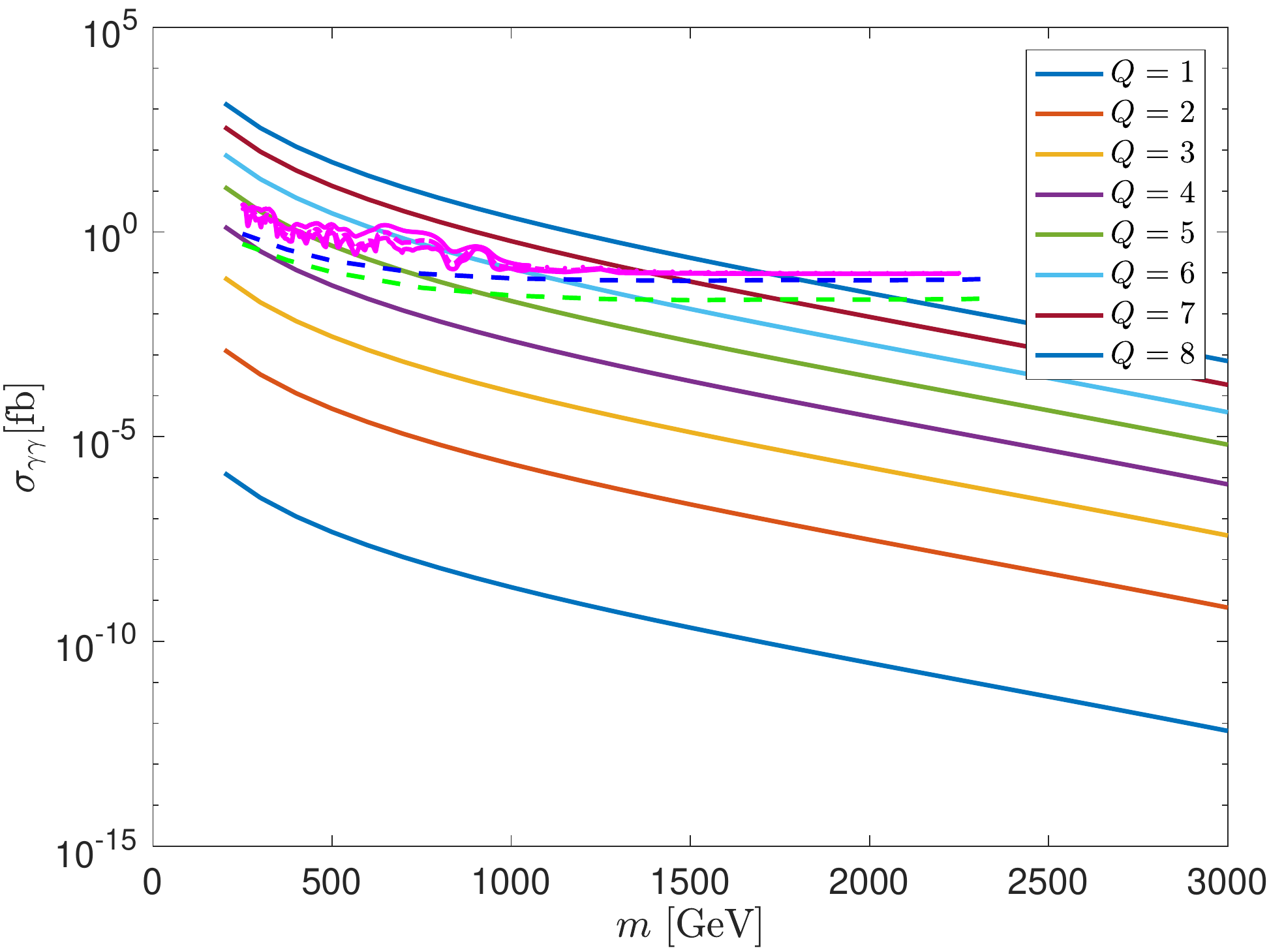}}
	\caption{Diphoton resonant production cross sections, given by a bound state of mass $2m$ at $\sqrt{s}=13$~TeV. \textit{Magenta} -- upper-limits observed at $\mathcal{L}=35.9$~$\text{fb}^{-1}$~\cite{diphotoncms13}, (\textit{solid} -- narrow , \textit{dashed} -- mid-width, \textit{dash-dotted} -- wide). \textit{Dashed blue} -- upper limits expected at $\mathcal{L}=100$~$\text{fb}^{-1}$~\cite{DiphotonProjection}. \textit{Dashed green} -- upper-limits expected at $\mathcal{L}=300$~$\text{fb}^{-1}$~\cite{DiphotonProjection}. }
	\label{cross_diphoton_plots}
\end{figure}

The diphoton production cross sections, from a bound state resonance, with observed and future-projected upper limits at $\sqrt{s}=13$~TeV are presented in Fig.~\ref{cross_diphoton_plots}.


\end{document}